\documentclass[showpacs,aps,superscriptaddress,letterpaper,nofootinbib
]{revtex4}

\usepackage{graphicx}
\usepackage{epsfig}
\usepackage{float}

\newcommand{\be}{\begin{equation}}
\newcommand{\ee}{\end{equation}}
\newcommand{\ba}{\begin{eqnarray}}
\newcommand{\ea}{\end{eqnarray}}

\def\vec#1{{\mbox{\boldmath$#1$}}}

\newcommand{\ep}{\epsilon}

\def\sss{\scriptscriptstyle}

\begin{document}


\vspace{0.6cm}

\title{ Spin determination of single-produced resonances at hadron colliders}
\author{Yanyan Gao \thanks{e-mail:  ygao@fnal.gov}}
\affiliation{Department of Physics and  Astronomy, Johns Hopkins University, Baltimore, MD, USA}
\affiliation{Fermi National Accelerator Laboratory (FNAL), Batavia, IL, USA}
\author{Andrei V. Gritsan \thanks{e-mail:  gritsan@pha.jhu.edu}}
\author{Zijin Guo \thanks{e-mail:  guozj@pha.jhu.edu}}
\author{Kirill Melnikov \thanks{e-mail:  melnikov@pha.jhu.edu}}
\author{Markus Schulze \thanks{e-mail:  schulze@pha.jhu.edu}}
\author{Nhan V. Tran  \thanks{e-mail:  ntran@pha.jhu.edu}}
\affiliation{Department of Physics and  Astronomy, Johns Hopkins University, Baltimore, MD, USA}

\date{submitted on January 19, 2010; revised on March 12, 2010}

\begin{abstract}
\vspace{2mm}
We study the production of a single resonance at the LHC and its decay into a pair of $Z$ bosons. 
We demonstrate how  full reconstruction of the final states allows us to determine the spin and parity of  
the resonance and restricts its coupling to vector gauge bosons. Full angular 
analysis is illustrated with the simulation of the production and decay chain including all spin 
correlations and the most general couplings of spin-zero, -one, and -two resonances to Standard Model 
matter and gauge fields.  We note implications for analysis of a resonance decaying to other final
states.
\end{abstract}

\pacs{12.60.-i, 13.88.+e, 14.80.Bn}

\maketitle

\thispagestyle{empty}

\section{Introduction}
\label{sect1}

Physics beyond the Standard Model (SM), to be probed at the LHC~\cite{lhc-paper, atlas-paper, cms-paper}, 
will manifest itself through observations of new 
particles. Such observations are  instrumental for establishing the 
existence of New Physics, though more effort is required to 
understand these observations in detail. It will be crucial to determine  the
quantum numbers of the new particles, their masses, and their 
couplings to SM fields as accurately as possible.

Measuring masses, coupling constants, and quantum numbers 
at a hadron collider is difficult, though many techniques for doing so 
were put forward recently.  Some of those   techniques evolved 
remarkably over time. For example, top quark mass determinations  
at the Tevatron~\cite{top} started out from measurements 
of the $t \bar t$ production cross-section 
and establishing the value of $m_t$ which fits the cross-section best.  
A more recent technique -- ``the matrix element method'' --  performs a  
likelihood fit  on an event-by-event basis. Since more information about 
the event is used,  more efficient separation of signal and background 
is accomplished  and a higher accuracy of the top quark mass measurement 
is achieved.
 
The idea that matrix elements, or multivariate per-event likelihoods, 
can guide us in maximizing the amount of information that can be extracted 
from a given event is appealing; 
but, to the best of our knowledge, it has not been widely used in  
hadron collider physics beyond top mass determinations.
On the other hand, these techniques are not new to experimental analyses 
since they were used in many $B$-physics measurements~\cite{polarization}.

The goal of this paper is to apply the multivariate likelihood method to the 
determination of a spin of a resonance, produced in hadron collisions. 
Many extensions of the SM postulate the existence of (elementary) particles 
of different spins that can be single-produced at the LHC.  
Once produced, these  resonances decay into SM particles whose angular 
distributions contain information about couplings, spins, and other quantum 
numbers of their parents.

Spin determination is  often discussed in the context of a particular angular distribution; 
the challenge is  to find a distribution that exhibits maximal sensitivity to the spin of 
a resonance. Single-observable distributions may be viable spin-analyzers but, 
as we illustrate with some examples in this paper,  loss of statistical power and 
certain information is inevitable.  Construction of the likelihood of the 
hypothesis that a given event, with its complete kinematic dependence, 
comes from the production and decay of a resonance with a particular spin
is the most efficient way to analyze the events. Testing this approach
in a realistic hadron collider setting is what we would like to do in this paper. 

It is best to pursue this program in a situation where the final state 
is reconstructed fully and accurately. For this reason we exclude the final states 
with missing energy and jets and examine the pure leptonic final  states.  
It follows that we can either consider direct decays of resonances to a lepton pair
or we can look at the decays of such resonances into neutral gauge bosons that 
subsequently decay into leptons. 
There are three reasons for us to choose the second option.
First, more information can be extracted from a fully-reconstructed 
four-body final state~\cite{soni, zerwas, bij, spin1, antipin, hagiwara, cao}
than from a two-body final state; 
second, direct decays to $l^+l^-$ are studied well in the literature\footnote{In 
the appendix we present angular distributions for $X \to l^+l^-$ that generalize 
results in the literature.}\cite{rosner1996, Davoudiasl2000, Allanach2000, 
gtopp, Cousins2005, Osland2008};
and, third, it is reasonable to assume that the
decay of a single-produced resonance
to $Z$ bosons is sizable, if not altogether dominant.
Recall that  this happens with the SM Higgs boson if its mass exceeds  $2 m_{\sss Z}$~\cite{higgs}.
It may also occur in well-motivated scenarios 
of Beyond the Standard Model (BSM) physics. 
For example,  in the extra-dimensional model~\cite{rsgraviton}
discussed in Refs.~\cite{fitzpatrick, agashe, atwood} 
KK graviton decays into pairs of gauge bosons 
are enhanced relative to direct decays into leptons.
Similar situations may occur in ``hidden-valley''-type models~\cite{Juknevich:2009ji}.
An example of a "heavy photon" is given in Ref.~\cite{Barger:2007}.

\begin{figure}[t!]
\centerline{
\setlength{\epsfxsize}{0.5\linewidth}\leavevmode\epsfbox{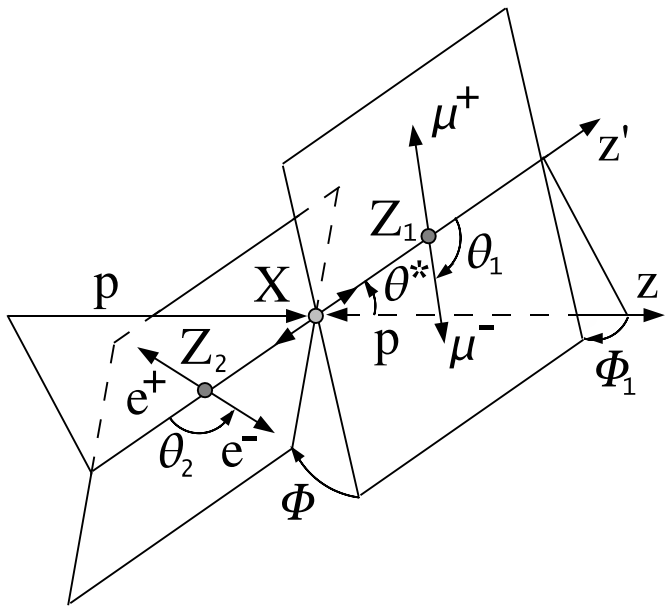}
}
\caption{
Illustration of an exotic $X$ particle production and decay 
in $pp$ collision
$gg$ or $q\bar{q}\to X\to ZZ\to 4l^\pm$. Six angles fully characterize
orientation of the decay chain:  
$\theta^*$ and $\Phi^*$ of the first $Z$ boson in the $X$ rest frame,
two azimuthal angles $\Phi$ and $\Phi_1$ between the three planes defined
in the $X$ rest frame, and two $Z$-boson helicity angles
$\theta_1$ and $\theta_2$ defined in the corresponding $Z$ rest frames.
The offset of angle $\Phi^*$ is arbitrarily defined and therefore 
this angle is not shown.
}
\label{fig:decay}
\end{figure}

Motivated by this, we consider  the production of a resonance $X$ at the LHC  
in gluon-gluon and quark-antiquark partonic collisions, with the subsequent 
decay of $X$ into two $Z$ bosons which, in turn, decay leptonically.
In Fig.~\ref{fig:decay}, we show  the decay chain $X\to ZZ \to e^+e^-\mu^+\mu^-$. 
However, our analysis is equally applicable to any combination of decays 
$Z\to e^+e^-$ or $\mu^+\mu^-$. It may also be applicable to $Z$ decays into 
$\tau$ leptons since  $\tau$'s from $Z$ decays will often be highly boosted and 
their decay products collimated.  
We study how the spin and parity of $X$, as well as information on its  
production and decay mechanisms, can be extracted from angular 
distributions of four leptons in the final state.

There are a few things that need to be noted. First, we obviously 
assume that the resonance production and its decays into four leptons 
are observed. Note that, because of a relatively small branching fraction 
for leptonic $Z$ decays, this assumption implies  
a fairly large production cross-section for $pp \to X$ and 
a fairly large branching fraction for the decay $X \to ZZ$. 
As we already mentioned, there are well-motivated scenarios of BSM physics 
where those requirements are satisfied.  

Second, having no bias towards any particular model of BSM  physics, 
we consider the most general couplings 
of the particle $X$ to relevant SM fields.
This approach has to be contrasted with typical  studies of 
e.g. spin-two particles at hadron colliders where such an exotic 
particle is often  identified with a  massive graviton that 
couples to SM fields through the energy-momentum tensor.
We will refer to this case as the ``minimal coupling'' of 
the spin-two particle to SM fields. 

The minimal coupling scenarios are well-motivated within 
particular models of New Physics, but they are not sufficiently general. 
For example, such a minimal coupling may restrict partial waves that 
contribute to the production and decay of a spin-two particle. 
Removing such restriction opens an interesting possibility to understand
the couplings of a particle $X$ to SM fields by means of partial 
wave analyses, and we would like to set a stage for doing that in this  paper. 
To pursue this idea in detail, the most general parameterization 
of the $X$ coupling to SM fields is required. 
Such parameterizations are known for spin-zero, spin-one, and spin-two 
particles interacting  with the SM gauge bosons~\cite{bij, zerwas}
and we use these parameterizations in this paper.
We also note that the model recently discussed in Refs.~\cite{fitzpatrick, agashe, atwood}
requires couplings beyond the minimal case in order to produce 
longitudinal polarization dominance.

Third, we note that while we concentrate on the decay $X\to ZZ \to l_1^+ l_1^- l_2^+ l_2^-$,
the technique discussed in this paper is more general  and can, in principle,  
be applied to final states with jets and/or missing energy by studying such processes  
as $X \to ZZ \to l^+l^- jj$, $X\to W^+W^- \to l^+ \nu jj$, etc.  
In contrast with pure leptonic final states, higher statistics, larger 
backgrounds, and a worse angular  resolution must be expected once final 
states with jets and missing energy are included.  We plan to perform detailed 
studies of these, more complicated final states, in the future. However, we note
that many results in this paper are applicable to these final states as well.

The remainder of the paper is organized as follows. 
In Section~\ref{sect2}, we describe the parameterization of 
production and decay amplitudes that is employed in our analyses.  
In Section~\ref{sect3}, we calculate helicity amplitudes for the decay of a resonance 
into a pair of gauge bosons or into a fermion-antifermion pair; 
helicity amplitudes for resonance production  are obtained by crossing.  
In Section~\ref{sect4},  angular distributions for 
$pp \to X \to ZZ \to f_1 \bar f_1 f_2 \bar f_2$
for resonances with spins zero, one, and two  are presented.
This is followed by detailed Monte Carlo simulation
which includes all spin correlations and main experimental effects
and which is shown in Section~\ref{sect5}.
Analysis using the multivariate maximum likelihood technique is applied 
to several key scenarios to illustrate separation power 
of different helicity amplitudes for all spin hypotheses 
and in both production and decay, as discussed in Section~\ref{sect6}.
For completeness, angular distributions, including distributions for other
decay channels, are given in the appendix.


\section{Interactions of an exotic particle with Standard Model fields}
\label{sect2}

In this section, the interaction of  a color- and charge-neutral 
exotic particle $X$ with two spin-one bosons $V$
(such as gluons, photons, $Z$, or $W$ bosons) 
or a fermion-antifermion pair (such as leptons or quarks) is summarized. 
The spin of $X$ can be zero, one, or two. We construct the most general 
amplitudes consistent with Lorentz invariance and Bose-symmetry,
as well as gauge-invariance with respect to unbroken subgroups 
of $SU(3) \times SU(2)_L \times U(1)_R$ of the SM.

The four-momentum of the particle $X$ is denoted by $q$ and the four-momenta 
of the gauge bosons or fermions by $q_{1,2}$. 
The polarization vectors of gauge bosons are denoted by $\epsilon_{1,2}$; we 
assume them to be transverse $q_i \epsilon_i = 0$. Fermion 
wave functions are conventional Dirac spinors.
We employ the field strength tensor of a gauge boson 
with momentum $q_i$ and polarization vector $\epsilon_i$ as 
$f^{(i),{\mu \nu}} = \epsilon_i^{\mu}q_i^{\nu} - \epsilon_{i}^\nu q_i^{\mu} $,
and  the conjugate field strength tensor as
${\tilde f}^{(i)}_{\mu \nu}  = 
1/2\; \epsilon_{\mu \nu \alpha \beta} f^{(i),\alpha \beta} 
= \epsilon_{\mu \nu \alpha \beta} \epsilon_i^{\alpha} q_i^{\beta}$. 
We use ${\tilde q} = q_1 - q_2$ to denote the particular combination of the momenta
of the two final state particles.

\subsection{Spin-zero $X$ and  two gauge bosons} 

The invariant amplitude that describes the interaction between a spin-zero 
particle $X$ of arbitrary parity 
and two spin-one gauge bosons reads 
\begin{equation}
A(X \to VV) = v^{-1} \left ( 
  g^{(0)}_{\sss 1} m_{\sss V}^2 \epsilon_1^* \epsilon_2^* 
+ g^{(0)}_{\sss 2} f_{\mu \nu}^{*(1)}f^{*(2),\mu \nu}
+ g^{(0)}_{\sss 3} f^{*(1),\mu \nu} 
f^{*(2)}_{\mu \alpha}\frac{ q_{\nu} q^{\alpha}}{\Lambda^2} 
+ g^{(0)}_{\sss 4}  f^{*(1)}_{\mu \nu} {\tilde f}^{*(2),\mu  \nu}
\right ).
\label{eq1}
\end{equation}
In Eq.~(\ref{eq1}), $f^*$ denotes the complex conjugate field strength tensor, 
$v$ is the SM vacuum expectation value of the Higgs field, 
and $\Lambda$ is the mass scale associated with BSM physics. 
The ``couplings'' $g^{(0)}_{\sss 1,..,4}$ are invariant form-factors; 
the upper index  reflects the $X$ spin.
Since we consider on-shell decays of the particle $X$ to two 
on-shell gauge-bosons, $g^{(i)}_{ j}$ can be thought of as effective 
dimensionless coupling constants which can, in general, be complex.

We note that, as written, Eq.~(\ref{eq1}) does not use  the
minimal set of independent variables since it uses {\it both}, 
the field strength tensors and polarization vectors for gauge 
bosons in the final state. However, we 
write Eq.~(\ref{eq1}) in that  particular way because 
it can be applied to $X$ decays into both massive and massless 
gauge bosons and because it has the simplest possible connection to 
SM couplings at tree level. Indeed, if we identify $X$ 
with the Higgs boson of the SM, the proper tree-level 
amplitude for $H \to ZZ$ 
is obtained by setting $g^{(0)}_{ j \sss > 1} = 0$ 
and  $g^{(0)}_{\sss 1}=2i$. 
To describe the coupling of the 
spin-zero particle to massless gauge bosons (gluons or photons), 
we simply set\footnote{For X coupling to two gluons, a trivial 
color factor needs to be introduced in Eq.~(\ref{eq1}).}
$m_{\sss V} = 0$ in Eq.~(\ref{eq1}). 
Clearly, the coefficients $g_{ j}^{(0), gg}$ for interaction with
gluons, for example, do not need to be equal to the coefficients 
for interaction with the Z bosons $g_{j}^{(0),\sss ZZ}$
or the photons $g_{j}^{(0),\gamma\gamma}$.
In fact, Eq.~(\ref{eq1}) is sufficiently general 
to accommodate all radiative corrections to Higgs interactions with 
gauge bosons, massive or massless, in the SM,
including $C\!P$-violating form  factors that appear at the 
three loop level~\cite{soni}.

In spite of the fact that there are four  form-factors required 
to describe the interaction of the spin-zero boson with two massive or 
massless spin-one bosons, there are only three 
independent structures in the 
scattering amplitude.  To see this, we rewrite  Eq.~(\ref{eq1}) 
through polarization vectors 
\begin{equation}
A(X \to VV)  = 
 v^{-1} \epsilon_{1}^{*\mu} \epsilon_{2}^{*\nu}
\left (
  a_{1} g_{\mu \nu} m_{\sss X}^2 
+ a_{2} \,q_\mu q_\nu 
+ a_{3} \epsilon_{\mu\nu\alpha\beta}\,q_1^\alpha q_2^\beta
\right),
\label{eq:ampl-spin0} 
\end{equation}
and find  the coefficients $a_{1,2,3}$ to be
\begin{equation}
a_{1} =  g^{(0)}_{\sss 1}\frac{m_{\sss V}^2}{m_{\sss X}^2} 
+ g^{(0)}_{\sss 2} \frac{2s}{m_{\sss X}^2} + g^{(0)}_{\sss 3} \kappa\frac{s}{m_{\sss X}^2}\,,
\;\;\;\;
a_{2} = -2 g^{(0)}_{\sss 2} - g^{(0)}_{\sss 3}\kappa\,,\;\;\;\; 
a_{3} = -2 g^{(0)}_{\sss 4}.
\label{eq13}
\end{equation}
We have defined the parameters
$s = q_1 q_2 = (m_{\sss X}^2 - 2m_{\sss V}^2)/2$ and $\kappa = s /\Lambda^2$. 
The amplitude for $X$ decay into two massless gauge bosons is obtained from 
Eqs.~(\ref{eq:ampl-spin0}) and~(\ref{eq13}) by setting $m_{\sss V}$  to zero.


\subsection{Spin-one $X$ and  two gauge bosons} 

We consider the case when the exotic particle $X$ has spin one and 
arbitrary parity. 
As a consequence of the Landau-Yang theorem,  the spin-one 
particle $X$ cannot interact with two massless identical 
gauge bosons.  
For this reason, a spin-one color-singlet particle cannot be produced
in gluon fusion, or decay to two photons.
The phenomenology of spin-one decays into two $Z$ bosons 
was recently discussed in Ref.~\cite{spin1}. Following that 
reference,  we  consider the amplitude for the decay 
to two identical massive gauge bosons $X \to ZZ$. This amplitude 
depends 
on  two independent form factors
%
\begin{eqnarray}
A(X \to ZZ)  = 
 g^{(1)}_{\sss 1} \left[  (\epsilon_1^* q) (\epsilon_2^* \epsilon_X) 
  +  (\epsilon_2^* q) (\epsilon_1^* \epsilon_X)  \right]
+g^{(1)}_{\sss 2} \epsilon_{\alpha\mu\nu\beta} \epsilon_X^{\alpha} 
\epsilon_1^{*,\mu} \epsilon_2^{*,\nu} {\tilde q}^\beta.
\label{eq:ampl-spin1} 
\end{eqnarray}
%
Similar to the spin-zero case, $g^{(1)}_{\sss 1}$ 
and $g^{(1)}_{\sss 2}$ are dimensionless 
effective coupling constants.
We note that these coupling constants are, in general, complex with 
absorptive parts  that may arise  from quantum loop effects.
This possibility was not considered in Ref.~\cite{spin1} 
where the case of zero complex phase difference between 
the two coupling constants was studied. 
In the case when $X$ has positive parity ($J^P=1^+$), the first term violates and the 
second term conserves parity. 
Alternatively, the two terms correspond to  parity-conserving and parity-violating
interactions of the $1^-$ particle, respectively.


\subsection{Spin-two $X$ and  two gauge bosons} 

We turn to the spin-two case and construct the most 
general amplitude for the decay of a spin-two particle $X$
into two identical vector gauge bosons.
The $X$  wave function is given by a symmetric
traceless tensor $t_{\mu \nu} $, transverse to its 
momentum $t_{\mu \nu} q^\nu = 0$. Since we would like to apply 
the formula for the amplitude to describe interactions of $X$ with 
massive and massless gauge bosons,  
we consider the possible dependence of the 
amplitude on both the  field strength 
tensor and the polarization vectors
%
\begin{eqnarray}
\label{eq:ampl-spin2-a} 
&&  A(X \to VV)  = \Lambda^{-1} \left [ 
2 g^{(2)}_{\sss 1} t_{\mu \nu} f^{*1,\mu \alpha} f^{*2,\nu \alpha} 
+ 2 g^{(2)}_{\sss 2} t_{\mu \nu} \frac{q_\alpha q_\beta }{\Lambda^2} 
f^{*1,\mu \alpha}  f^{*2,\nu,\beta}
\right.  \nonumber \\
&& \left. + g^{(2)}_{\sss 3} 
 \frac{{\tilde q}^\beta {\tilde q}^{\alpha}}{\Lambda^2}
t_{\beta \nu} ( 
f^{*1,\mu \nu} f^{*2}_{\mu \alpha} + f^{*2,\mu \nu} f^{*1}_{\mu \alpha} 
) + g^{(2)}_{\sss 4}\frac{{\tilde q}^{\nu} {\tilde q}^\mu}{{\Lambda^2} } 
t_{\mu \nu} f^{*1,\alpha \beta} f^{*(2)}_{\alpha \beta}
\right. \nonumber \\
&& \left. + m_{\sss V}^2  \left ( 
2 g^{(2)}_{\sss 5}  t_{\mu\nu} \ep_1^{*\mu} \ep_2^{*\nu} 
+2 g^{(2)}_{\sss 6} \frac{{\tilde q}^\mu q_\alpha}{\Lambda^2}  t_{\mu \nu}
\left ( \ep_1^{*\nu} \ep_2^{*\alpha} - 
\ep_1^{*\alpha} \ep_2^{*\nu} \right ) 
+g^{(2)}_{\sss 7}
 \frac{{\tilde q}^\mu {\tilde q}^\nu}{\Lambda^2}  
t_{\mu \nu} \ep^*_1 \ep^*_2
\right) 
\right.  \nonumber \\
&& \left. 
+g^{(2)}_{\sss 8} \frac{{\tilde q}_{\mu} {\tilde q}_{\nu}}{\Lambda^2} 
 t_{\mu \nu} f^{*1,\alpha \beta} {\tilde f}^{*(2)}_{\alpha \beta}
+ g^{(2)}_{\sss 9} t_{\mu \alpha} {\tilde q}^\alpha \epsilon_{\mu \nu \rho \sigma} \epsilon_1^{*\nu} 
\epsilon_2^{*\rho} q^{\sigma} 
+\frac{g^{(2)}_{\sss 10} t_{\mu \alpha} {\tilde q}^\alpha}{\Lambda^2}
\epsilon_{\mu \nu \rho \sigma} q^\rho {\tilde q}^{\sigma} 
\left ( \epsilon_1^{*\nu}(q\epsilon_2^*)+
\epsilon_2^{*\nu}(q\epsilon_1^*) \right )
\right ] \,.
\end{eqnarray}
%
As in the spin-zero and spin-one cases, 
$g^{(2)}_{\sss 1,..,10}$ are dimensionless effective coupling constants which are, in general, 
complex numbers. They are different for different gauge bosons $V$.
The first seven constants $g^{(2)}_{\sss 1,..,7}$ correspond to the $J^P=2^+$
particle parity-conserving interaction, while the last three terms
with $g^{(2)}_{\sss 8, 9, 10}$ correspond to its parity-violating interaction.
Alternatively, they correspond to parity-violating and parity-conserving 
interactions of the $2^-$ particle, respectively.

We can now write the amplitude through polarization vectors 
%
\begin{eqnarray}
&&  A(X \to ZZ)  =   \Lambda^{-1} e_1^{*\mu} \, e_2^{*\nu} 
\left[  \, c_{ 1} \, (q_1 q_2) t_{\mu\nu}
+  c_{ 2} \, g_{\mu\nu} t_{\alpha\beta} {\tilde q}^\alpha {\tilde q}^\beta
 +  c_{ 3} \, \frac{q_{2\mu} q_{1\nu}}{m_{\sss X}^2} 
t_{\alpha\beta} {\tilde q}^\alpha {\tilde q}^\beta
+  2 c_{ 4} \, (  q_{1\nu} q_{2}^{\alpha} t_{\mu\alpha} 
\right. \nonumber \\
&& \left. 
+ q_{2\mu} q_{1}^{\alpha} t_{\nu\alpha} )
+ c_{ 5} t_{\alpha \beta} \frac{{\tilde q}^\alpha {\tilde q}^\beta}{m_{\sss X}^2}
 \epsilon_{\mu \nu \rho \sigma} q_1^{\rho} q_2^\sigma
+  c_{ 6} t^{\alpha \beta} {\tilde q}_\beta
 \epsilon_{\mu \nu \alpha  \rho } q^\rho
+\frac{c_{ 7} t^{\alpha \beta} {\tilde q}_\beta}{m_{\sss X}^2}
\left ( 
\epsilon_{\alpha \mu \rho \sigma} q^\rho {\tilde q}^{\sigma}q_\nu
+ \epsilon_{\alpha \nu \rho \sigma} q^\rho {\tilde q}^{\sigma}q_\mu
\right )
\right ]. 
\label{eq:ampl-spin2} 
\end{eqnarray}
%

The coefficients $c_{ 1-7}$ can be expressed through $g^{(2)}_{\sss 1,..,10}$  
\begin{eqnarray}
&&c_{ 1} = 2  g^{(2)}_{\sss 1} +  2 g^{(2)}_{\sss 2}\kappa\left (1+ \frac{m_{\sss V}^2}{s} \right )^2 +2 g^{(2)}_{\sss 5} \frac{m_{\sss V}^2}{s}\,,\;\;\; 
\nonumber \\
&&
c_{ 2} = -\frac{g^{(2)}_{\sss 1}}{2} +g^{(2)}_{\sss 3} \kappa\left ( 1 - \frac{m_{\sss V}^2}{s} \right ) + 2 g^{(2)}_{\sss 4} \kappa + g^{(2)}_{\sss 7}\kappa\frac{m_{\sss V}^2}{s}\,, \nonumber \\
&&c_{ 3} = - \left ( \frac{g^{(2)}_{\sss 2}}{2} 
+ g^{(2)}_{\sss 3} + 2 g^{(2)}_{\sss 4} \right ) 
\kappa \frac{m_{\sss X}^2}{s}\,,\;\;\;\;\;\;\; 
\nonumber \\
&&
c_{ 4} = -g^{(2)}_{\sss 1} - g^{(2)}_{\sss 2}\kappa -  (g^{(2)}_{\sss 2}+g^{(2)}_{\sss 3}+g^{(2)}_{\sss 6})\kappa\frac{m_{\sss V}^2}{s}\,,  \nonumber \\
&&c_{ 5} = 2 g^{(2)}_{\sss 8} \kappa \frac{m_{\sss X}^2}{s}\,,\;\;\;\;\;\; 
c_{ 6} = g^{(2)}_{\sss 9}\,,\;\;\;\;\;\;\;\;
c_{ 7} = g^{(2)}_{\sss 10} \kappa \frac{m_{\sss X}^2}{s}.
\label{eq:ampl-spin2-c} 
\end{eqnarray}

To describe production of the particle $X$ in hadron 
collisions, we need to know the $X$'s coupling to 
gluons. The corresponding amplitude can be obtained 
from the case $A(X \to VV)$ that we just considered 
by crossing transformation and setting $m_{\sss V} = 0$,  
$g^{(2)}_{\sss 9} = 0$. Also, because $e_1 q_2 = e_2 q_1 = 0$ in the massless 
case, we find that terms proportional to $c_3$ and $c_4$ 
do not contribute when an analog of Eq.~(\ref{eq:ampl-spin2}) 
is written for massless gauge bosons.


\subsection{$X$ and  two fermions}

For completeness, we also give here the general couplings
of the particle $X$ to  two fermions.
We denote fermion masses as $m_q$. We assume
that the chiral symmetry is exact in the limit when fermion
masses vanish.  We obtain the following amplitudes
%
\begin{eqnarray}
&& A(X_{J=0} \to q\bar q) = \frac{m_q}{v}
\bar u_{q_1} \left ( \rho^{(0)}_{\sss 1} + \rho^{(0)}_{\sss 2} \gamma_5 \right ) v_{q_2},
\label{eq:ampl-spin0-qq}
\\
&& A(X_{J=1} \to q\bar q) = \epsilon^\mu
\bar u_{q_1}
\left (
\gamma_\mu  \left (  \rho^{(1)}_{\sss 1} + \rho^{(1)}_{\sss 2} \gamma_5 \right )
+ \frac{m_q {\tilde q}_\mu}{\Lambda^2} \left ( \rho^{(1)}_{\sss 3} +
\rho^{(1)}_{\sss 4} \gamma_5
\right ) \right ) v_{q_2},
\label{eq:ampl-spin1-qq}
\\
&& A(X_{J=2} \to q\bar q) = \frac{1}{\Lambda} t^{\mu \nu}
\bar u_{q_1} \left (
 \gamma_\mu {\tilde q}_\nu \left ( \rho^{(2)}_{\sss 1} + \rho^{(2)}_{\sss 2}
\gamma_5 \right )
+ \frac{ m_q {\tilde q}_\mu {\tilde q}_\nu}{\Lambda^2}
\left (
 \rho^{(2)}_{\sss 3} + \rho^{(2)}_{\sss 4} \gamma_5
\right )
\right ) v_{q_2},
\label{eq:ampl-spin2-qq}
\end{eqnarray}
%
where $m_q$ is the fermion mass and $\bar u$ and $v$ are the Dirac
spinors.  It follows that, in the case when fermions are massless,
the minimal couplings are also the most general ones and no new structures
appear.


\section{Helicity amplitudes} 
\label{sect3}

We are now in position to compute helicity amplitudes for the 
production and decay processes. Helicity amplitudes 
are important because, as we will see in the following discussion, 
those amplitudes parameterize angular distributions and, 
hence, can be directly   extracted from data. By knowing  how these 
amplitudes are expressed through effective couplings introduced in 
the previous section, we can constrain those couplings through 
measurements of angular distributions. 

To compute the helicity amplitudes 
$A_{\lambda_1 \lambda_2}$ for the decay $X \to VV$, we 
calculate  amplitudes presented in the previous section for polarization 
vectors that correspond to $\lambda_1, \lambda_2$. We begin with 
the description of the polarization 
vectors that we use in the  analysis.   
Consider the decay $X \to VV$ in the rest frame of $X$. The momenta of the 
two $V$'s are parameterized as
$q_{1,2} = (m_{\sss X}/2, 0, 0, \pm \beta m_{\sss X}/2)$, where 
$\beta = (1-4m_{\sss V}^2/m_{\sss X}^2)^{1/2}$ is the velocity of gauge bosons 
in the $X$ rest frame. The polarization vectors for the two $Z$-bosons read
\begin{eqnarray}
& & e_{1,2}^{\mu}(0) = m_{\sss V}^{-1} (\pm \beta m_{\sss X}/2, 0, 0, m_{\sss X}/2)\,,\;\;\; 
e_1^{\mu}(\pm) =  e_2^{\mu}(\mp) = \frac{1}{\sqrt{2}}(0, \mp 1, -i, 0).
\label{eq:polarization} 
\end{eqnarray}

The polarization vectors of the particle $X$
 are defined as follows. For the spin-one boson, $J_X = 1$, we use 
\begin{eqnarray}
&& e_X(0) = (0,0,0,1)\,,\;\;\;\;\;
e_X(\pm) = \frac{1}{\sqrt{2}} \left ( 0, \mp1 , -i, 0 \right ).
\label{eq:polarization-X} 
\end{eqnarray}
For the spin-two boson, $J_X = 2$, the polarization vectors read
\begin{eqnarray}
&& t^{\mu\nu}(\pm 2)=e_X^{\mu}(\pm) e_X^{\nu}(\pm)\,,\;\;\;\; 
t^{\mu\nu}(\pm 1)=\frac{1}{\sqrt{2}}\left[ e_X^{\mu}(\pm)
  e_X^{\nu}(0)+e_X^{\mu}(0) e_X^{\nu}(\pm) \right]\,, \nonumber \\
&& t^{\mu\nu}(0)= \frac{1}{\sqrt{6}}\left[e_X^{\mu}(+)e_X^{\nu}(-) + e_X^{\mu}(-)e_X^{\nu}(+)\right]
+\sqrt{\frac{2}{3}}e_X^{\mu}(0) e_X^{\nu}(0) \,.
\label{eq:tensor-polar} 
\end{eqnarray}

It is straightforward to establish general properties 
of the helicity amplitudes $A_{\lambda_1 \lambda_2}$.
In general there are nine complex amplitudes $A_{\lambda_1\lambda_2}$,
since conservation of the angular momentum component along the 
decay axis fixes  the spin projection of the  $X$ particle to
$\lambda_X = \lambda_1 - \lambda_2$. 
Because of this identity, we do not reference $\lambda_X$ in the notation for 
helicity amplitudes. Moreover, in the case of two identical vector
bosons, such as $ZZ$, $gg$, or $\gamma\gamma$,
the number of independent amplitudes is reduced from nine to six
due to the following identity~\cite{trueman, dell}
\begin{eqnarray}
A_{\lambda_1 \lambda_2} = (-1)^J A_{\lambda_2 \lambda_1}
\label{eq:amplitude-identical} \,,
\end{eqnarray}
where $J$ is the spin of the $X$ particle.
If parity is conserved, further constraints apply~\cite{trueman, dell}
\begin{eqnarray}
A_{\lambda_1 \lambda_2} = \eta_{\sss P} (-1)^J A_{-\lambda_1 -\lambda_2} \,,
\label{eq:amplitude-parity} 
\end{eqnarray}
where $\eta_{\sss P}$ is the parity of the $X$ particle.
We note that Eqs.~(\ref{eq:amplitude-identical}, \ref{eq:amplitude-parity})
depend on phase  conventions for the polarization vectors in 
Eqs.~(\ref{eq:polarization}, \ref{eq:polarization-X}).

For a {\it spin-zero} $X$ particle, only $\lambda_1-\lambda_2=0$ values 
are possible. Therefore only $A_{++}$, $A_{--}$, and $A_{00}$ contribute.
For a parity-even scalar with $J^P=0^+$, such as a SM Higgs, 
one has $A_{++}=A_{--}$ 
and for parity-odd pseudo-scalar with $J^P=0^-$ one has 
$A_{++}=-A_{--}$ and $A_{00}=0$.
For a {\it spin-one} $X$ particle, Bose symmetry
prohibits $A_{++}$, $A_{--}$, and $A_{00}$ amplitudes,
as also evident from Eq.~(\ref{eq:amplitude-identical}).
Therefore we are left with only two independent contributions
$A_{+0}=-A_{0+}$ and $A_{-0}=-A_{0-}$.
Furthermore, in the 
vector case $J^P=1^-$, one has $A_{+0}=A_{-0}=-A_{0+}=-A_{0-}$,
and for an axial vector $J^P=1^+$ the amplitudes are related as
$A_{+0}=-A_{-0}=-A_{0+}=A_{0-}$
in the case of parity-conserving interactions.
For a {\it spin-two}
 $X$ particle, there are generally six independent contributions
$A_{00}$, $A_{++}$, $A_{--}$, $A_{+-}=A_{-+}$, $A_{+0}=A_{0+}$, and $A_{-0}=A_{0-}$.
With parity conservation, there are additional constraints for the case $J^P=2^+$:
$A_{++}=A_{--}$ and $A_{+0}=A_{0+}=A_{-0}=A_{0-}$;
and for the case $J^P=2^-$:
$A_{++}=-A_{--}$ and $A_{+0}=A_{0+}=-A_{-0}=-A_{0-}$, 
while $A_{00}=0$ and $A_{+-}=A_{-+}=0$.
We now use explicit expressions for the amplitudes constructed 
in the previous section to illustrate 
these assertions and compute the independent helicity amplitudes 
explicitly.


\subsection{Helicity amplitudes for spin-zero decay}

We use Eq.~(\ref{eq:ampl-spin0}) to compute the helicity amplitudes
and obtain for the decay to two massive vector bosons
\begin{eqnarray}
\label{eq:helicity-ampl0} 
&& A_{00}  =  -\frac{m_{\sss X}^4}{4vm_{\sss V}^2} 
\left ( a_1 (1 + \beta^2) +a_2 \beta^2
\right )\,, \nonumber \\
&& A_{++}  =  \frac{m_{\sss X}^2}{v} \left ( 
a_1 + \frac{ia_3\beta}{2} 
\right )\,,\;\;\;\;
 \nonumber \\
&& A_{--}  =  
\frac{m_{\sss X}^2}{v} \left ( 
a_1 - \frac{ia_3\beta}{2} \right )\,.
\end{eqnarray}
As expected, these amplitudes satisfy
Eq.~(\ref{eq:amplitude-parity}) for parity-even 
($a_1$ and $a_2$) and parity-odd ($a_3$) contributions separately.

Helicity amplitudes for the decay of the $X$ particle into two massless 
gauge bosons can be obtained from Eq.~(\ref{eq:helicity-ampl0}) 
by disregarding  $A_{00}$,
since massless gauge bosons cannot be longitudinally polarized,  
and by setting $\beta = 1$ in $A_{++}$ and $A_{--}$. 
The relationship between coefficients $a_1,a_3$ and 
the fundamental couplings $g^{(0)}_{i}$ , in this case, can be read off from 
Eq.~(\ref{eq13}), where the mass of the vector boson $m_{\sss V}$ should be 
set to zero.

It is interesting to point out some 
features  of Eq.~(\ref{eq:helicity-ampl0}) for the case 
when $X$ is the SM Higgs boson, sufficiently 
heavy to decay into two $Z$ bosons. Then, as follows from  
Eq.~(\ref{eq:helicity-ampl0}) the longitudinal amplitude $A_{00}$
dominates in the limit that $m_{\sss H} \gg m_{\sss Z}$. To see this, 
recall that at tree level in the SM only 
the term with $a_1$ in Eq.~(\ref{eq:ampl-spin0}) contributes. Then, 
$A_{++}=A_{--}=-{A_{00}}/\gamma$, 
where $\gamma$ is the boost between the rest frames of the two 
$Z$ boson 
\begin{eqnarray}
\gamma = \frac{m_{\sss X}^2}{2m_{\sss Z}^2}-1 = \frac{1+\beta^2}{1-\beta^2}\,.
\label{eq:boost}
\end{eqnarray}

Note also that additional contributions to the helicity amplitudes 
are present even in the SM, beyond the tree level.
For example,  both $a_2$ and $a_3$ may appear from radiative
corrections. The natural scale for $a_2$, generated radiatively 
in the SM, is ${\cal O}(\alpha_{\rm  EW}) \sim 10^{-2}$,
while the SM contribution to $a_3$ appears only at three loops 
and therefore is tiny ${\cal O}(10^{-11})$~\cite{soni}. 
In general, all three coefficients $a_1$, $a_2$, and $a_3$
are complex numbers with a priori unknown relative
complex phases between them.
In the context of the SM, the
measurement of the radiatively induced contributions   $a_{2,3}$  is 
a non-trivial test  of the  Higgs couplings to 
gauge bosons  at the quantum level.


\subsection{Helicity amplitudes for  spin-one decays}

Using Eq.~(\ref{eq:ampl-spin1}) we find the following helicity amplitudes
\begin{eqnarray}
A_{+0} = - A_{0+} & = & 
  \frac{\beta  m_{\sss X}^2}{2 m_{\sss Z}} \left( g^{(1)}_{\sss 1} 
+i \beta g^{(1)}_{\sss 2} \right)\,, \nonumber\\
A_{-0} = - A_{0-} & = & 
  \frac{\beta m_{\sss X}^2}{2 m_{\sss Z}} \left( g^{(1)}_{\sss 1} 
- i \beta  g^{(1)}_{\sss 2} \right)\,. 
\label{eq:helicity-ampl1} 
\end{eqnarray}
These amplitudes  satisfy Eqs.~(\ref{eq:amplitude-identical},
\ref{eq:amplitude-parity}) for  parity-even ($g^{(1)}_{\sss 1}$) and 
parity-odd ($g^{(1)}_{\sss 2}$) contributions separately.
Note that the Landau-Yang theorem forbids decays of  spin-one particles  
into a pair of massless identical bosons. This feature is apparent from 
Eq.~(\ref{eq:helicity-ampl1}) which shows that  
one of the vector bosons $V$  in the decay of the spin-one particle $X$ 
should be longitudinally polarized. Since massless vector bosons cannot 
be polarized longitudinally, the decay of $X$ into a pair of massless 
identical vector bosons cannot occur.


\subsection{Helicity amplitudes for spin-two decays}

Using explicit parameterization for the decay $X \to VV$ in Eq.~(\ref{eq:ampl-spin2}), we 
obtain the helicity amplitudes
%
\begin{eqnarray} 
\label{eq35}
&& A_{+-} = A_{-+} = \frac{m_{\sss X}^2}{4\Lambda} c_1  \left ( 1 + \beta^2 \right )\,, 
 \nonumber \\
&& A_{++} =
\frac{m_{\sss X}^2}{\sqrt{6} \Lambda} \left [ 
\frac{c_1}{4}  \left (1 + \beta^2 \right )
+ 2 c_2  \beta^2  +i \beta ( c_5 \beta^2 - 2 c_6) \right ]\,,\nonumber \\
&& A_{--} = \frac{m_{\sss X}^2}{\sqrt{6}\Lambda} \left [ 
\frac{c_1}{4}  \left (1 + \beta^2 \right )
+ 2 c_2  \beta^2 -i \beta ( c_5 \beta^2 - 2 c_6)  \right ]\,, \nonumber \\
&& A_{+0} = A_{0+} = \frac{m_{\sss X}^3}{m_{\sss V} \sqrt{2} \Lambda}
\left [ \frac{c_1}{8} \left ( 1+\beta^2 \right ) +  \frac{c_4}{2}\beta^2 -\frac{c_6+c_7 \beta^2}{2}i \beta \right ]\,, \nonumber \\
&&  A_{-0} = A_{0-} = \frac{m_{\sss X}^3}{m_{\sss V} \sqrt{2} \Lambda}
\left [ \frac{c_1}{8} \left ( 1+\beta^2 \right ) +  \frac{c_4}{2}\beta^2 + \frac{c_6+c_7 \beta^2}{2}i \beta \right ]\,,
\nonumber \\
&& A_{00} =
\frac{m_{\sss X}^4}{m_{\sss V}^2\sqrt{6} \Lambda}
\left [  
         \left (  1+ \beta^2 \right ) \left ( \frac{c_1}{8}
       - \frac{c_2}{2}  \beta^2  \right )
       - \beta^2 \left ( \frac{c_3}{2}   \beta^2 
       - c_4 \right )
\right ]\,.
\end{eqnarray} 
%
As expected, we find six independent helicity amplitudes for the 
most general case with two identical massive bosons. 
Note that only two independent combinations of $c_{5,6,7}$
constructed from $g^{(2)}_{\sss 8,9,10}$ enter Eq.~(\ref{eq35}),
and therefore only those combinations are accessible in a measurement.

To describe decays of  $X$ into two massless bosons, all helicity 
amplitudes with a longitudinal polarization in Eq.~(\ref{eq35}) 
should be disregarded and $\beta = 1$ should be substituted everywhere. 
The relation between the coefficients $c_j$ and couplings $g^{(2)}_{i}$ 
in this case is found from Eq.~(\ref{eq:ampl-spin2-c}) where $m_{\sss V}=0$ 
should be substituted.  We note that  a peculiar feature emerges 
as the result of this procedure in the case when  the $X$ coupling 
to massless gauge fields is {\it minimal}. The minimal coupling 
corresponds to $g^{(2)}_{\sss 1} = 1$ and $g^{(2)}_{\sss j > 1} = 0$. 
 From Eq.~(\ref{eq:ampl-spin2-c}), it follows that in that case 
$c_2 = -c_1/4=c_4/2$ and all other coefficients are zero. 
Eq.~(\ref{eq35}) then implies that, 
in the case of the minimal coupling of the particle $X$ to  massless   
gauge bosons, $A_{++} = A_{--}=0$. 
Hence, only projections $J_z=\pm 2$ of the  $X$ spin 
on the collisions axis are allowed in that case. 
However, for more general couplings,  $A_{++} = A_{--}$ 
amplitudes do not vanish and, therefore, zero projection of 
the $X$ spin on the collision axes is allowed. 


\subsection{Helicity amplitudes for decays into  two fermions}

In this subsection, we write down helicity amplitudes for $X$ 
coupling to two quarks.  We need those helicity amplitudes to describe 
production of the resonance $X$ in $q \bar q$ annihilation. 
The helicity amplitudes for the spin-zero case read
\ba
&&
A_{++} = \frac{m_q}{v} m_{\sss X} \left ( \rho^{(0)}_{\sss 2} - \beta \rho^{(0)}_{\sss 1} \right )\,,
\;\;\;\;
~~~~~~~~~~~~~~~~~~~~~~~~~~~~
\nonumber \\
&& 
A_{--} = \frac{m_q}{v} m_{\sss X} \left ( \rho^{(0)}_{\sss 2} + \beta \rho^{(0)}_{\sss 1} \right )\,.
\label{eq:helicity-ff-ampl0} 
\ea

The helicity amplitudes for the spin-one case read
\ba
&& 
A_{++} = -2m_q \left ( \rho^{(1)}_{\sss 1} +
\frac{\beta m_{\sss X}^2}{2\Lambda^2}
\left ( \rho^{(1)}_{\sss 4} -  \beta\rho^{(1)}_{\sss 3} \right ) \right )\,,\;\;\;
~~~~~~~~~~
\nonumber \\
&& 
A_{--} = -2m_q \left ( -\rho^{(1)}_{1} + \frac{\beta m_{\sss X}^2}{2\Lambda^2}
\left ( \rho^{(1)}_{\sss 4} + \beta \rho^{(1)}_{\sss 3}\right ) \right ),\nonumber \\
&& 
A_{+-} = \sqrt{2} m_{\sss X} \left (\rho^{(1)}_{\sss 1}+\beta \rho^{(1)}_{\sss 2} \right)\,,
\;\;\;\;
\nonumber \\
&& 
A_{-+} = -\sqrt{2} m_{\sss X} \left (\rho^{(1)}_{\sss 1}-\beta \rho^{(1)}_{\sss 2} \right)\,.
\label{eq:helicity-ff-ampl1} 
\ea

The helicity amplitudes for the spin-two case read
\ba
&& A_{++} = \frac{2\sqrt{2}\,m_qm_{\sss X} \beta}{\sqrt{3}\Lambda}
 \left ( \rho^{(2)}_{\sss 1} + \frac{\beta m_{\sss X}^2}{2\Lambda^2}
\left ( \rho^{(2)}_{\sss 4} - \beta \rho^{(2)}_{\sss 3} \right ) \right ),\nonumber \\
&& A_{--} = \frac{2\sqrt{2}\,m_qm_{\sss X} \beta}{\sqrt{3}\Lambda}
 \left ( -\rho^{(2)}_{\sss 1} + \frac{\beta m_{\sss X}^2}{2\Lambda^2}
\left ( \rho^{(2)}_{\sss 4} + \beta\rho^{(2)}_{\sss 3}  \right ) \right )\,,\nonumber \\
&& A_{+-} = -\frac{m_{\sss X}^2 \beta}{\Lambda}
 \left (\rho^{(2)}_{\sss 1}+\beta \rho^{(2)}_{\sss 2} \right)\,,\;\;\;\;\; \nonumber \\
 &&
A_{-+} = \frac{m_{\sss X}^2\beta}{\Lambda}
 \left (\rho^{(2)}_{\sss 1}-\beta \rho^{(2)}_{\sss 2} \right)\,.
\label{eq:helicity-ff-ampl2} 
\ea


\section{Angular distributions} 
\label{sect4}

Information about the quantum numbers of $X$ and its couplings to the SM
fields can be extracted from the angular distributions. In general, there 
are five  angles that can be studied, see Fig.~\ref{fig:decay}. 
The two production angles, $\theta^*$ and $\Phi_1$, are defined relative 
to the parton collision axis; distributions of those angles depend on 
the production mechanism. 
The three helicity angles, $\theta_1$, $\theta_2$, and $\Phi$,  
are sensitive to the structure of the interactions with the resonance decay products,  
but they do not depend on the production mechanism.
These angles are  illustrated in Fig.~\ref{fig:decay} for the process 
of the type $2\to4$, where the decay chain $X\to Z_1Z_2\to(f_1\bar{f}_1)(f_2\bar{f}_2)$ 
is considered and the production mechanism is either $gg \to X $ or $q\bar q\to X$. 

The exact definition of the five angles shown in 
Fig.~\ref{fig:decay} is as follows: 
$\theta^*$ is the angle between the parton collision axis $z$ and 
the $X\to Z_1Z_2$ decay axis $z^\prime$, both defined in the $X$ rest frame;
$\Phi_1$ is the angle between the $zz^\prime$ plane and the plane 
of the $Z_1 \to (f_1\bar{f}_1)$  decay in the $X$ rest frame;
$\theta_i$ is the angle between the direction of the fermion $f_i$  
from $Z_i\to (f_i\bar{f}_i)$ and 
the direction opposite the $X$ in the $Z_i$ rest frame, where index $i=1,2$ 
refers to the first or second $Z$ boson; finally,  
$\Phi$ is the angle between the decay planes of the two $Z$ systems
in the $X$ rest frame. 
The sixth angle $\Phi^*$ is the azimuth angle 
of the $z^\prime$ axis with respect
to $z$.  It can be  arbitrarily defined and it does not 
carry  any  information about the process. 
However, $\Phi^*$ can be used for bookkeeping purposes to differentiate between 
the first and second $Z$ in the decay in the case that both have identical decay channels.
{Discussion of the small possible difference between 
the $z$ axis and the beam collision axis follows below.}

Distributions of the five angles described above allow one 
to determine the spin of the $X$ boson
and measure contributions 
of the different helicity amplitudes $A_{\lambda_1\lambda_2}$ 
in both production and decay. 
The $X$ boson couplings to the SM fields 
can  be deduced from the helicity amplitudes, cf. 
Eqs.~(\ref{eq:helicity-ampl0} $-$\ref{eq:helicity-ff-ampl2}). 
We therefore proceed to the general expression for 
the angular distributions in the process 
$pp \to X\to V_1 V_2\to (f_1f^\prime_1)(f_2f^\prime_2)$ and illustrate 
the general formula by considering a number of specific examples. 
We use the  helicity formalism as described in Refs.~\cite{trueman, dell}; 
for recent examples of applications of the helicity formalism, 
see Refs.~\cite{spin1, bij, hagiwara,Buckley:2008eb,Murayama:2009jz, btovv, phik2}.
Our results are consistent with those references but are, typically, more general.

Let us first consider the $1\to 2$ decay process. The helicity 
amplitudes are defined through a matrix element of the scattering 
matrix between states with definite projections of angular momenta 
on a chosen quantization axis. Hence, 
\begin{eqnarray}
\langle \Omega, \lambda_1, \lambda_2 | S | Jm\rangle
=\sqrt\frac{(2J+1)}{4\pi} D^{J\ast}_{m,\lambda_1-\lambda_2}(\Omega) A_{\lambda_1\lambda_2}\,,
\label{eq:1to2}
\end{eqnarray}
where $\Omega$ describes the polar and the 
azimuthal angles of one of the final state particles  and $D$ denotes the 
corresponding  Wigner function. Viewing 
the collision process of two partons $a$ and $b$, 
$a\; b \to X\to V_1 V_2\to (f_1f^\prime_1)(f_2f^\prime_2)$,
as a sequence of $1 \leftrightarrow 2$ elementary 
processes, we   describe it  by the following formula
%
\begin{eqnarray}
&& A_{ab}(p_a,p_b;
\chi_1,\chi_2, m, \lambda_1,\lambda_2, \mu_1,\mu_2,\tau_1,\tau_2;\Omega^*,\Omega,\Omega',\Omega'')=
\frac{(2J+1)}{4\pi}\sqrt{\frac{(2s_1+1)}{4\pi}\frac{(2s_2+1)}{4\pi}} \\
&& 
\times
D^{J\ast}_{\chi_1-\chi_2, m}(\Omega^{\ast})B_{\chi_1\chi_2}
\times D^{J\ast}_{m,\lambda_1-\lambda_2}(\Omega)A_{\lambda_1\lambda_2}
\times D^{s_1\ast}_{\lambda_1,\mu_1-\mu_2}(\Omega')T(\mu_1,\mu_2)
\times D^{s_2\ast}_{\lambda_2,\tau_1-\tau_2}(\Omega'')W(\tau_1,\tau_2)\,.
\nonumber
\label{eq:a-15}
\end{eqnarray}
%
Here  the $A_{\lambda_1\lambda_2}$ and $B_{\chi_1\chi_2}$ amplitudes correspond 
to the $X$ decay and production  processes, respectively;
$T(\mu_1,\mu_2)$ and $W(\tau_1,\tau_2)$ describe   the
decays of the first and the second 
$X$ daughter to pairs of fermions;
$\lambda_{1,2}$ are the helicities of the two $X$ daughters;
$\chi_{1,2}$ are the helicities of two gluons or quarks in the initial state;
$\mu_{1,2}$ and $\tau_{1,2}$  are the helicities of fermions in the $Z$ decays;
and $m$ denotes the helicity of $X$.
We keep the notation general enough with $J$ denoting the spin of 
the $X$ boson and 
$s_{1,2}$ referring to the spins of its decay products, e.g. the $Z$ bosons.
We choose the convention where $\Omega = (0,0,0)$, which by conservation 
of angular momentum effectively sets $m=\lambda_1-\lambda_2$, 
$\Omega^\ast = (\Phi_1, \theta^{\ast}, -\Phi_1)$, 
$\Omega^\prime= (0, \theta_1, 0)$ and $\Omega''= (\Phi, \theta_2, -\Phi)$.
Summing over the  
helicities of all particles, we obtain the differential cross-section
for the process $a\; b \to X\to V_1 V_2\to (f_1f^\prime_1)(f_2f^\prime_2)$ as 
\begin{eqnarray}
\frac{d\sigma_{ab}(p_a,p_b,\theta^\ast, \Phi_1, \theta_1,\theta_2, \Phi)
}
{{\rm d} Y_{\sss X} d\cos\theta^\ast d\Phi_1  d\cos\theta_1 d\cos\theta_2 d\Phi}
= {\cal N}_{ab} \delta(s_{ab} - m_{\sss X}^2) \delta(Y_{ab} - Y_{\sss X})
\sum_{\{\chi,\mu,\tau\}}
\Bigl|\sum_{\{\lambda,  m\}}A_{ab}(p_a,p_b;\{\chi, \lambda; m,\mu,\tau\};\{\Omega\})\Bigr|^2 \,,
\label{eq:rate-amplitude}
\end{eqnarray}
where $s_{ab}$ is the partonic center-of-mass energy squared,
$Y_{ab}$ and $Y_{\sss X}$ are the rapidities of the colliding partons $ab$ and the resonance
$X$,  and ${\cal N}_{ab}$ is the normalization factor
\be
{\cal N}_{ab} = 
\frac{\pi \beta_Z}{8 m_{\sss X}^3 \Gamma_X c_{ab} (32\pi^2)^3 (2 m_{\sss Z} \Gamma_Z)^2 }\,,
\label{eq:norm}
\ee
with $\beta_Z = \sqrt{1-4m_{\sss Z}^2/m_{\sss X}^2}$, $c_{q \bar q} = 3$, $c_{gg} = 8$, and 
$\Gamma_{X,Z}$ being the decay widths of the resonance $X$ and the $Z$ boson, respectively. 
The relevant differential cross-section for hadron collisions is  obtained 
by convoluting parton cross-sections with parton distribution functions
\be
\frac{{d\sigma_{pp}}(\theta^\ast, \Phi_1, \theta_1, \theta_2, \Phi)}
{d\cos\theta^\ast d\Phi_1  d\cos\theta_1 d\cos\theta_2 d\Phi}
= \sum_{ab} \int  {\rm d} Y_{\sss X} \; {\rm d} x_1 {\rm d} x_2\; 
\tilde f_a(x_1) \; \tilde f_b (x_2)\;\;
\frac{d\sigma_{ab}(x_1 p_1,x_2 p_2,
\theta^\ast, \Phi_1, \theta_1, \theta_2, \Phi)}
{{\rm d} Y_{\sss X} d\cos\theta^\ast d\Phi_1  d\cos\theta_1 d\cos\theta_2 d\Phi}
\Big|_{Y_{ab} = \frac{1}{2}\ln \frac{x_1}{x_2}}\,,
\label{eq:rate-amplitude-h}
\ee
where  $p_{1,2}$ are the momenta of the two protons. Note that 
the angles $\cos\theta^\ast, \Phi_1, \cos\theta_1, \cos\theta_2, \Phi$
are defined in the $X$ rest frame and are not affected by the integrations
over $x_{1}, x_{2},$ and $Y_{\sss X}$. Convolution 
with the parton distribution functions results in the rapidity distribution 
of the $X$ boson and may affect angular distributions  of decay products 
on an event-by-event basis due to the detector acceptance, discussed later.
However, it does not affect angular distributions measured with the 
ideal detector since in this case, integrations over $x_1, x_2,$ and $Y_{\sss X}$ 
in Eq.~(\ref{eq:rate-amplitude-h}) factorize from all angular
dependences. As the result, we may rewrite Eq.~(\ref{eq:rate-amplitude-h})
in the form of the angular distribution of the decay 
products of the {\it polarized} particle $X$ 
\be
\frac{1}{\sigma_{pp,\rm tot}}
\frac{{d\sigma_{pp}}}
{d\cos\theta^\ast d\Phi_1  d\cos\theta_1 d\cos\theta_2 d\Phi}
= \frac{1}{\Gamma_X} \frac{{\rm d} \Gamma_X}{d\cos\theta^\ast d\Phi_1  d\cos\theta_1 d\cos\theta_2 d\Phi}\,.
\label{eqaa}
\ee
Since we are interested in the normalized distributions, many 
normalization factors, as in Eq.~(\ref{eq:norm}), drop out.
However, ratios of amplitudes squared $|A_{ab}|^2$ 
and ratios of partonic luminosities for different production channels appear
in the normalized angular distributions and contribute to the degree 
of the polarization of the particle X. 
Note that if the $X$ particle can only  be produced 
in a single partonic channel, all dependence on the partonic luminosities 
cancels out. This is relevant for both a spin-zero and spin-one $X$ 
which are produced in $ab=gg$ and $q \bar q$ collisions, respectively. 
For a spin-two $X$ particle, the two parton channels contribute and their
relative partonic luminosities affect the relative fraction of spin-projection-one polarization
as we note below. Apart from this relative normalization, which is incorporated
into notation discussed below, we can derive angular distributions
using the sum of $|A_{ab}|^2$ over the helicity states, as indicated in 
Eq.~(\ref{eq:rate-amplitude}), for each parton channel $ab$ independently.

To define easy-to-measure parameters related to the helicity amplitudes, we consider 
 the case of a spin-two resonance as an example. 
The decay of a spin-two particle to two vector bosons is characterized by
nine complex helicity amplitudes  $A_{\lambda_1\lambda_2}$.
When vector bosons in the final state are identical, only six helicity amplitudes 
remain independent. Six complex amplitudes are parameterized by twelve real numbers.
However, since we are interested in the normalized angular distribution, 
two of them, the normalization and the overall phase, are unobservable. Hence, angular 
distributions of the spin-two particle decay to two identical vector bosons can 
be parameterized by ten real parameters. We can choose them to be  
\begin{eqnarray}
 && f_{\sss \lambda_1\lambda_2}=|A_{\lambda_1\lambda_2}|^2/ \!\!\!\! \sum_{k,l=\pm,0} \!\!\!\! |A_{kl}|^2 \,,
\label{eq:fractions} \\
 && \phi_{\sss \lambda_1\lambda_2}={\rm arg}(A_{\lambda_1\lambda_2}/A_{00}) \,,
\label{eq:phases}
\end{eqnarray}
with $(\lambda_1,\lambda_2)=(++), (--), (+-), (+0), (0-)$.  
The remaining  (dependent) amplitude parameters $f_{kl}$ and $\phi_{kl}$
can be expressed either using Eq.~(\ref{eq:amplitude-identical}) 
or the relationship 
\begin{eqnarray}
f_{00} = 1 - \!\!\!\! \sum_{k,l=\pm,0}  \!\!\!\!  f_{kl}\,,\;\;\;\;\;\; \phi_{00} = 0\,.
\end{eqnarray}
For the  $X$ decays to massless vector bosons or if its spin is less than two, 
the number of independent non-vanishing  parameters is further reduced. 
Our notation for those (simpler) cases are similar to what is described above. 
For example, the spin-zero case is parameterized with four real parameters
for $(\lambda_1,\lambda_2)=(++), (--)$,
and the spin-one case with two real parameters for $(\lambda_1,\lambda_2)=(+0)$
after re-defining the phase convention with respect to the $A_{0-}$ amplitude.

Equivalent parameters can be defined for the helicity 
amplitudes $B_{\chi_1\chi_2}$ describing the production mechanism.
However, not all parameters would enter the angular distributions. 
We choose the three real parameters  $f_{z0}$, $f_{z1}$, and $f_{z2}$ to describe
the fraction of spin-two $X$ resonance production with spin $z$-projections 
$0$, $\pm1$, and $\pm2$, respectively, where $f_{z0}+f_{z1}+f_{z2}\equiv 1$,
leaving only two independent parameters.
The fractions $f_{z0}$ or  $f_{z2}$ arise from the gluon fusion mechanism
with amplitudes $B_{++}$ and $B_{--}$ or $B_{+-}$ and $B_{-+}$ in Eq.~(\ref{eq35}), 
respectively, where we now use the $B_{\chi_1\chi_2}$  notation in place of
$A_{\chi_1\chi_2}$ to distinguish the process of production from decay.
The fraction $f_{z1}$ originates from $q\bar{q}$ annihilation
with amplitudes $B_{+-}$ and $B_{-+}$ in Eq.~(\ref{eq:helicity-ff-ampl2}).
Note that because we assume that chirality is a good quantum number,
only $\pm1$ spin projections on the $z$ axis are possible in the annihilation 
of massless quarks. On the other hand, both $J_z = \pm2$ and $J_z = 0$ spin 
projections of a spin-two $X$ are possible in gluon fusion. 
The latter possibility is often ignored in the literature because, 
accidentally, this contribution vanishes in the case 
of the minimal coupling of the spin-two particle to massless gauge bosons.
Finally we note that, since $f_{z1}$ and $f_{z0},f_{z2}$ are produced by different partonic
initial states, those quantities are proportional to corresponding 
partonic luminosities, in addition to the production helicity amplitudes $B_{\chi_1\chi_2}$.
Below we discuss simplified versions of the general angular distribution which 
are obtained upon integrating over some of the angles in Eq.~(\ref{eq:rate-amplitude-h}).


\subsection{Distributions of production angles}

We consider the $gg$ and $q\bar{q}$ production of an exotic resonance $X$
and its subsequent decay  $X\to Z_1Z_2\to(f_1\bar{f}_1)(f_2\bar{f}_2)$.
Angular distributions for other decay channels such as $X \to \gamma \gamma$, 
$X \to gg$, and $X \to f \bar f$ are given in the appendix.

The production angle $\theta^*$ is shown  in Fig.~\ref{fig:decay}
for an $X$ decaying to two $Z$ bosons, as an example. It 
is defined as the angle between the parton collision axis $z$ and the
$X$ decay axis in the $X$ rest frame. Determination of the $z$ axis
requires care since  it may differ somewhat from the beam
collision axis which we discuss  below. 
The distribution of the angle  $\theta^*$ is the only
angular observable that  contains information about the spin and parity
properties of the particle $X$, unless decay chains of the $X$
daughters are  analyzed. The latter is possible in the decay to
two massive vector bosons $ZZ$ or $W^+W^-$ for certain cases, but is challenging or
practically impossible in other cases. The production angle distributions 
have been extensively studied  in the literature~\cite{rosner1996, Davoudiasl2000, 
Allanach2000, gtopp, Cousins2005, Osland2008}.
However, for the spin-two resonance, we point to several modifications 
of the standard formulas due to generally ignored $J_z = 0$  
projection on the collision axis.

Transverse momentum of the $X$ particle introduces uncertainties
in the production angle determination. The Collins-Soper
frame~\cite{Collins1977} is designed to minimize the effect of $X$
transverse momentum by placing the $z$ axis half way between axes 
of two beams in the $X$ rest frame. Note that if $X$ has transverse 
momentum, the two beams are no longer collinear in the $X$ rest frame. 
The uncertainty introduced in the $\theta^*$ measurement by 
the non-vanishing transverse momentum of the $X$ particle is
expected to be relatively small~\cite{Cousins2005}, at least when 
compared to the statistical uncertainties in early LHC measurements. 
With a larger number of events becoming available later, one can, on an
event-by-event basis, find and disregard events where $X$ recoils 
against hadronic jets with a large transverse momentum.

Note that there is an ambiguity in the direction of the $z$ axis and one
cannot distinguish  $\theta^*$ from  $(\pi-\theta^*)$ when two identical
particles are involved in either the production or decay of the $X$ particle.
In particular, this happens if $X$ is produced in gluon fusion 
or when $X$ decays to two identical bosons, such as $ZZ$, $gg$,
or $\gamma\gamma$. In those cases, terms with an odd power in $\cos\theta^*$
drop out from the angular distributions and we do not consider them further.
Only when $X$ is produced in $q\bar{q}$ annihilation and decays into a particle-antiparticle
pair, such as $W^+W^-$ or $l^+l^-$, can one try to deduce the $\cos\theta^*$ sign.

The case of a {\it spin-zero particle} is very simple. 
Since no spin correlations are involved,  the production 
angles distributions  are flat.   However, determination of these
angles is still relevant for the analysis of a spin-zero particle
because it is needed to discriminate against backgrounds or
resonances with non-zero spin.
Normalized distributions of production angles 
for {\it spin-one} and {\it spin-two} resonances that 
decay as  $X\to ZZ\to(f_1\bar{f}_1)(f_2\bar{f}_2)$ can be written 
as follows
\begin{eqnarray}
\frac{32\pi\, d\Gamma_{J=1}}{3\, \Gamma\, d\cos\theta^*d\Phi_1} 
&&   \!\!\!\! = 2(1+\cos^2\theta^\ast) 
- \sqrt{2f_{+0}(1-2f_{+0})}\,\sin^2\theta^\ast\cos (2\Phi_1 - [\phi_{+0} - \phi_{0-}]) \,,
\label{eq:zprime-ang12d} 
\end{eqnarray}

\begin{eqnarray}
&& \frac{32\pi\, d\Gamma_{J=2}}{5\, \Gamma\, d\cos\theta^*d\Phi_1} 
 = (2-2f_{z1}+f_{z2}) -6(2 -4f_{z1} -f_{z2})\cos^2\theta^\ast +3(6 -10f_{z1}-5f_{z2})\cos^4\theta^\ast 
 \nonumber \\
&& +f_{+-}\Bigl\{(2+2f_{z1}-7f_{z2})+6(2-6f_{z1}+f_{z2})\cos^2\theta^\ast-5(6-10f_{z1}-5f_{z2})\cos^4\theta^\ast \Bigr\} 
\nonumber \\
&& -2(f_{+0}+f_{0-})\,\Bigl\{(2-4f_{z1}-f_{z2})-6(4-7f_{z1}-3f_{z2})\cos^2\theta^\ast+5(6-10f_{z1}-5f_{z2})\cos^4\theta^\ast \Bigr\}  
\nonumber \\
&& - 2\sqrt{f_{+0}f_{0-}}\,\Bigl\{ (f_{z1}-f_{z2}) + (6-10f_{z1}-5f_{z2})\cos^2\theta^\ast \Bigr\} \sin^2\theta^\ast \cos(2\Phi_1 + \phi_{+0} - \phi_{0-}) 
\nonumber \\
&& - \frac{1}{2}\,\Bigl\{ (2-2f_{z1}-3f_{z2}) - (6-10f_{z1}-5f_{z2}) \cos^2\theta^\ast \Bigr\} \sin^2\theta^\ast 
\nonumber \\
 && ~~ ~~ \times \Bigl[ \sqrt{6f_{++}f_{+-}}\cos(2\Phi_1 - \phi_{++} + \phi_{+-}) +  \sqrt{6f_{--}f_{+-}}\cos(2\Phi_1 + \phi_{--} - \phi_{+-}) \Bigr]
\nonumber \\
&& +\frac{3\pi R_1}{4}   \Bigl\{ (2-4f_{z1}-f_{z2})-(6-10f_{z1}-5f_{z2})\cos^2\theta^\ast \Bigr\}  \cos\theta^\ast \sin\theta^\ast 
\nonumber \\
&& ~~ ~~ \times \Bigl[ \sqrt{3f_{+0}(1- f_{++}-f_{--}-2f_{+-}-2f_{+0}-2f_{0-})} \cos(\Phi_1-\phi_{+0})  + \sqrt{3f_{+0}f_{++}} \cos(\Phi_1+\phi_{+0}-\phi_{++}) 
\nonumber \\
&& ~~ ~~ ~~ ~~ -  \sqrt{3f_{0-}(1- f_{++}-f_{--}-2f_{+-}-2f_{+0}-2f_{0-})}  \cos(\Phi_1+\phi_{0-}) - \sqrt{3f_{0-}f_{--}} \cos(\Phi_1-\phi_{0-}+\phi_{--}) \Bigr]
\nonumber \\
&& +\frac{3\pi R_1}{8}  \Bigl\{( 6-6f_{z1}-9f_{z2}) -(6-10f_{z1}-5f_{z2})\cos^2\theta^\ast \Bigr\} \cos\theta^\ast \sin\theta^\ast 
\nonumber \\
&& ~~ ~~ \times \Bigl[  \sqrt{2f_{+-}f_{+0}} \cos(\Phi_1+\phi_{+-}-\phi_{+0}) -  \sqrt{2f_{+-}f_{0-}} \cos(\Phi_1-\phi_{+-}+\phi_{0-})    \Bigr] 
\,.
\label{eq:grav-ang2mix2d}
\end{eqnarray}

Note that the right hand side of  Eq.~(\ref{eq:grav-ang2mix2d}) 
is written as a linear combination of terms 
of the form $(\alpha_1 +\alpha_2 f_{z1} + \alpha_3 f_{z2})$,
 multiplied by other
parameters. It is peculiar that in all such  terms, 
except in the very first, $\theta^*$- and 
$\Phi_1$-independent one, there is a relationship between the $\alpha$-coefficients
$\alpha_1 = -0.4\times(\alpha_2 + \alpha_3)$.  This means  that 
if we choose $f_{z1}=f_{z2}=0.4$ in Eq.~(\ref{eq:grav-ang2mix2d}), the 
production angle distribution for the spin-two particle becomes flat!
To understand this, note that those helicity fractions imply 
production of an  unpolarized $X$-boson in which case the production angle 
distribution {\it must} be constant. 

It is interesting to point out in this regard that observation of the flat production 
angle distribution in a two-body decay $X\to P_1 P_2$ does not mean that  
the $X$'s spin is zero. 
In fact, $X$ can have {\it any spin} $J$ but, for the flat distribution,  it must 
be produced unpolarized.  In turn, this implies that all helicity fractions 
must be equal $2 f_{z0} = f_{z1} = f_{z2} = ....f_{zJ}$.  In general, 
each helicity fraction $f_{zm}$ is  a sum of many terms
\be
f_{zm} \sim \sum_{ab} |g_{ab \to X,m}|^2 \tilde f_{a} \tilde f_{b}\,,
\ee
where $a,b$ are the two partons whose collision produces the resonance $X$, 
$g_{ab \to X, m}$ are the couplings of the two partons to the resonance 
$X$ in  the helicity state $m$ 
and $\tilde f_{a,b}$ are the parton distribution functions. It follows 
that the equality of {\it all} helicity fractions requires an unnatural
tuning  between the coupling constants and the parton distribution 
functions. For example, a spin-one resonance can only be single-produced 
in $q \bar q$  collisions and,  in general,  there are two helicity fractions 
$f_{z0}$ and $f_{z1}$. Since, as follows from Eq.~(\ref{eq:helicity-ff-ampl1}),  
$f_{z0}$ is proportional to the quark mass squared, it is very unnatural
to expect unpolarized production in the $J=1$ case. For an $X$ particle 
of spin two, unpolarized production requires tuning of the coupling 
constants {\it and} parton distributions since gluon collisions are responsible 
for the $f_{z0}$ and $f_{z2}$ helicity fractions and $q \bar q$ collisions  -- for the $f_{z1}$ helicity fraction.
It is interesting that for a fixed mass $X$ resonance, 
such tuning can only be argued for a particular energy of the
hadron collider; changing the collider energy, at least as a matter of principle, 
 will clearly destroy the tuning and, if $J > 0$, will turn flat 
$\cos \theta^*$ distributions into non-flat.
 While the above discussion 
shows that unpolarized production of the  resonance with non-vanishing 
spin requires a high degree of tuning between coupling constants and 
parton distribution functions, we emphasize that, as a matter of principle, 
observation of a flat production angle distribution does not immediately 
imply that the spin of the resonance is zero. As we discuss next,  
the analysis of helicity angle distributions helps in  distinguishing 
the different spin scenarios.


\subsection{Distributions of helicity angles} 

It follows from Eq.~(\ref{eq:rate-amplitude}) that the 
most general angular distribution depends on  five angles. 
Such a distribution contains information about production of the resonance 
$X$ and its decay into $ZZ \to(f_1\bar{f}_1)(f_2\bar{f}_2)$. 
In this section, we restrict the presentation to angular distributions which are averaged 
over the production angle $\Phi_1$. 
The most general angular distributions are given in the appendix.
To obtain distributions differential in the three helicity angles, 
one can easily integrate over the $\cos\theta^*$ in the formulas 
below. In a similar manner, 
it is easy to obtain one-dimensional projection of any of the five angles that 
describe the decay. In this subsection, we present those distributions for 
spin-zero, spin-one, and spin-two resonances that decay as
$X\to ZZ\to(f_1\bar{f}_1)(f_2\bar{f}_2)$.  Because $Z$ decays 
are involved, distributions depend 
on the parameters $R_{1,2}=2r_{1,2}/(1+r_{1,2}^2)$, where 
$r_{1,2}$ is the ratio of axial to vector couplings of the 
fermions $f_{1,2}$. Specifically, 
$r = {c_A}/{c_V}={t_{\sss 3L}}/({t_{\sss 3L}-2q\sin^2\theta_W})$, 
where $q$ is the fermion charge and 
$t_{\sss 3L}$ is its weak isospin. In this paper we mostly 
consider $Z$ decays to 
charged leptons. For them,  $q=-1$, $t_{\sss 3L}=-1/2$, and $R \simeq0.15$.   
Note that if the first $Z$ decays to charged leptons and the second $Z$ decays to quark 
and antiquark jets,  one needs to average the value of $R_2$ over contributing quark 
flavors since $R$ is different for up and down quarks, 
$R_{\rm up} \simeq0.67,\; R_{\rm down}\simeq 0.94$. 
However, if the quark and antiquark jets cannot be distinguished in the $Z$ decay, 
this is equivalent to $R_2=0$ in the measured angular distributions.

The angular distribution for a {\it spin-zero} resonance is independent of the
production angles. It depends on four free parameters and reads 
%
\begin{eqnarray}
\frac{128\pi \, d\Gamma_{J=0}}{9\,\Gamma\,d\cos\theta_1d\cos\theta_2 d\Phi}
&&  \!\!\!\!   = 4\,(1- f_{++}-f_{--})\,\sin^2\theta_1\sin^2\theta_2   \nonumber \\
&&  
+(f_{++}+f_{--})\left((1+\cos^2\theta_1)(1+\cos^2\theta_2)+4R_{1}R_{2}\cos\theta_1\cos\theta_2\right)  \nonumber \\
 && 
- 2\,(f_{++}-f_{--})\left(R_{1}\cos\theta_1(1+\cos^2\theta_2) + R_{2}(1+\cos^2\theta_1)\cos\theta_2\right) \nonumber \\
&& 
+4 \sqrt{f_{++}(1-f_{++}-f_{--})}\,(R_{1}-\cos\theta_1) \sin\theta_1(R_{2}-\cos\theta_2) \sin\theta_2 \cos(\Phi + \phi_{++}) \nonumber \\
&& 
+4 \sqrt{f_{--}(1-f_{++}-f_{--})}\,(R_{1}+\cos\theta_1) \sin\theta_1(R_{2}+\cos\theta_2) \sin\theta_2 \cos(\Phi - \phi_{--}) \nonumber \\
&&
+2\sqrt{f_{++}f_{--}}\sin^2\theta_1\sin^2\theta_2  \cos (2\Phi + \phi_{++} - \phi_{--}) \,.
\label{eq:higgs-ang1}
\end{eqnarray}
%
We point out that non-zero values of $R_i$ are reflected in preferential 
directions of fermions in $Z$ decays, see e.g. 
terms  $R_1 \cos \theta_1$ and $R_2 \cos \theta_2$ that are present 
in Eq.~(\ref{eq:higgs-ang1}) if parity is violated with $f_{++} \ne f_{--}$.
Equation~(\ref{eq:higgs-ang1}) is the most general angular distribution 
of the decay of a spin-zero particle and, as such,  generalizes many 
similar results presented in the literature. 

We note, however, that in specific cases the number of 
parameters in Eq.~(\ref{eq:higgs-ang1}) 
can be reduced. For example, considering the {\it tree-level coupling} of the 
SM Higgs boson to two $Z$ bosons, we find the relation 
between three helicity amplitudes to be 
$A_{++}=A_{--}=-{A_{00}}/\gamma$, where $\gamma$ is defined in Eq.~(\ref{eq:boost}).
This leads to
\begin{eqnarray}
f_{++}=f_{--} = \frac{1}{\gamma^2+2}\,, ~~~\phi_{++}=\phi_{--} = \pi \,.
\label{eq:higgsfraction-sm}
\end{eqnarray}
It is easy to account for changes in $H \to ZZ \to 4l$ angular distributions 
caused by the radiative corrections in the SM, 
if $a_2^{\rm SM}$ and $a_1^{\rm SM}$ are
 known. Similarly, for a parity-conserving interaction of 
a $J^P=0^-$ particle,  one has $f_{++}=f_{--}=1/2$ and $\phi_{++}-\phi_{--}=\pi$.

The angular distribution for a 
{\it spin-one} particle decaying to $X\to ZZ$ is determined by 
two parameters; we choose them to be 
$f_{+0}$ and $[\phi_{+0} - \phi_{0-}]={\rm arg}(A_{+0}/A_{0-})$.
The angular distribution reads 
%
\begin{eqnarray}
\label{eq:zprime-ang2}
\frac{512\pi \, d\Gamma_{J=1}}{27\, \Gamma\, d\cos\theta^*d\cos\theta_1d\cos\theta_2 d\Phi}
&&  \!\!\!\! = \Bigl\{1+\cos^2\theta^*\Bigr\}  \times\Bigl\{ 1-\cos^2\theta_1\cos^2\theta_2 
+(1-4f_{+0}) (R_1\cos\theta_1\sin^2\theta_2+R_2\sin^2\theta_1\cos\theta_2)
\Bigr. \nonumber\\
&& 
\Bigl.
+ \sqrt{8f_{+0}(1-2f_{+0})}\sin\theta_1 \sin\theta_2 
( R_1R_2 - \cos\theta_1\cos\theta_2) \cos(\Phi+  [\phi_{+0} - \phi_{0-}]) \Bigr\} \,.
\end{eqnarray}
%
This distribution differs from a recent result presented in Ref.~\cite{spin1}. 
The origin of the difference is the reality of the two effective couplings 
$g^{(1)}_{\sss 1,2}$ assumed in that reference. Indeed, if the two couplings 
are real then, independent of their actual values, $f_{+0}=1/4$, 
as follows from Eq.~(\ref{eq:helicity-ampl1}). 
Then, the terms 
in Eq.~(\ref{eq:zprime-ang2}) that are 
{\it linear} in $R_i$ do not contribute.  
In the more general case of two complex couplings,  
the preferential polarization of the $Z$ boson appears and gets reflected 
in the preferential direction of, say, the negatively charged lepton in $Z$-decays. 
We also note that the angular dependence in Eq.~(\ref{eq:zprime-ang2})
is a product of the production angle distribution and the 
distribution of three helicity angles.

Finally, we consider a {\it spin-two} resonance.
The normalized angular distribution is
%
\begin{eqnarray}
&&\frac{2048\pi \; d \Gamma_{J=2}}{45\, \Gamma d\cos\theta^\ast d\cos\theta_1d\cos\theta_2 d\Phi}
= 
\Bigl\{ (2-2f_{z1}+f_{z2})-6(2-4f_{z1}-f_{z2})\cos^2\theta^\ast+3(6-10f_{z1}-5f_{z2})\cos^4\theta^\ast \Bigr\}  \nonumber \\ 
\nonumber \\
&& ~~ \times  \Bigl\{  4\,(1- f_{++}-f_{--}-2f_{+-}-2f_{+0}-2f_{0-})\,\sin^2\theta_1\sin^2\theta_2  \nonumber \\
&& ~~ ~~ +(f_{++}+f_{--})\left((1+\cos^2\theta_1)(1+\cos^2\theta_2)+4R_{1}R_{2}\cos\theta_1\cos\theta_2\right)   \nonumber \\
&& ~~ ~~ -2\,(f_{++}-f_{--}) \left(R_{1}\cos\theta_1(1+\cos^2\theta_2) + R_{2}(1+\cos^2\theta_1)\cos\theta_2\right)   \nonumber \\
&& ~~ ~~ +4 \sqrt{f_{++}(1-f_{++}-f_{--}-2f_{+-}-2f_{+0}-2f_{0-})} 
(R_1-\cos\theta_1)\sin\theta_1(R_2-\cos\theta_2) \sin\theta_2\cos ( \Phi + \phi_{++} ) \nonumber \\
&& ~~ ~~ +4 \sqrt{f_{--}(1-f_{++}-f_{--}-2f_{+-}-2f_{+0}-2f_{0-})} 
 (R_1+\cos\theta_1) \sin\theta_1(R_2+\cos\theta_2)\sin\theta_2\cos ( \Phi - \phi_{--} ) \nonumber \\
&& ~~ ~~ +2\sqrt{f_{++}f_{--}} \sin^2\theta_1 \sin^2\theta_2 \cos(2\Phi+\phi_{++}-\phi_{--})  \Bigl\} \nonumber \\
&& +8 \Bigl\{ (f_{z1} + f_{z2}) + 3(2-3f_{z1}-2f_{z2}) \cos^2\theta^\ast - (6-10f_{z1}-5f_{z2})  \cos^4\theta^\ast \Bigr\} \nonumber \\ 
&& ~~ \times \Bigl\{ (f_{+0}+f_{0-})(1-\cos^2\theta_1\cos^2\theta_2) 
   -(f_{+0} - f_{0-})( R_1 \cos\theta_1 \sin^2\theta_2 + R_2 \sin^2\theta_1 \cos\theta_2)  \nonumber \\
&&  ~~ ~~ + 2  \sqrt{f_{+0}f_{0-}} \sin\theta_1 \sin\theta_2(R_1R_2 - \cos\theta_1\cos\theta_2) \cos (\Phi+[\phi_{+0}-\phi_{0-}]) \Bigr\} \nonumber \\ 
&&  + \Bigl\{ (6-2f_{z1}-5f_{z2})-6(2-2f_{z1}-3f_{z2})\cos^2\theta^\ast+(6-10f_{z1}-5f_{z2})\cos^4\theta^\ast \Bigr\}   \nonumber \\ 
&& ~~ \times f_{+-}  \Bigl\{ (1+\cos^2\theta_1)(1+\cos^2\theta_2) - 4R_1R_2\cos\theta_1\cos\theta_2  \Bigr\} \,,
\label{eq:mixedgravitontotal}
\end{eqnarray}
%
where the dependent parameter $f_{00}$ is expressed as $(1-f_{++}-f_{--}-2f_{+-}-2f_{+0}-2f_{0-})$.

There are two ways to obtain an
angular distribution in the three helicity angles from
Eq.~(\ref{eq:mixedgravitontotal}). One way is to 
integrate over the $\cos\theta^\ast$ in Eq.~(\ref{eq:mixedgravitontotal}); 
note that Eq.~(\ref{eq:mixedgravitontotal}) is written in such a way 
that the dependence on $\cos \theta^*$ in each term is factored out.
Upon integrating  over $\cos \theta^*$, the dependence on 
$f_{z1}$ and $f_{z2}$ must disappear since the distribution over 
helicity angles is independent of the production mechanism. We note that 
this independence suggests an alternative way to obtain the distribution 
of the helicity angles from Eq.~(\ref{eq:mixedgravitontotal}). 
Imagine that $X$ is produced unpolarized, which amounts to setting 
$f_{z1}=f_{z2}=0.4$ in Eq.~(\ref{eq:mixedgravitontotal}). In this case 
coefficients of all $\cos \theta^*$-dependent terms 
in Eq.~(\ref{eq:mixedgravitontotal}) vanish and integration over that 
angle becomes trivial.

Even if $X$ is produced unpolarized, it is still possible to separate spin hypotheses 
by analysis of the helicity angles. The opposite situation is also possible when 
helicity angles provide no separation while productions angle distributions are
different. In fact, joint analysis of production and helicity angular distributions 
is the most efficient way to separate different  hypotheses.
However, there is a special case of the helicity fractions $f_{+-}=f_{+0}=f_{0-}=0$
and production polarization $f_{z1}=f_{z2}=0.4$ where  the angular distributions 
in Eqs.~(\ref{eq:higgs-ang1}) and~(\ref{eq:mixedgravitontotal}) become identical.
The situation $f_{+-}=f_{+0}=f_{0-}=0$ is realized when only terms with $c_2$, 
$c_3$, and $c_5$ contribute to helicity amplitudes in Eq.~(\ref{eq:ampl-spin2}), 
which becomes equivalent to Eq.~(\ref{eq:ampl-spin0}).
This is possible, though very unnatural, situation where the $A_{+-}$ amplitude
contributes in production but vanishes in the decay mechanism. 
However, this possibility cannot be ruled out experimentally unless 
different spin projections in the production are probed.

It is a general feature of Eq.~(\ref{eq:mixedgravitontotal}) written
for spin-two decays that it includes both Eqs.~(\ref{eq:higgs-ang1}) 
and~(\ref{eq:zprime-ang2}) written for spin-zero and spin-one, respectively.
Note that Eq.~(\ref{eq:mixedgravitontotal}) factorizes
into the product of a function of $\cos\theta^\ast$ and 
a function of three helicity angles under the following three conditions.
In the case $f_{+-}=f_{+0}=f_{0-}=0$, the helicity angular distribution is given
by Eq.~(\ref{eq:higgs-ang1}) and the production angle distribution 
appears in Eq.~(\ref{eq:grav-ang2mix2d}).
This trivially reduces to the case of unpolarized production discussed above.
In the other case $f_{0-}+f_{+0}=1/2$, and therefore the other $f_{kl}=0$,
the helicity angular distribution is given by Eq.~(\ref{eq:zprime-ang2}) 
and the production angle distribution can be read from 
Eq.~(\ref{eq:grav-ang2mix2d}).
In this case it happens that, under the condition $f_{z1}=0.6$ and $f_{z2}=0$, the
function of the production angles is proportional to ($1+\cos^2\theta^\ast$)
and Eqs.~(\ref{eq:zprime-ang2}) and~(\ref{eq:mixedgravitontotal}) 
become identical. However, this is only a coincidence and the full 
angular distributions which also include angle $\Phi_1$ do not
match, as for example Eqs.~(\ref{eq:zprime-ang12d}) 
and~(\ref{eq:grav-ang2mix2d}) are not identical in this case.
Finally, the last term in Eq.~(\ref{eq:mixedgravitontotal}) corresponds to
the non-zero $A_{+-}=A_{-+}$ amplitudes and is unique to the spin-two case,
though its helicity angular distribution is very similar to the other transverse
term which remains in the case $f_{++}=f_{--}=1/2$.


\section{Monte Carlo Simulation} 
\label{sect5}

We have written a Monte Carlo program to simulate the production 
and decay of spin-zero, spin-one, and spin-two resonances 
in hadron-hadron collisions, including all spin correlations,
in the processes $gg(q\bar q) \to X\to ZZ\to 4l$. 
A special feature of our program, that distinguishes it from 
other recent implementations \cite{Hagiwara:2008jb}, is that it 
includes the most general couplings of the $X$ particle 
to gluons and  fermions in production and to $Z$ bosons in decay, 
as explained in the  previous sections. Extension 
to other final states, including hadronic $Z$ decays, is 
straightforward.

The spin-zero resonance is produced in collisions of gluons, and 
the spin-one resonance is produced in $q \bar q$ collisions. 
For the spin-two 
resonance, both partonic channels are included in the simulation. 
  Since, in the general case, 
the relative strength of $gg X$ and $q \bar q X$ couplings is not 
known, the program allows the request that the spin-two $X$ particle is produced 
with fixed relative frequency in $q \bar q$ and 
$gg$ collisions. We note that both the resonance $X$ and 
the two $Z$ bosons are considered to be on-shell in our program; all 
the off-shell effects are neglected. The narrow width approximation 
is a reasonable assumption for the $Z$ bosons, but it is not 
possible to say a priori  if this also is a good 
approximation for the resonance $X$. 
The applicability of the narrow width approximation 
for the particle $X$ is the assumption built into our program,  but 
it is relatively straightforward  to remove this restriction in 
the future.

The program can output both weighted and 
unweighted events, depending on the requested mode of operation.
 Weighted events are typically  used for fast calculations  of simple 
one-dimensional  
distributions and for debugging purposes. 
Unweighted events, on the other hand, are used 
to interface the results of our program  to programs that simulate 
realistic detector response.

We now turn to the discussion of how backgrounds of the resonance $X$ for 
four-lepton production in hadron collisions are simulated. Such backgrounds arise 
from $Z(\gamma)Z(\gamma)$, $Z b \bar b$, $t \bar{t}$, $W^+W^- b \bar b$, 
$WWZ$, $t \bar t Z$, $4b$ 
etc, where we assume that $b$-quarks decay semileptonically. 
Backgrounds that involve $b$-decays into leptons can be controlled 
by requiring that leptons are isolated. Other backgrounds can be 
strongly suppressed  by requiring that the invariant masses of lepton 
pairs are close to the $Z$-boson mass, that there is no 
missing energy in the event, and that the four tracks originate
from the same vertex near the interaction region.
As a result, $pp \to ZZ \to 4l$ is the major irreducible background 
that  survives all the possible selection requirements~\cite{exphiggs1, exphiggs2}.  
We emphasize that, while it is possible to understand gross features 
of the backgrounds, it is conceivable that subtle (but relevant) details 
of the angular distributions  due 
to background processes may  weakly depend on the {\it exact} background 
composition.  Such an exact composition is hard to predict 
theoretically with any degree of confidence but, fortunately, this is not 
necessary.  Indeed, this problem 
can be solved by using sideband analysis. This possibility is 
incorporated into our analysis discussed in the next section, 
but its detailed study is beyond the scope of the present paper. 
Here, we restrict ourselves to $pp \to ZZ \to 4l$ as the only 
background source and  simulate it using Madgraph~\cite{madgraph}.

\begin{figure}[t]
\centerline{
\setlength{\epsfxsize}{0.25\linewidth}\leavevmode\epsfbox{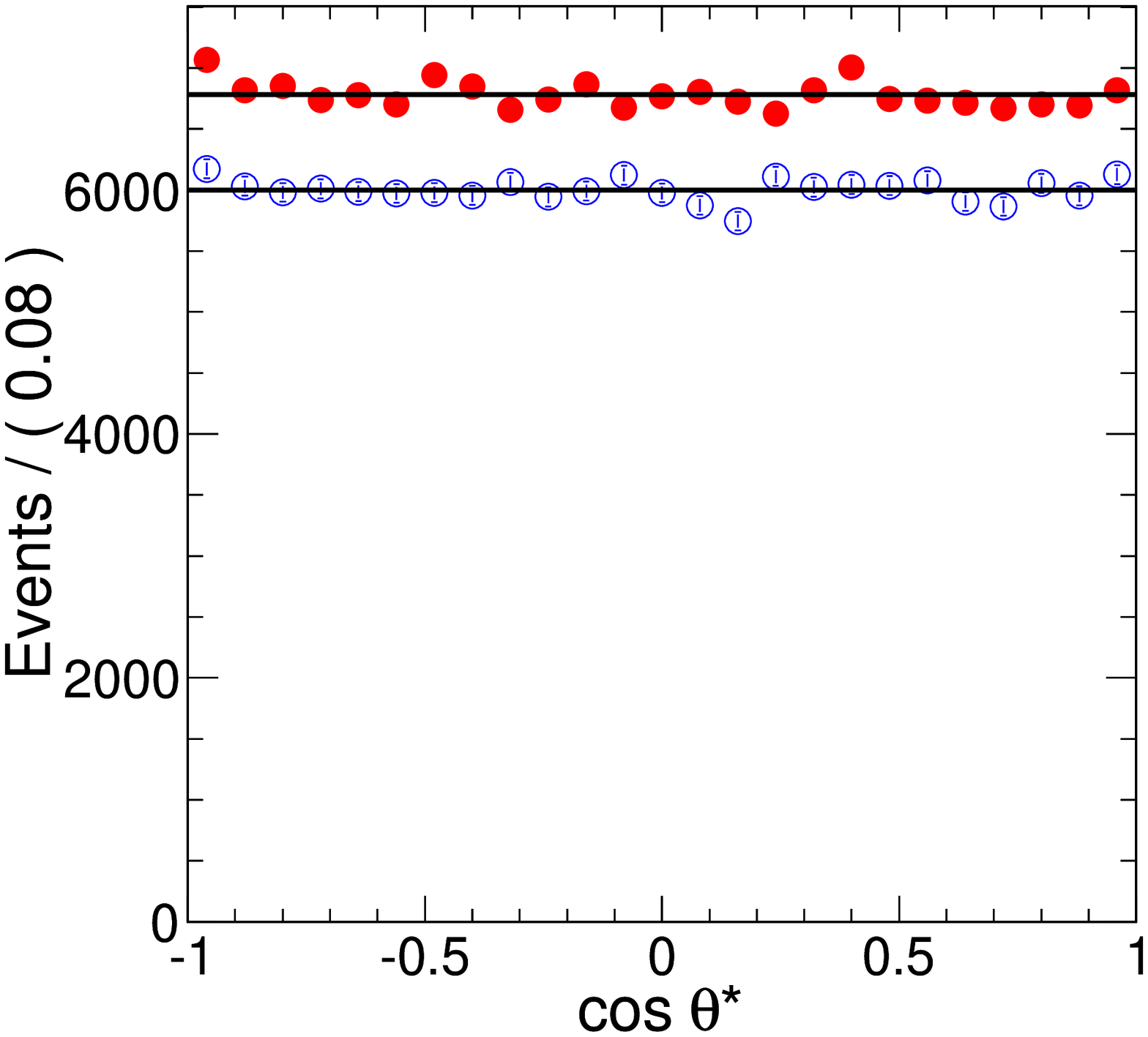}
\setlength{\epsfxsize}{0.25\linewidth}\leavevmode\epsfbox{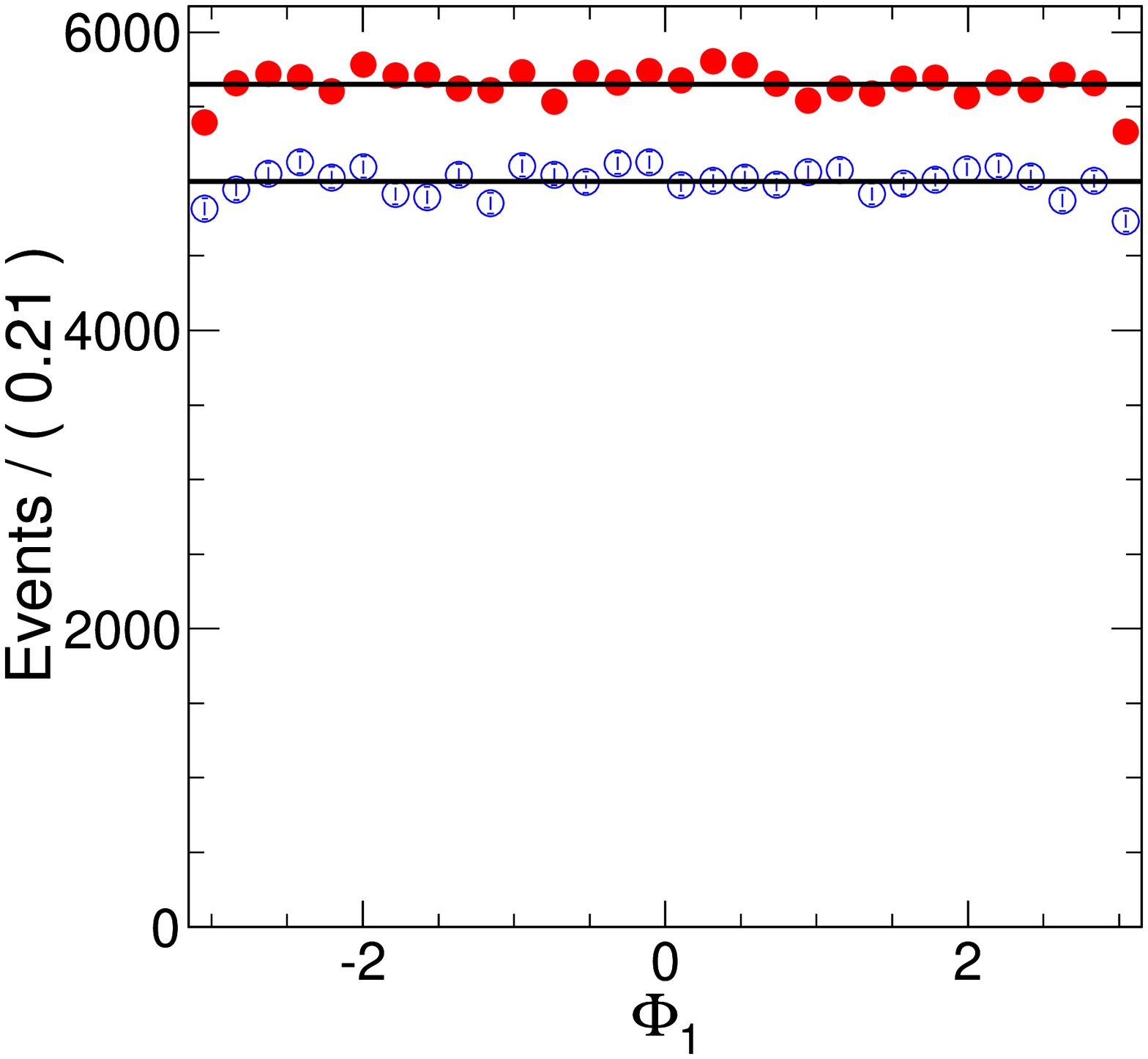}
\setlength{\epsfxsize}{0.25\linewidth}\leavevmode\epsfbox{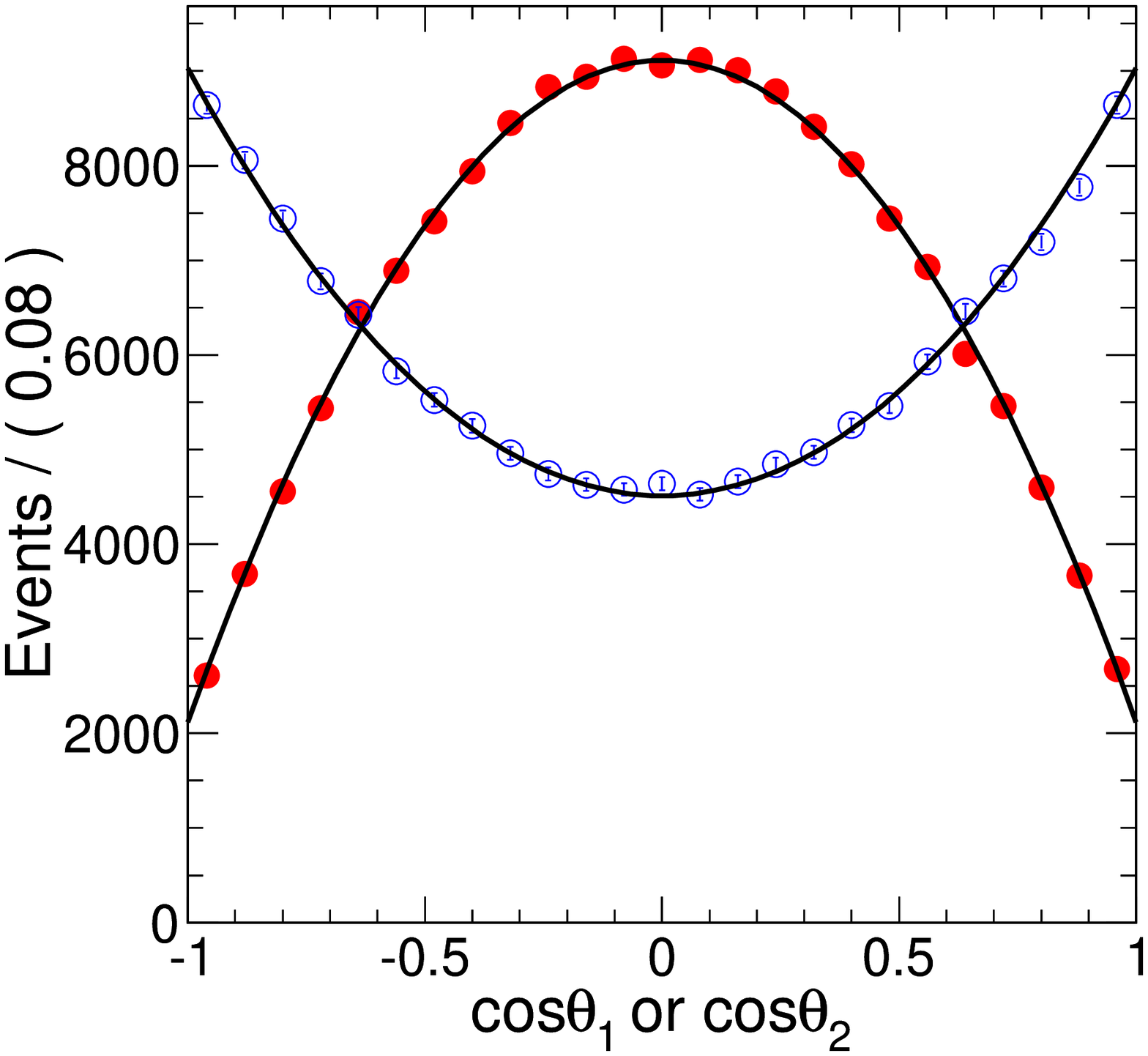}
\setlength{\epsfxsize}{0.25\linewidth}\leavevmode\epsfbox{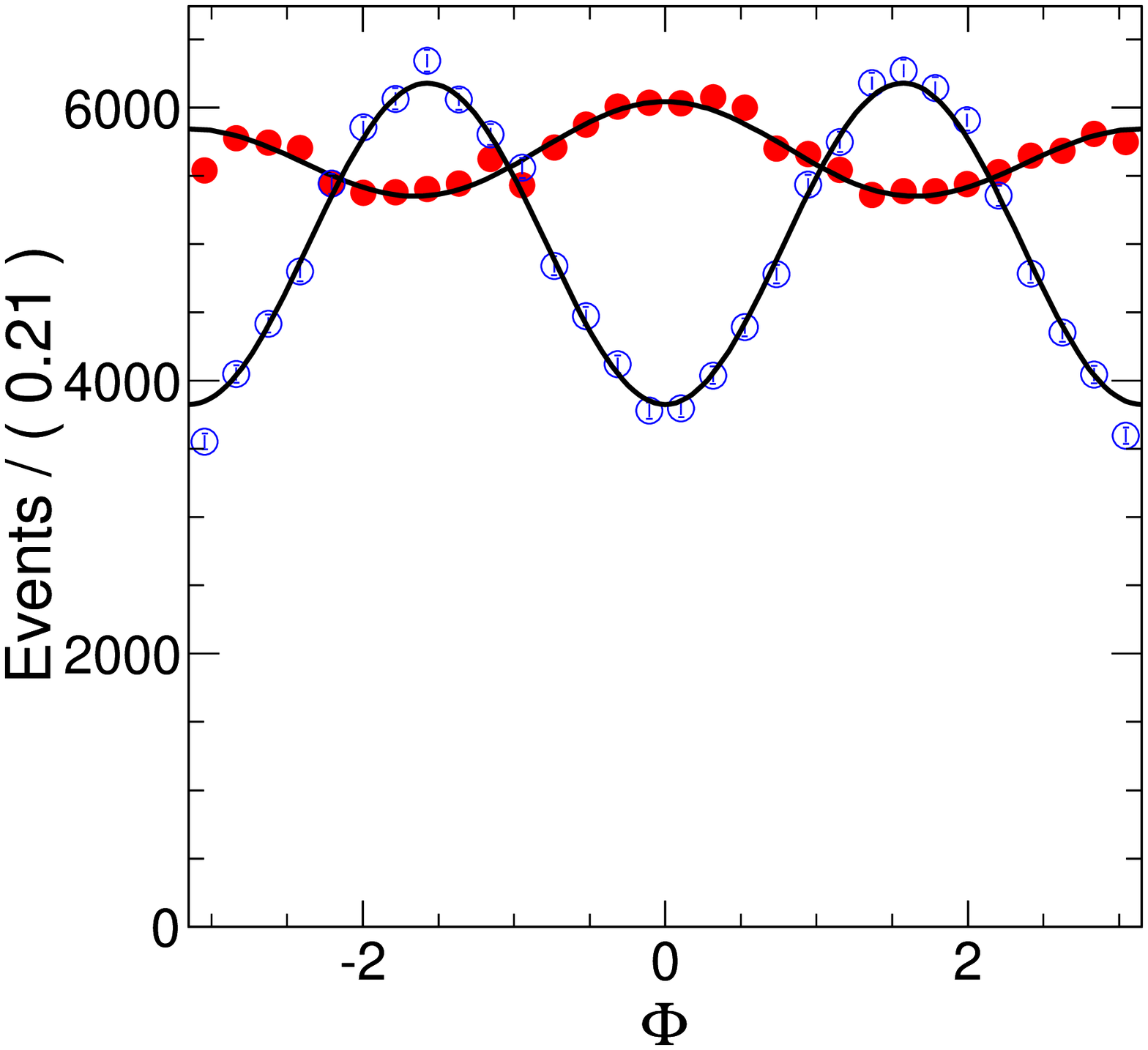}
}
\centerline{
\setlength{\epsfxsize}{0.25\linewidth}\leavevmode\epsfbox{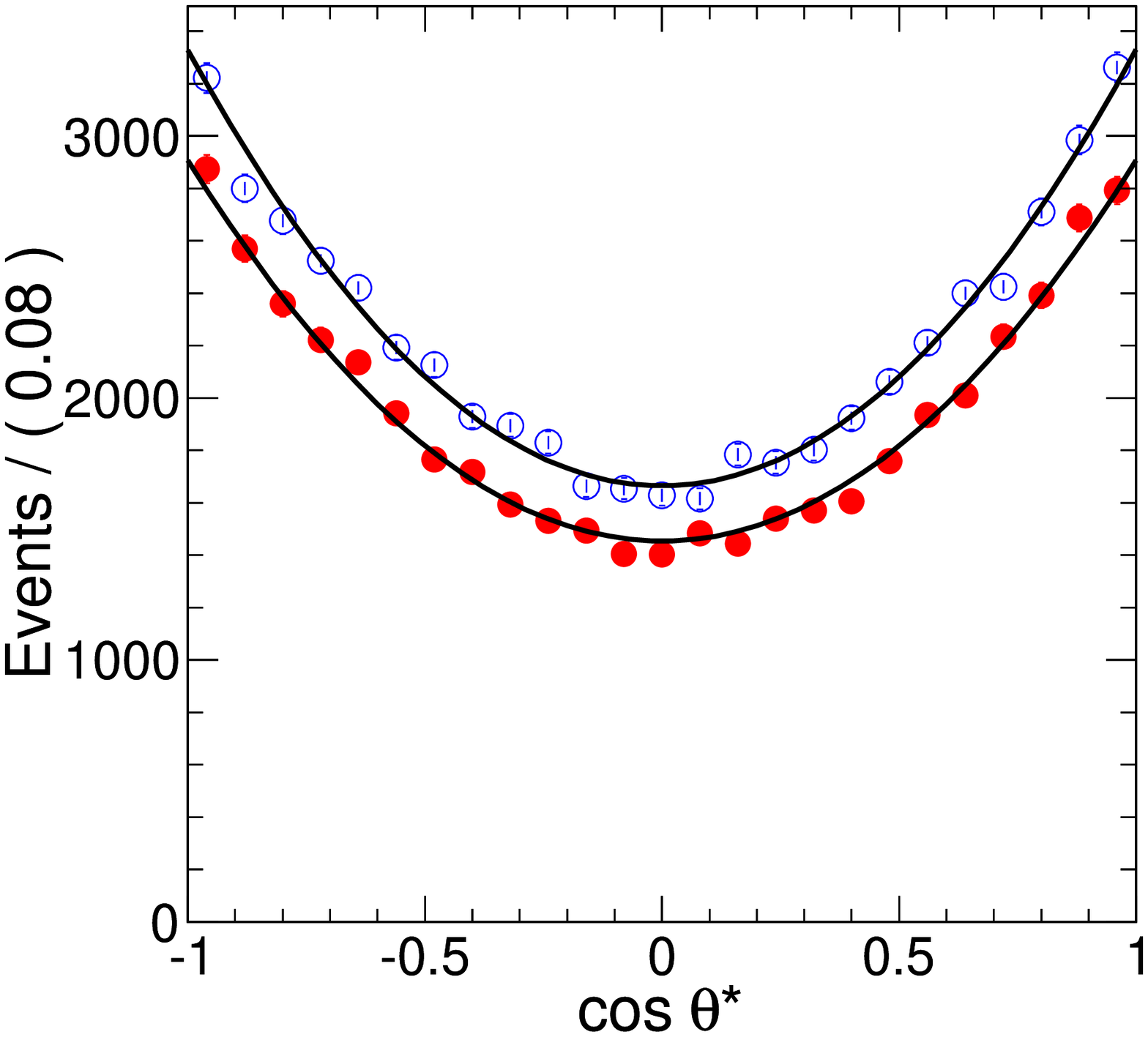}
\setlength{\epsfxsize}{0.25\linewidth}\leavevmode\epsfbox{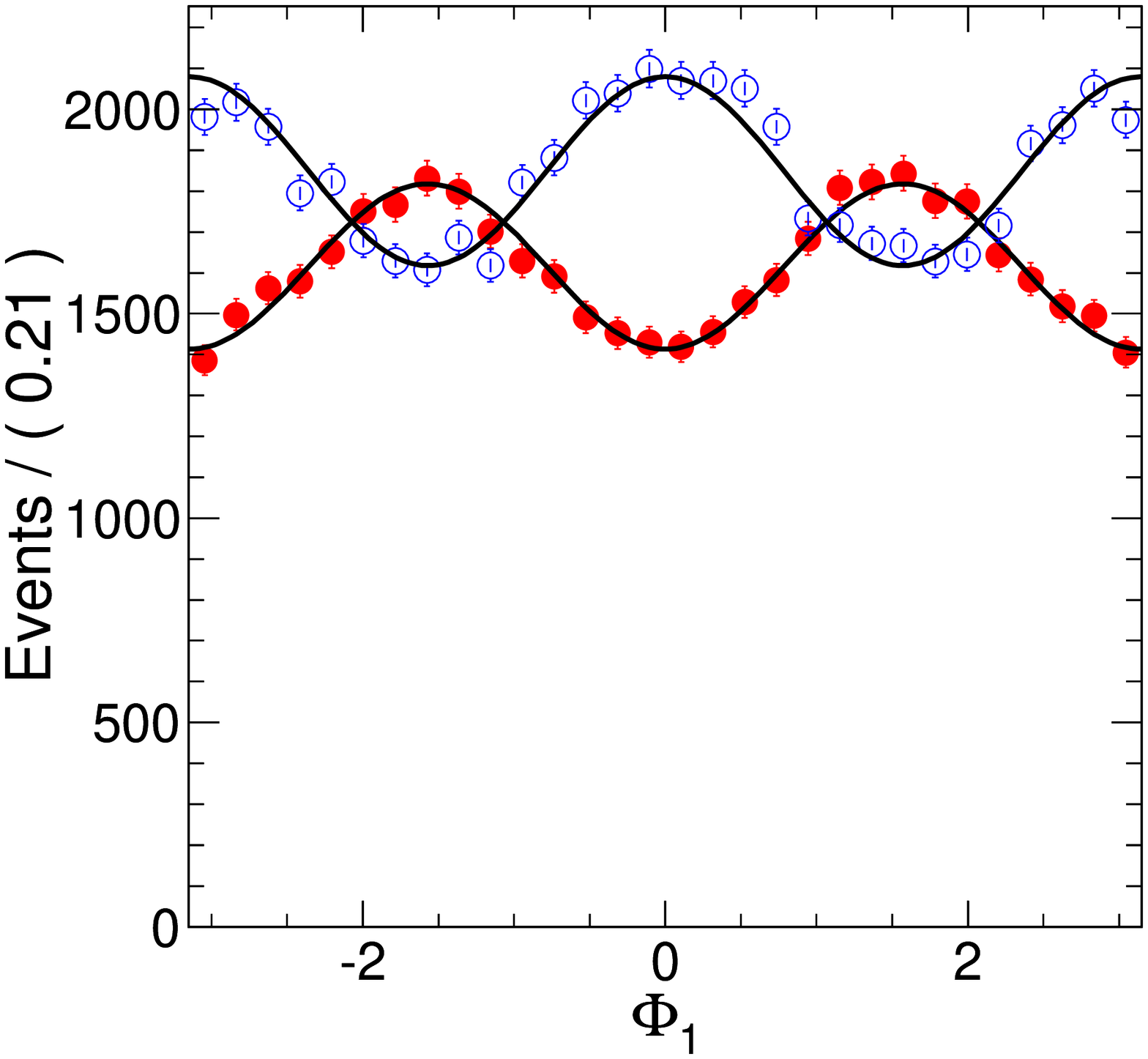}
\setlength{\epsfxsize}{0.25\linewidth}\leavevmode\epsfbox{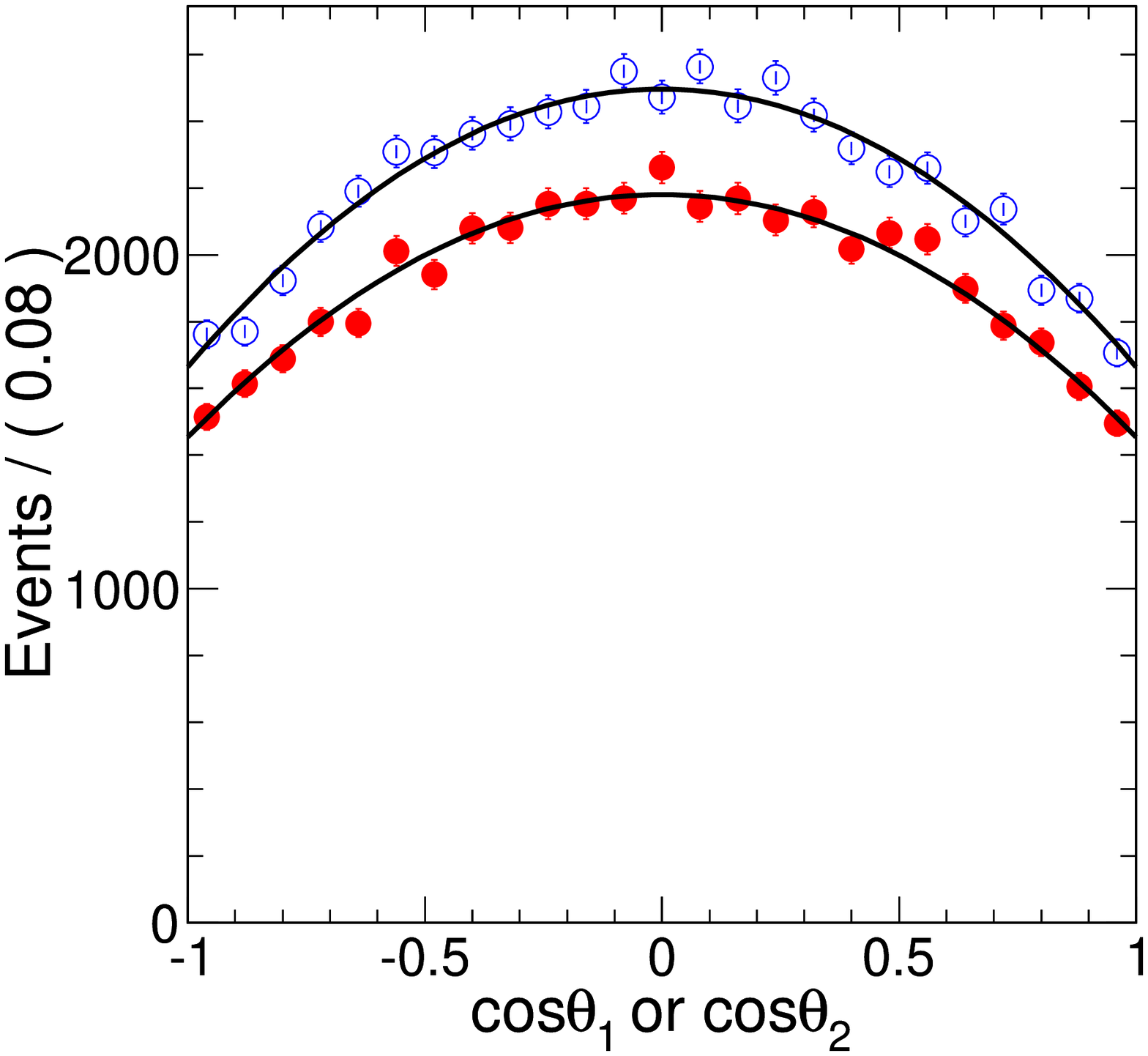}
\setlength{\epsfxsize}{0.25\linewidth}\leavevmode\epsfbox{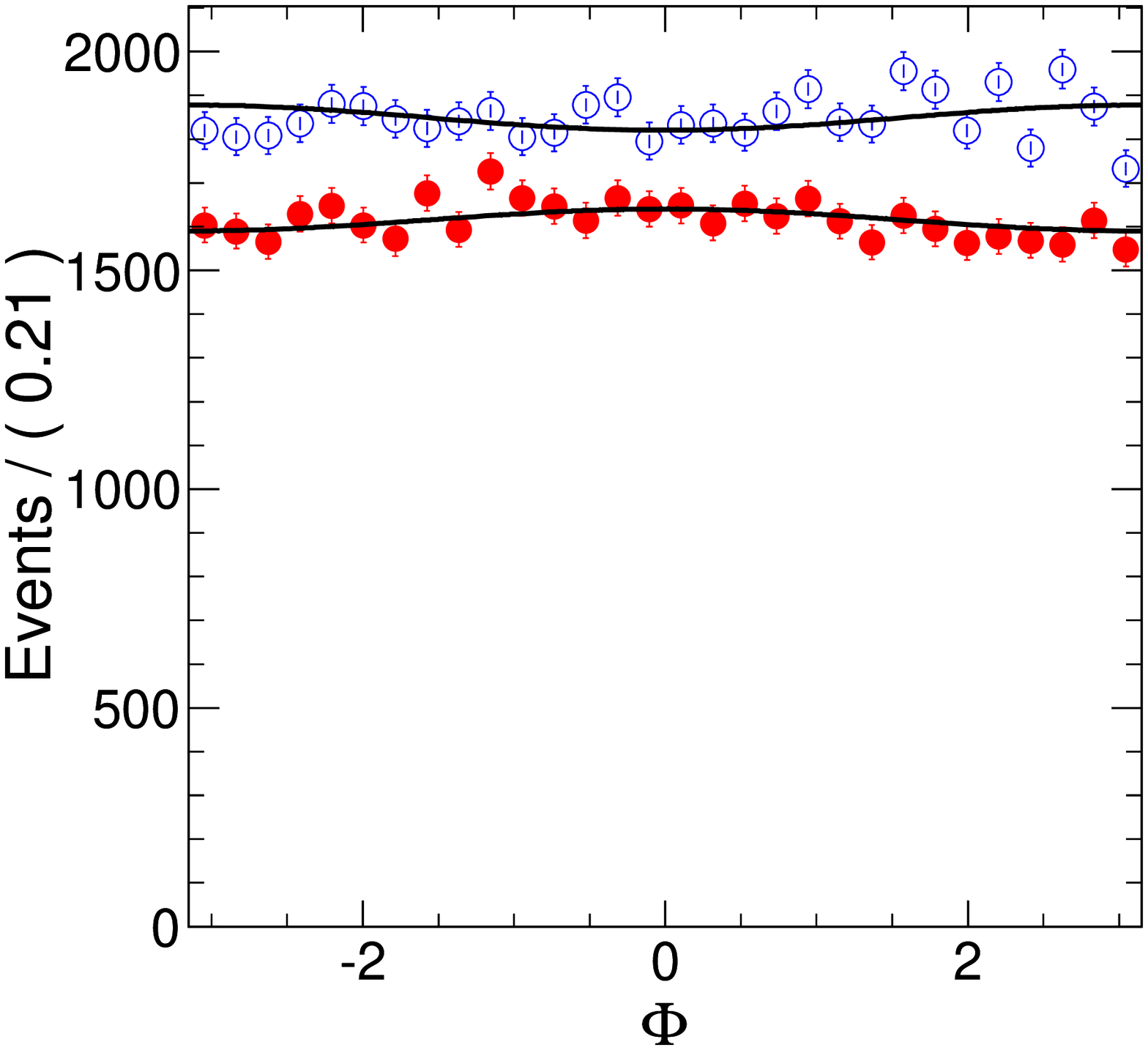}
}
\centerline{
\setlength{\epsfxsize}{0.25\linewidth}\leavevmode\epsfbox{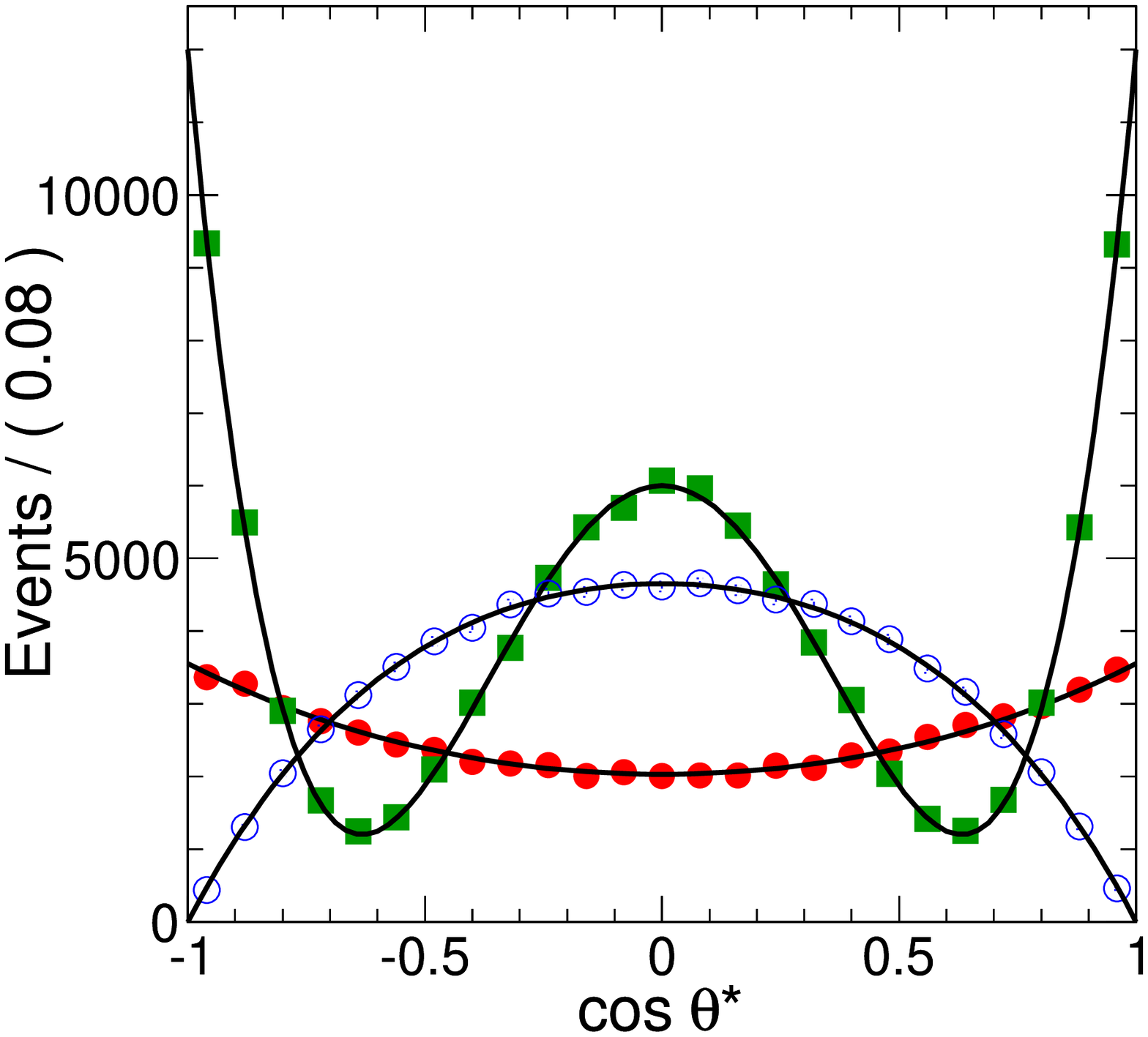}
\setlength{\epsfxsize}{0.25\linewidth}\leavevmode\epsfbox{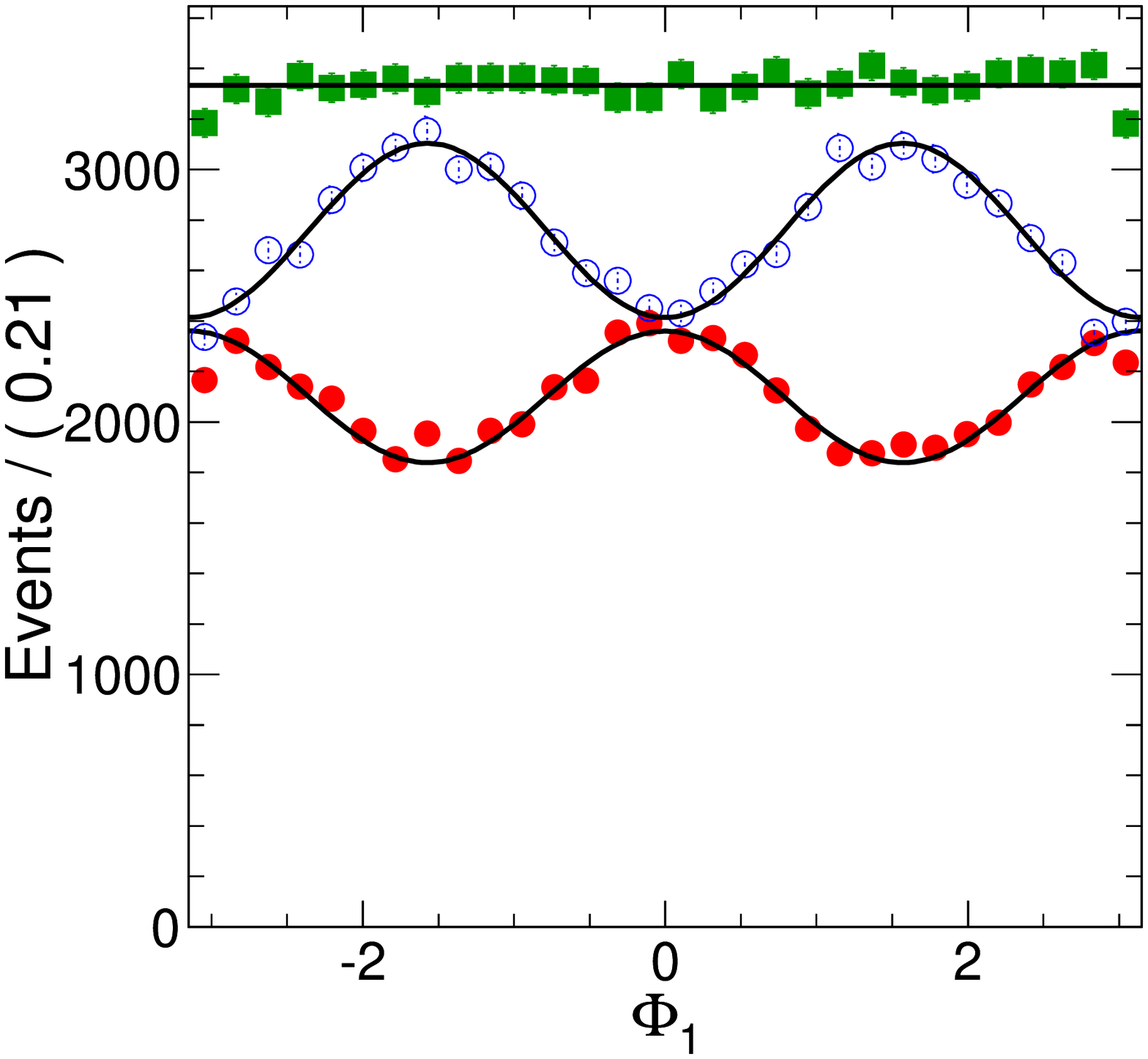}
\setlength{\epsfxsize}{0.25\linewidth}\leavevmode\epsfbox{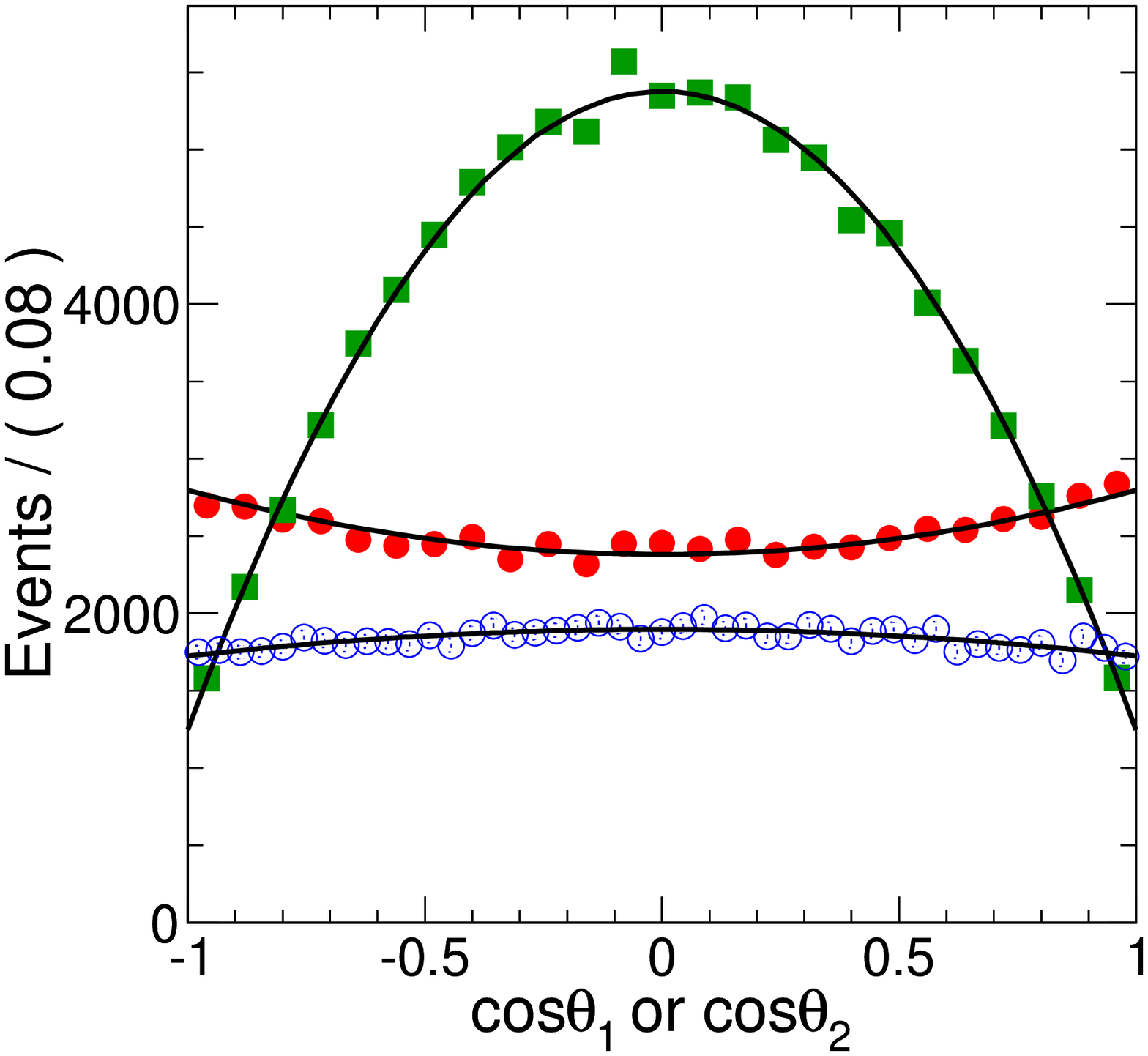}
\setlength{\epsfxsize}{0.25\linewidth}\leavevmode\epsfbox{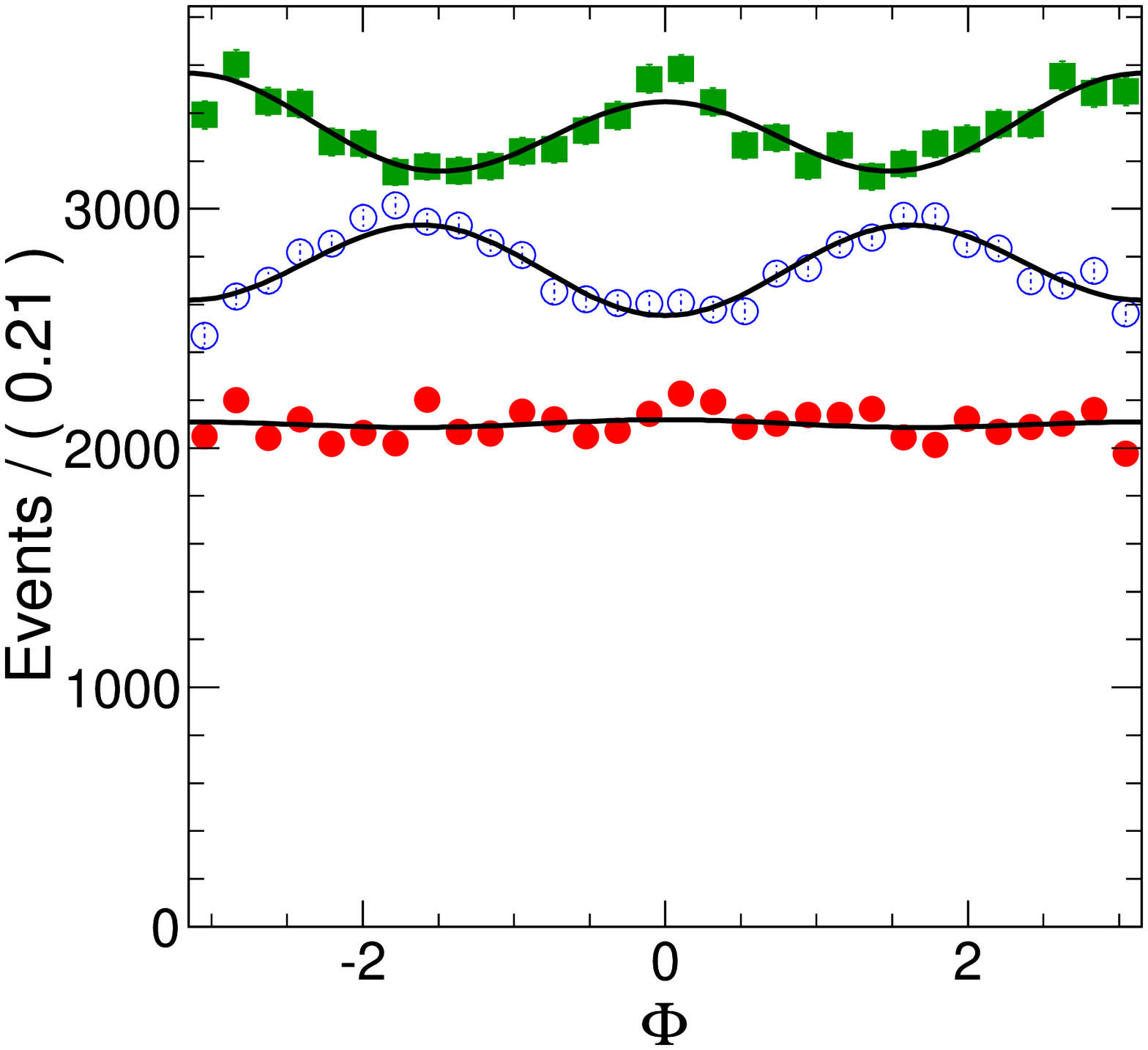}
}
\centerline{
\setlength{\epsfxsize}{0.25\linewidth}\leavevmode\epsfbox{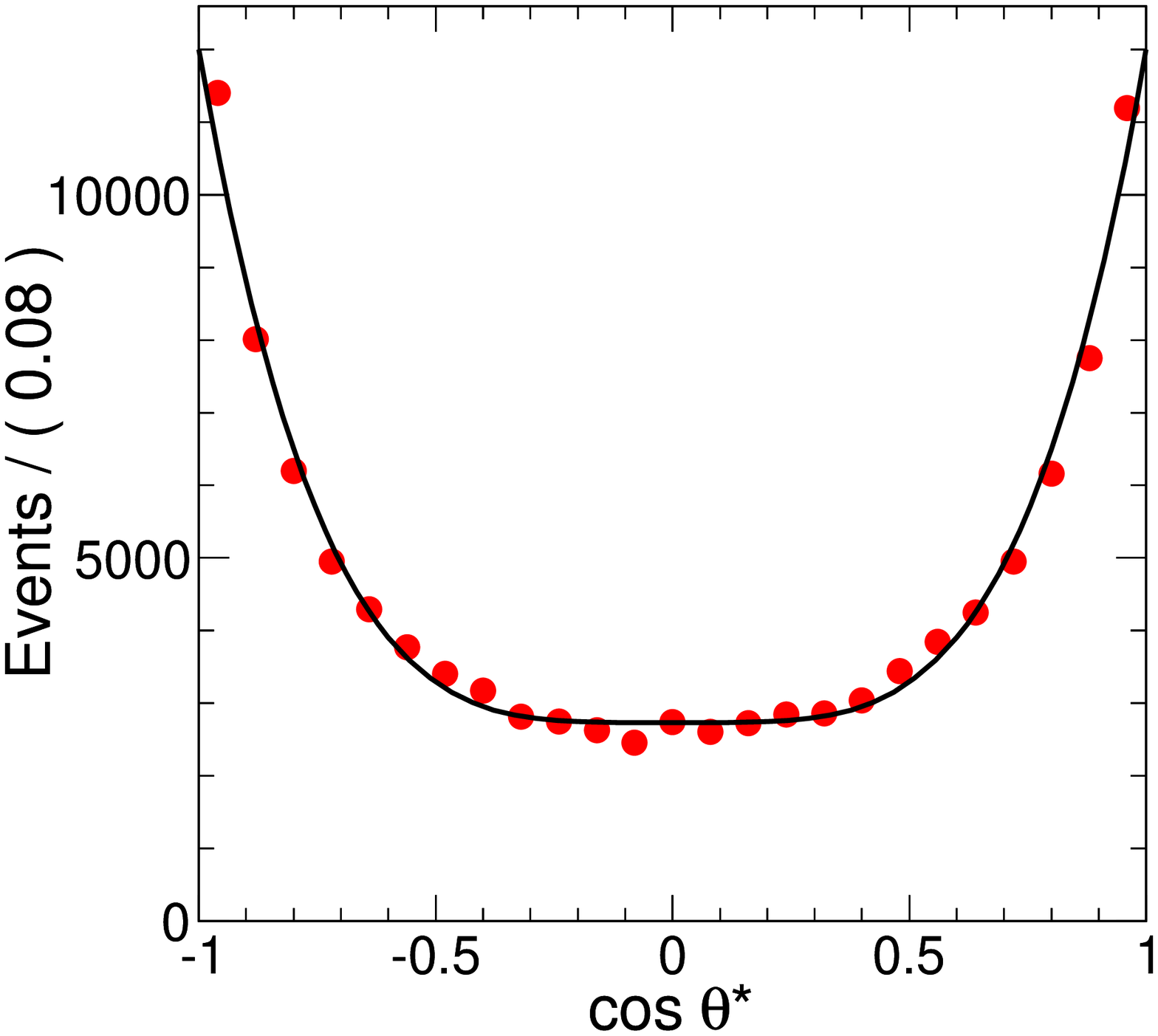}
\setlength{\epsfxsize}{0.25\linewidth}\leavevmode\epsfbox{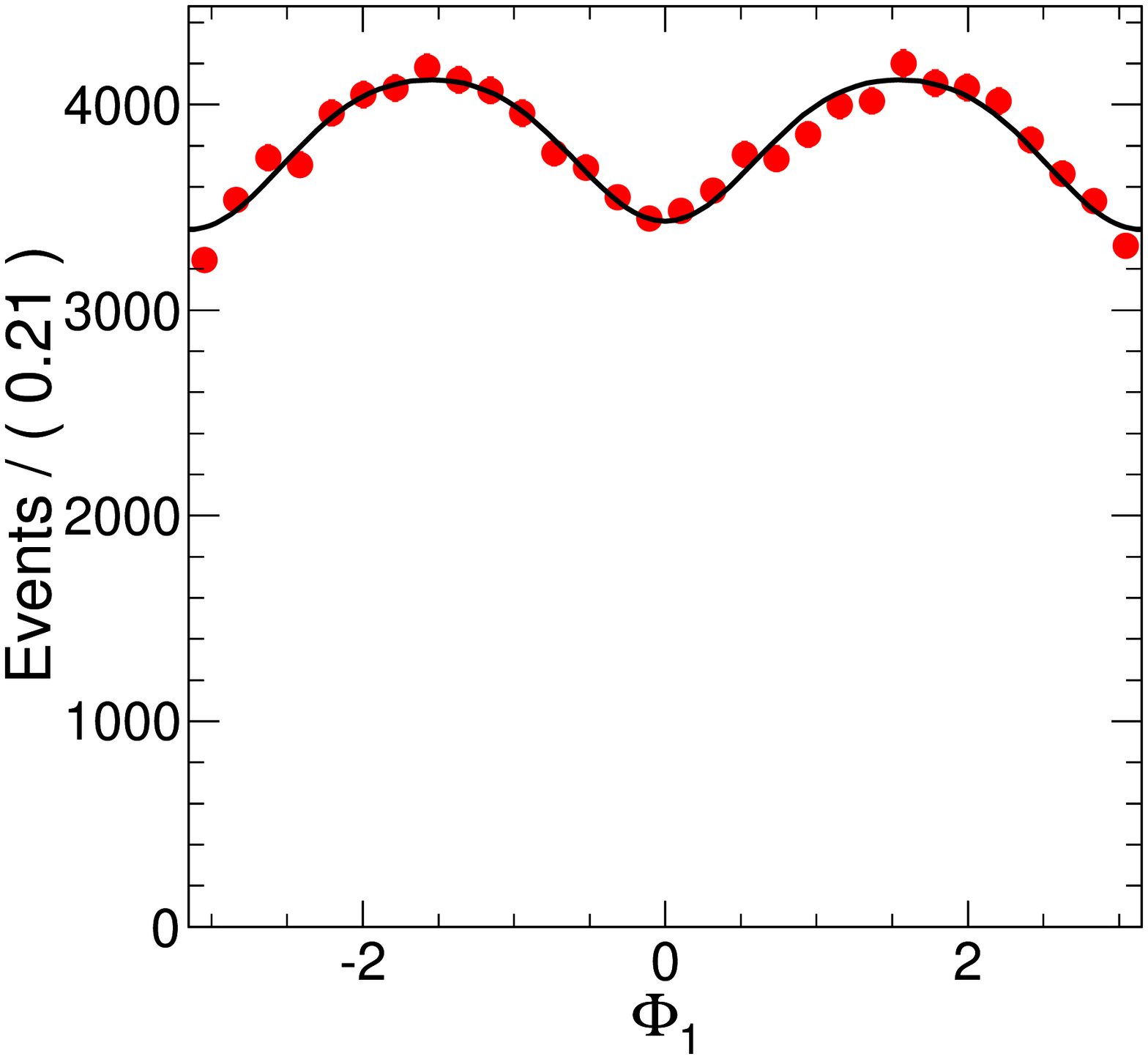}
\setlength{\epsfxsize}{0.25\linewidth}\leavevmode\epsfbox{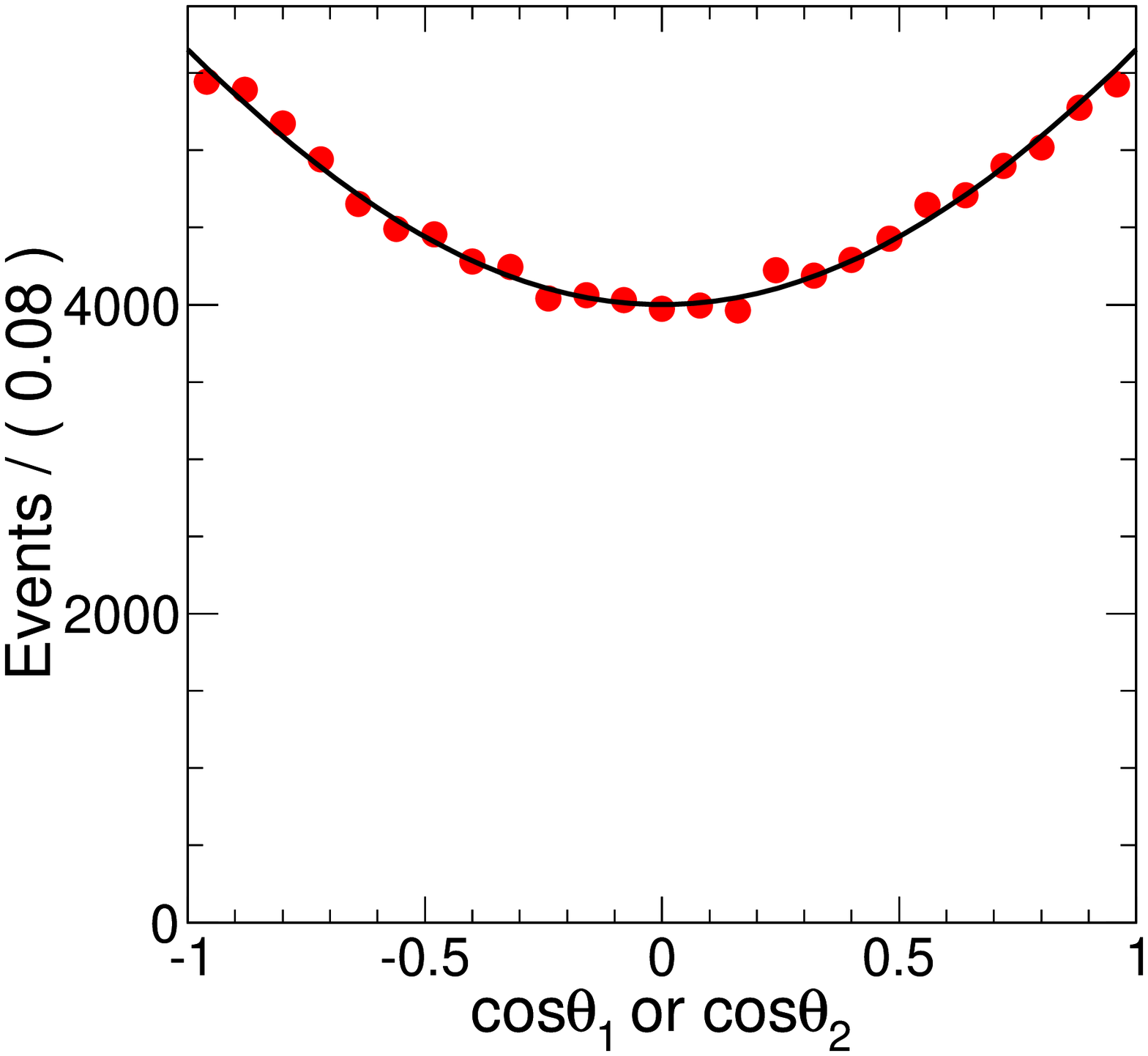}
\setlength{\epsfxsize}{0.25\linewidth}\leavevmode\epsfbox{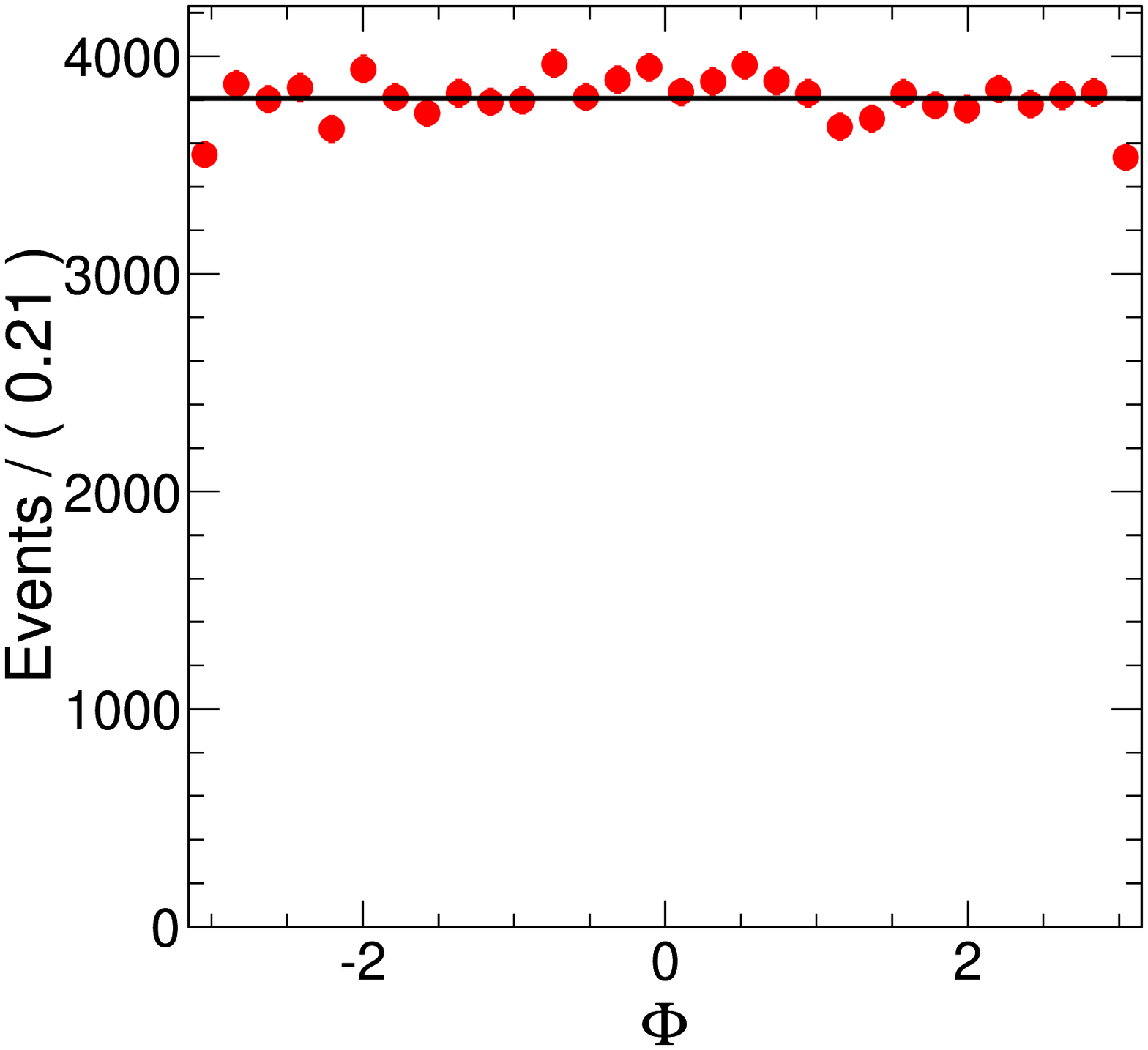}
}
\caption{
Distribution of the $\cos\theta^*$ (left), $\Phi_1$ (second from the left),
$\cos\theta_1$ and $\cos\theta_2$ (second from the right), and $\Phi$ (right)
generated for $m_{\sss X}=250$ GeV with the program discussed in the text
(unweighted events shown as points with error bars)
and projections of the ideal angular distributions given  in 
the text (smooth lines).
The four sets of plots from top to bottom show the models discussed in
Table~\ref{table-scenarios} for spin-zero $0^+$ and $0^-$ (top),
spin-one $1^+$ and $1^-$ (second row from top),
spin-two $2_m^+$, $2_L^+$, and $2^-$ (third row from top), and
the bottom row shows distributions in background generated with Madgraph
(points with error bars) and empirical shape (smooth lines).
The $J^+$ distributions are shown with solid red points and
$J^-$ distributions are shown with open blue points, while the
$2_m^+$ and $2_L^+$ are shown with red circles and green squares, respectively.
}
\label{fig:generated-angles}
\end{figure}

In order to illustrate MC simulation, 
compare it to the derived angular distributions,
provide examples of the data analysis techniques, 
and understand the statistical power of the
proposed approaches; we choose seven scenarios
which cover all spin and parity combinations.
They are described in Table~\ref{table-scenarios}. Note that 
for the $2^+$ graviton-like resonance, 
we consider two models with different couplings to matter and gauge fields.
The distributions of all five angles --
$\cos\theta^*$, $\Phi_1$, $\cos\theta_1$, $\cos\theta_2$, and $\Phi$ -- 
for the seven models in Table~\ref{table-scenarios}
and the $pp \to ZZ \to 4l$ background process  
are shown  in Fig.~\ref{fig:generated-angles}.
These distributions were generated with our program assuming 
that the resonance   mass  is $m_{\sss X}=250$~GeV. Throughout 
the paper we consider $\sqrt{s}= 14~{\rm TeV}$ proton-proton collisions 
and use the CTEQ6L1    parton distribution functions \cite{cteq1, cteq2}.
Note that the distributions of $\cos\theta_1$ and $\cos\theta_2$ are identical 
and are combined in one plot in each case.  
Projections of the ideal angular distributions 
derived in the previous sections  agree
well with simulated distributions. A glance at Fig.~\ref{fig:generated-angles} 
suggests that different hypothesis about resonance quantum numbers 
can be efficiently separated if all five angles are analyzed simultaneously.
Of course,  correlations in the multi-dimensional space of all angles 
are important for full separation power and those correlations cannot be easily 
illustrated. We fully exploit those correlations in the angular analysis
discussed in the next section.
\begin{table}[t]
\caption{
The list of scenarios chosen for the analysis of 
the production and decay of an exotic $X$ particle
with quantum numbers $J^P$. 
For the
two $2^+$ cases,
the superscripts $m$ (minimal)  and $L$ (longitudinal) 
distinguish  two scenarios, 
 as discussed in the last column. 
When relevant, the relative fraction of 
$gg$ and $q\bar{q}$ production is taken to be
1:0 at $m_{\sss X}=250$ GeV and 3:1 at  $m_{\sss X}=1$ TeV. 
The spin-zero $X$ production mechanism does not affect the angular distributions
and therefore is not specified.
}
\begin{tabular}{cccc}
\hline\hline
\vspace{0.1cm}
scenario ($J^P$) & $X\to ZZ$ decay parameters & $X$ production parameters  & comments \\
\hline
$0^+$ & $a_1\ne0$ in Eq.~(\ref{eq:ampl-spin0})  &  $gg\to X$ & SM Higgs-like scalar \\
$0^-$ & $a_3\ne0$ in Eq.~(\ref{eq:ampl-spin0})  &  $gg\to X$ & pseudo-scalar  \\
$1^+$ & $g_{\sss 12}\ne0$ in Eq.~(\ref{eq:ampl-spin1}) 
      &  $q\bar{q}\to X$:  $\rho_{\sss 11}$, $\rho_{\sss 12}\ne0$ in Eq.~(\ref{eq:ampl-spin1-qq}) 
      & exotic pseudo-vector \\
$1^-$ & $g_{\sss 11}\ne0$ in Eq.~(\ref{eq:ampl-spin1})  
      &  $q\bar{q}\to X$:  $\rho_{\sss 11}$, $\rho_{\sss 12}\ne0$ in Eq.~(\ref{eq:ampl-spin1-qq})
      & exotic vector \\
\vspace{-0.1cm}
$2_m^+$ & $g^{(2)}_{\sss 1}=g^{(2)}_{\sss 5}\ne0$ in Eq.~(\ref{eq:ampl-spin2-a}) 
      & $gg\to X$: $g^{(2)}_{\sss 1}\ne0$ in Eq.~(\ref{eq:ampl-spin2-a})  
      & Graviton-like tensor with minimal couplings\\
& & $q\bar{q}\to X$:  $\rho_{\sss 21}\ne0$ in Eq.~(\ref{eq:ampl-spin2-qq}) \\
\vspace{-0.1cm}
$2_L^+$ & $c_2\ne0$ in Eq.~(\ref{eq:ampl-spin2}) 
      & $gg\to X$:  $g^{(2)}_{\sss 2}=g^{(2)}_{\sss 3}\ne0$ in Eq.~(\ref{eq:ampl-spin2-a}) 
      & Graviton-like tensor longitudinally polarized \\
& & $q\bar{q}\to X$:  $\rho_{\sss 21}$, $\rho_{\sss 22}\ne0$ in Eq.~(\ref{eq:ampl-spin2-qq}) 
      & and with $J_z=0$ contribution \\
\vspace{-0.1cm}
$2^-$ & $g^{(2)}_{\sss 8}=g^{(2)}_{\sss 9}\ne0$ in Eq.~(\ref{eq:ampl-spin2-a})
      &  $gg\to X$: $g^{(2)}_{\sss 1}\ne0$ in Eq.~(\ref{eq:ampl-spin2-a})  & 
``pseudo-tensor''\\
\vspace{0.1cm}
  & & $q\bar{q}\to X$:  $\rho_{\sss 21}$, $\rho_{\sss 22}\ne0$ in Eq.~(\ref{eq:ampl-spin2-qq}) \\
\hline\hline
\end{tabular}
\label{table-scenarios}
\end{table}

Our MC generation is performed stand-alone but, since unweighted events 
are produced, it is easy to
incorporate it into a software framework  that includes full detector 
simulation. This is achievable in the same
way as for Madgraph interfaced through Pythia~\cite{pythia}. 
However,  to  illustrate effects of 
 realistic detector response, we employ a simplified
technique not attached to any particular experiment.

Note that, with our choice of the final state, we 
require measurements of the four-momenta of all  charged leptons for 
complete reconstruction of the event kinematics, including boosts 
to the rest frames of $X$ and $Z$s, where the production and helicity 
angles are defined. In an experiment like ATLAS or  CMS, 
the four-momentum of the charged lepton is reconstructed from 
its track in the detector and there are two main effects that 
influence these measurements: 
(1) deviation of the five measured track parameters,
transverse momentum ($p_T$), direction ($\phi$, $\theta$),
and distance to the origin ($d_{xy}$, $d_z$),
from their true values, and
(2) non-uniform efficiency of particle detection across geometric and
kinematic parameters. We model the detector response in the following way.
First, we assume that, for typical track momenta of order $50-250$~GeV/$c$
considered in our examples, the track parameter resolution can be identified 
with the typical ATLAS or CMS tracker system resolution. 
Throughout this paper, we model detector resolution using the recently 
reported CMS track resolution parameters \cite{cmsalign}, 
obtained from analysis of  cosmic ray data. For the three track parameters 
($p_T$, $\phi$, $\theta$), we apply conservative Gaussian random smearing  
with an rms $\Delta p_T=0.025\times p_T+0.0001\times p_T^2$ (GeV/$c$),
$\Delta\phi=\Delta\theta=0.001$ (rad), and neglect resolution effects on 
the track origin. Within this simplified model of detector response,  
we observe that production and helicity angles can be measured 
with a typical resolution of the order of $\sim\,$0.01 rad. 
This resolution is rather good and does not infringe
on our ability to perform angular analysis.
Note that because production and helicity angles are defined in 
Lorentz frames which differ from the laboratory frame, 
they are affected by uncertainties in all track parameters
and, most importantly, by $p_T$ uncertainties.

The second detector effect is the non-uniform reconstruction efficiency. 
We model it in a simple way assuming that only tracks produced centrally can 
be measured. We also assume that the detection efficiency does not change 
within the detector acceptance. Hence, the acceptance function is given by 
the step-function
\be
{\cal G}(\theta^\ast,\Phi_1,\theta_1,\theta_2,\Phi) = \prod_{i\,=1,2; ~q=\pm1} 
\theta(|\eta_{\rm max}| - |\eta_{i,q}(\theta^\ast,\Phi_1,\theta_1,\theta_2,\Phi)|)\,,
\label{eq:ac1}
\ee
where $\eta_{i,\pm}=\ln\cot(\theta_{i\pm}/2)$ is the pseudorapidity of a lepton
with the charge $q=\pm$ that originates from the decay 
of the $i$'th $Z$ boson, 
$|\eta_{\rm max}| = 2.5$ is the maximal pseudorapidity 
that the tracker can reconstruct,
and the product runs over the four leptons in the $X$ decay.

\begin{figure}[t!]
\centerline{
\setlength{\epsfxsize}{0.30\linewidth}\leavevmode\epsfbox{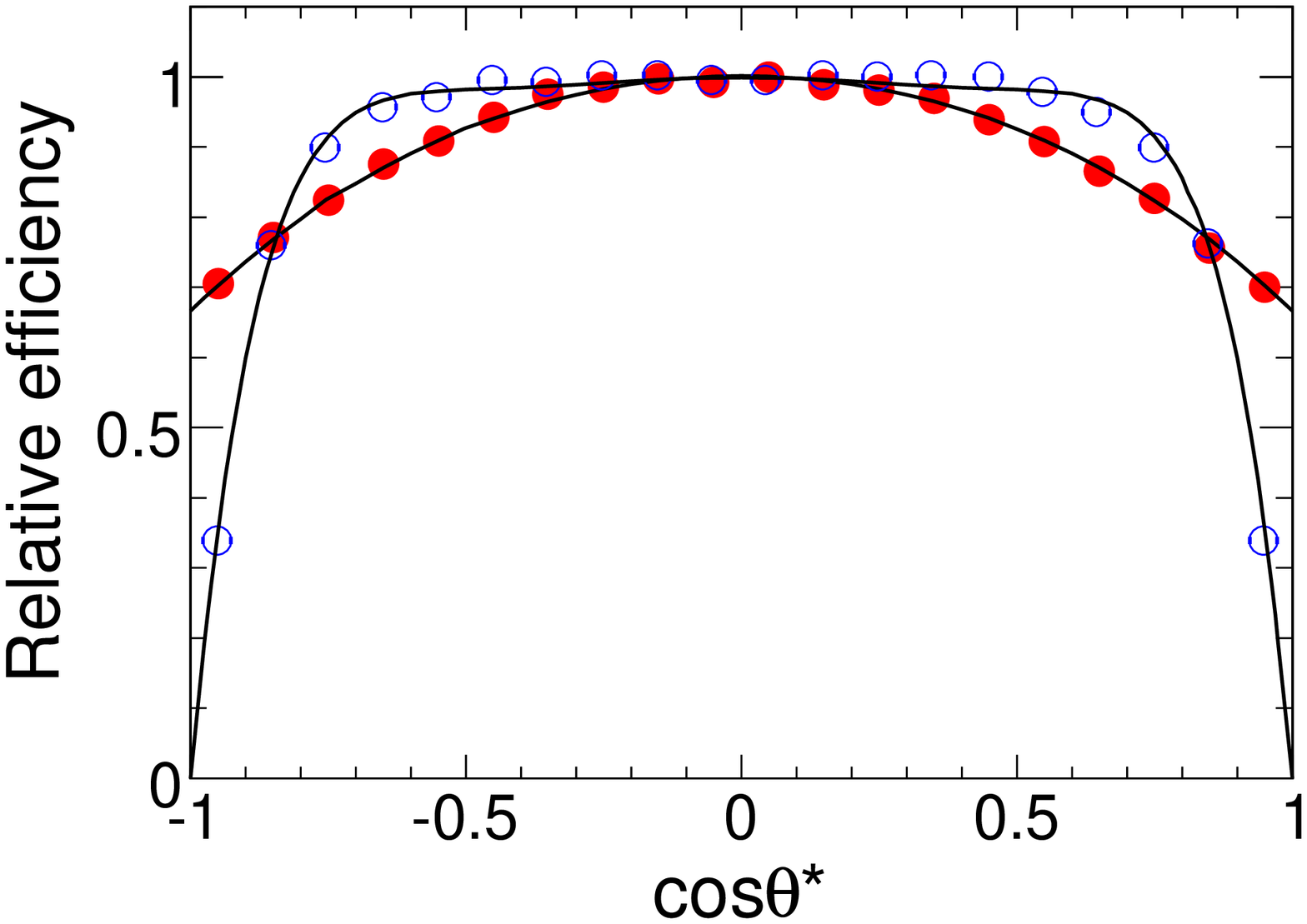}
~~~~~~~
\setlength{\epsfxsize}{0.30\linewidth}\leavevmode\epsfbox{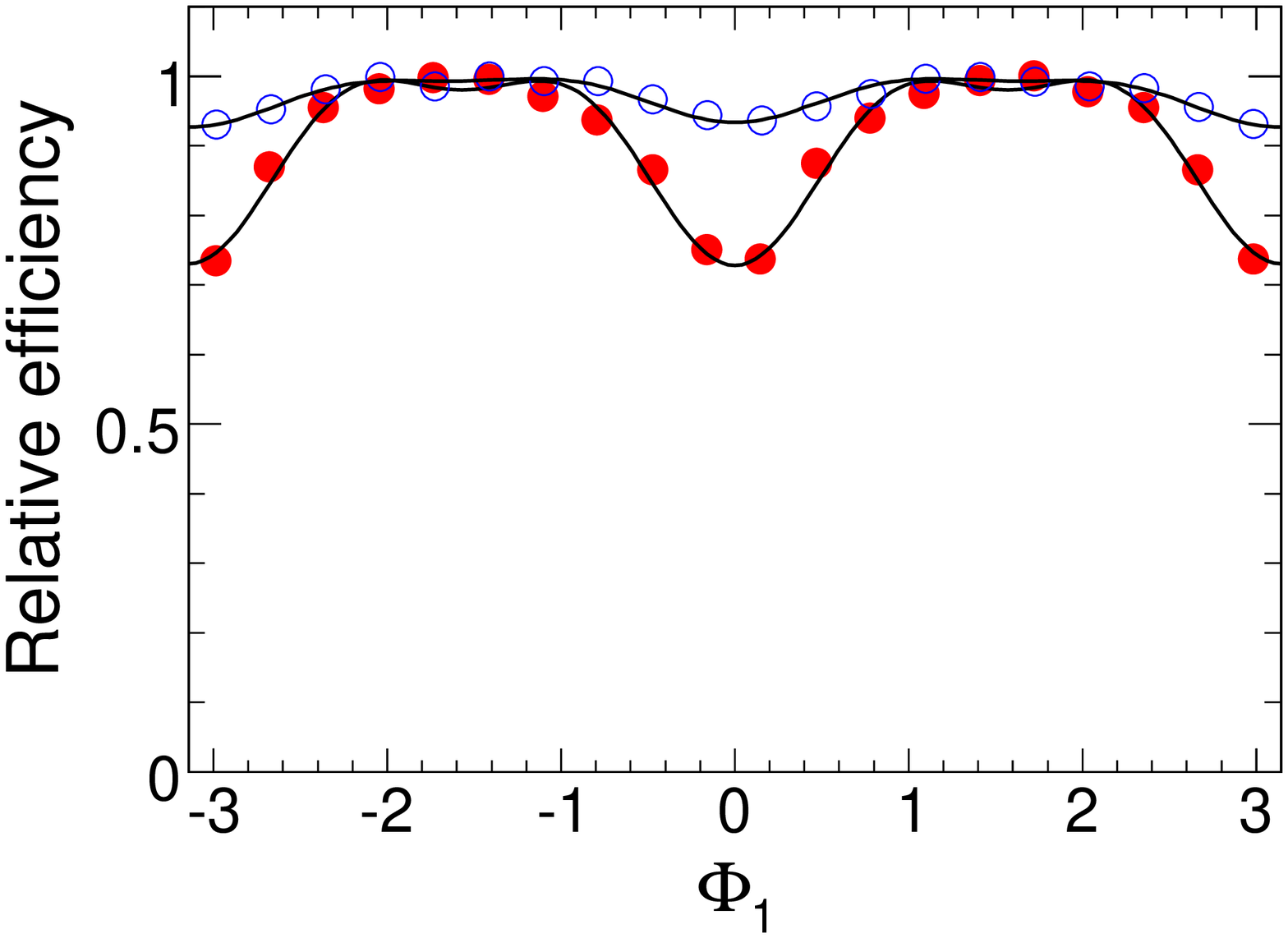}
}
\caption{
Distribution of the $\cos\theta^*$ (left) and $\Phi_1$ (right) for the case 
of   spin-zero 
resonance production $gg\to X\to ZZ$. The mass of the resonance 
is   $m_{\sss X}=250$ GeV (solid points) 
and 1 TeV (open points).  Detector acceptance effects 
are taken into account, see text for details.  
Lines show empirical parameterization. 
}
\label{fig:pdf_cstar}
\end{figure}

Given the simple form of the acceptance function, 
it is straightforward  to understand the effect of detector acceptance on 
angles measured in the laboratory frame, but their effect on 
the production and helicity angles is far less obvious. Indeed, the rapidity 
of the lepton 
$i$ is a function of all production and helicity angles; 
as the result, Eq.~(\ref{eq:ac1}) cuts out a complicated region in the 
five-dimensional space of production and helicity angles and makes the 
efficiency 
dependent on those angles. For example, if 
$X$ and the $Z$ bosons are produced at rest in the laboratory frame (this, 
of course, requires $m_{\sss X} = 2m_{\sss Z}$), 
the polar angles of
the four final state leptons $\theta_{1+,1-,2+,2-}$ that appear 
in Eq.~(\ref{eq:ac1}) can be expressed through
the five angles characterizing the decay as (with 
the convention $\Phi_2\equiv\Phi+\Phi_1$)
\begin{eqnarray}
\mp\cos\theta_{i\pm}=
\cos\theta^*\cos\theta_i+\sin\theta^*\sin\theta_i\cos\Phi_i\,.
\label{eq:accep-simple12}
\end{eqnarray}
However, in general, $X$ is boosted 
in the laboratory frame and the $Z$ bosons are boosted in the $X$ frame. 
This introduces an additional transformation of the four $\theta_{i\pm}$ angles.
We illustrate the effect of the lepton rapidity cut on production angles
in Fig.~\ref{fig:pdf_cstar}, where we plot 
the  distributions of $\theta^*$ 
and $\Phi_1$ production angles for the {\it spin-zero particle} $X$.
If these distributions are  measured with the 
``ideal'' ($4\pi$) detector, the results  are  flat. Hence,  
the non-trivial shapes of these distributions  shown in 
Fig.~\ref{fig:pdf_cstar} are {\it entirely} due to  an acceptance effect.

It is evident from Fig.~\ref{fig:pdf_cstar} that 
the acceptance effects are very important in the analysis of  data.
They have to be taken into account explicitly, 
otherwise the results of the analysis will be biased. 
This can be easily done in our MC simulation program  
on an event-by-event basis using the acceptance 
function in Eq.~(\ref{eq:ac1}), where we reject events if at 
least one lepton exceeds the maximal pseudorapidity.
It is also possible, but much harder, 
to incorporate this acceptance 
function into the likelihood function that is  discussed in the 
next section. However, as we explain now,  
certain simplifications in the  treatment of the acceptance 
function are possible. These simplifications allow us to derive a  
probability distribution that accounts for major acceptance effects.

The first of the simplifying assumptions is easy to understand. 
Suppose  $X$ is a heavy resonance. Then,  $Z$ bosons from $X$ decays 
are strongly boosted and  
decay into highly collimated ``leptonic jets''. 
Therefore, as the $X$ mass increases,   
in the laboratory  frame the decays  look
 exceedingly pencil-like, with all the leptons 
aligned with directions of their parent $Z$ bosons. It is then clear 
that  the detection efficiency of the leptons is  highly correlated 
with the $Z$-production angle $\theta^*$ {\it but not with the other 
angles}.  If the 
$X$ boson has an intermediate mass, the story 
 becomes  more complex and 
acceptance effects in other angles should be expected. 
However, it is reasonable to expect 
those effects to be moderate.   
We can check these features of the acceptance with the MC simulation. 
From Fig.~\ref{fig:pdf_cstar}, it is evident that with the increase in the 
mass  of the particle $X$ the acceptance effect in $\cos\theta^*$ gets more 
pronounced while it is indeed reduced in the other angles. 

The second simplifying assumption is based on the observation that 
distributions of the helicity angles 
$(\cos\theta_1,\cos\theta_2,\Phi)$ are independent of  the production 
mechanism. Hence, there are no acceptance effects in the 
helicity angles provided that $\cos\theta^*$ and $\Phi_1$ are random, 
as in the case of spin-zero particle production. In a more general 
case, acceptance effects on the helicity angles are expected to be 
rather mild. 

\section{Data Analysis} 
\label{sect6}

In this section, we  illustrate the application 
of the full angular analysis formalism by considering the production 
of a resonance $X$ and its subsequent decay to two $Z$ bosons.
We use  the MC 
simulation discussed in the previous section to generate both signal and
background events, and apply a multivariate fitting technique discussed later 
in this section to extract as much information 
about the produced resonances as possible. We point out that, because 
our analysis is  general and because it is not based on any particular 
model of BSM  physics, 
it is not possible to accurately predict how many $X$ particles are produced, 
for a given luminosity.  
Instead, we find it more reasonable to  {\it assume} a 
certain  number of signal and background events, reconstructed 
by  an LHC experiment. 
This number of events  should be 
sufficiently large to enable meaningful angular 
analysis, yet it should be small enough,  to be relevant for the  situation 
soon after the resonance observation. The statistical power of the method 
can be easily extrapolated to a different number of resonance events. 

As we explain below, we perform two different analyses in this paper,
for which we choose to have 30 and  150 signal events, respectively. 
Roughly, this corresponds~\cite{exphiggs1, exphiggs2} to the number of SM Higgs
boson  events  
$pp \to H\to ZZ\to 4l^\pm$, $l = e,\mu$, detected by one LHC experiment 
with 5 or 25~${\rm fb}^{-1}$ integrated luminosity, for a Higgs boson 
mass $m_{\sss H}=250$ GeV, at a proton-proton center-of-mass energy of 14 TeV. 
Of course, as we pointed out a few times already, the actual production 
rate of exotic particles may be larger or smaller; we choose 
$30$ and $150$ events merely for illustration purposes. 
 
We consider two mass points, the low mass $m_{\sss X}=250$ GeV and the 
relatively high mass $m_{\sss X}=1$ TeV. The background size is chosen according 
to the above LHC integrated luminosity convention, which leaves the high-mass 
scenario $m_{\sss X}=1$ TeV with essentially no background, and the low-mass 
scenario with  about 24 background events in the $m_{\sss X}$ mass 
window between 230 and 270 GeV~\cite{exphiggs1, exphiggs2}. The exact number of 
background events 
is not crucial  for  our analysis because we are 
mostly interested in  understanding  how well different signal 
hypotheses can be disentangled, rather than how well a signal
can be separated from the background. 

We now imagine that 30 non-SM resonance-like events 
have been experimentally  observed and we need to understand their 
origin. As the very first step, we may want to find out 
if the observed events are described by 
a resonance with particular quantum numbers and a particular structure of  
interactions with the SM particles. An efficient way to do this 
is to compute a confidence level at which one hypothesis 
about resonance quantum numbers can be separated from another hypothesis. 
For example, we may ask if an observed
resonance looks more like a  SM Higgs  boson or more like  a graviton
with minimal couplings to SM fields.  We use our MC event simulation 
to check how well separation of the two hypotheses can be done in practice.
Once we have a good idea about the nature of the resonance $X$, 
we  can proceed to a more statistically demanding analysis 
whose goal is  to determine {\it all} parameters which describe the angular 
distributions within a particular hypothesis. Then, one may 
attempt to interpret  those parameters within a particular  model of BSM physics.

Note that the second approach  clearly supersedes the first one and 
is the most optimal way to analyze  data. Indeed, in the extreme 
version of the second approach, {\it everything} --
mass, width, spin, parity, {\it and} polarization parameters 
of the particle $X$ -- is obtained  through a single fitting process.
The shortcoming of this approach is that it is very demanding statistically,
since  a large sample of events is required to enable 
simultaneous determination of a large number of signal parameters 
with decent precision. For the illustration of the second approach, 
we assume that 150 signal events are available.

We note that our methods  are general enough to implement either 
of the two approaches  described above.
The only practical difference in application of these methods 
is that the polarization parameters are either fixed 
or left as free parameters in the fit. 
Since it is not practical to cover the continuous spectrum 
of possible scenarios 
in this paper, we choose several examples to illustrate 
these approaches.  These scenarios are specified  in   
Table~\ref{table-scenarios}.

Our studies are  encouraged by the multivariate techniques
developed for polarization studies in  $B$-physics~\cite{phikst}, 
where more than two dozen parameters 
describing   angular and mass distributions are extracted in
a simultaneous fit to a sample of several hundred events.
We use an unbinned extended maximum-likelihood (ML) fit~\cite{phikst, root}
to extract simultaneously the signal and background yields, the background shape,  
and the parameters of the signal angular distributions.
The likelihood function for $N$ candidate events is written  as
%
\begin{equation}
{{\cal L}} =  \exp\left( - \sum_{J=1}^3n_{\sss J}-n_{\rm bkg}  \right) 
\prod_i^{N} \left( ~\sum_{J=1}^3
n_{\sss J} \times{\cal P}_{\sss J}(\vec{x}_{i};~\vec{\zeta}_{\sss J};~\vec{\xi})  
+n_{\rm bkg} \times{\cal P}_{\rm bkg}(\vec{x}_{i};~\vec{\xi})  
\right)\,,
\label{eq:likelihood}
\end{equation}
where $n_{\sss J}$ is the number of signal events for each 
resonance spin $J$, $n_{{\rm bkg}}$ is the number of background events and 
${\cal P}(\vec{x}_{i};\vec{\zeta};\vec{\xi})$ is the
probability density function for signal or background.
It is assumed that only one resonance is observed  in a given mass window and 
the three yields, $n_{\sss J}$, allow one to test different hypotheses.
For this reason, no interference between the different resonances is considered. 
Alternatively, we can consider a signal event yield $n_{\rm sig}$ 
in place of the three $n_{\sss J}$ if we assume that mass distributions of resonances 
with different spins are identical and a single angular distribution can be written 
that incorporates angular distributions for all three spins as limiting cases. 
An example of the latter approach is given in Eq.~(\ref{eq:mixedgravitontotal}) 
which describes the general spin-two distribution in four angles, 
but also includes both Eqs.~(\ref{eq:higgs-ang1}) and~(\ref{eq:zprime-ang2}) 
written for spin-zero and spin-one.

Each event candidate $i$ is characterized by a set of six observables
$\vec{x}_{i}=\{m_{\sss ZZ},~\cos\theta^*,~\Phi_1,~\cos\theta_1,~\cos\theta_2,~\Phi\}_i$.
The number of observables can be extended or reduced, depending 
on the desired fit. The signal polarization parameters 
$\{f_{\lambda_1\lambda_2},\phi_{\lambda_1 \lambda_2}, f_{zm} \}$
are collectively denoted by $\vec{\zeta}_{\sss J}$, 
and the remaining parameters by $\vec{\xi}$.
When several decay channels must be combined, the joint likelihood 
in Eq.~(\ref{eq:likelihood}) is the natural way to achieve this. 
In the current  analysis, a combination of all lepton channels with electrons and muons 
in the $ZZ$ decays is assumed, but it can  be easily extended to include hadronic 
decays of $Z$ bosons as well.

For the signal, the probability density function
${\cal P}_{\sss J}(\vec{x}_{i};\vec{\zeta}_{\sss J};\vec{\xi})$
is taken to be a product of a function of  the angles and 
a function of the resonance mass.
For the background, the probability density function
is a product of uncorrelated functions for each observable.
The assumption of small correlations 
among the discriminating variables, except for angles where relevant,
can be validated in the selected data sample 
by evaluating  correlation coefficients
and is further tested with generated experiments.
As long as the results  of the generated experiments are unbiased,
small correlations are acceptable.  
Since the parameterization of the background probability density function
also depends on detector effects, 
similarly to the signal, experiment-specific treatment will, eventually, 
be required. It is an approximation of our approach that the background
angular distributions are not correlated, which we find acceptable
for the current  analysis.

The angular probability density function for the signal is the
ideal distribution in Eq.~(\ref{eqaa}),
multiplied by an empirically determined acceptance function
${\cal{G}}(\cos\theta^*,\Phi_1,\cos\theta_1,\cos\theta_2,\Phi; Y_{\sss X})$,
which can be evaluated independently for each event 
with the $X$ particle rapidity $Y_{\sss X}$.
We approximate the acceptance function by 
${\cal{G}}={{\cal G}^{(1)}}(\cos\theta^*)\times{{\cal G}^{(2)}}(\Phi_1)$,
neglecting all correlations between different angles; 
the justification for this approximation was given in the previous section.
The parameterization of 
the signal acceptance functions ${\cal G}^{(1,2)}$ is compared with 
simulated distributions in Fig.~\ref{fig:pdf_cstar}.
The acceptance function constructed along these lines can be 
included in the overall probability density function.

It is important to point out that one can do a better job if 
all that is needed  is to understand how well a particular hypothesis 
describes experimental data. Since we test a {\it specific} hypothesis, 
ratios of helicity amplitudes are fixed. Hence, it is possible to determine 
the likelihood function numerically through the MC simulation of the full 
five-dimensional angular distribution, including the detector effects, 
though averaged over the $Y_{\sss X}$ values. Effectively, this  requires
storing a normalized five-dimensional histogram with fine binning. 
Dependence on $Y_{\sss X}$ could in principle be tabulated as well.
A similar approach can be used for background parameterization.

However, it is more difficult to employ MC computation of the signal 
probability density, if we want to treat all the parameters, 
including helicity fractions, as free parameters in the fit. 
A straightforward use of the MC program would require generating and 
storing probability  distributions for a large set of  independent 
polarization parameters. Clearly, for a large enough number of parameters 
to be fitted, this procedure  becomes impractical.
Instead, one could factorize the angular distribution into 
a sum of simple terms expressed as the product of a function of the angles 
and a function of the polarization parameters, as we have illustrated
for the spin-zero case in Eq.~(\ref{eq:higgs-ang1}).
A single independent MC simulation for the angular dependencies 
of each such term can be performed and the results can be combined 
to produce a normalized angular parameterization with free parameters.
However, for the purpose of this paper, we find the simplified 
approach to be acceptable since, as we have verified with our MC 
program, biases in the results are small. 

The only other kinematic observable that is used in the fit 
in addition to the angles is the $ZZ$ invariant mass $m_{\sss ZZ}$.
The background $m_{\sss ZZ}$ distribution is parameterized with a  polynomial
and its shape can be left unconstrained in the fit.
We make the following assumptions about the mass and the width 
of the $X$ resonance. 
We assume that its  width is negligible compared to the 
detector resolution effects, which are estimated to be about 3.5 GeV at
$m_{\sss X}=250$ GeV and 23 GeV at $m_{\sss X}=1$ TeV with our simulation.
The exact value of the resolution is only relevant for signal and background 
separation at the lower mass; the values quoted above are typical
for  both muon and electron final states. Hence, 
we describe the signal mass distribution
with a Gaussian function of a known width. Should the resonance have
a finite  width that is larger than detector resolution, we can 
incorporate it into our analysis by replacing the Gaussian function with 
a convolution of the Breit-Wigner and Gaussian distributions.
However, since the mass distribution has little, if any, influence 
on the resonance spin and its  couplings to SM fields, 
a detailed study of mass resolution issues is outside the
scope of this paper. 

The mass of the resonance $m_{\sss X}$ is assumed to be known and is
fixed in the fit. This assumption only affects the signal and  
background separation and requires certain care 
in the  interpretation of the discovery significance.  
Detailed study of this interpretation is also outside the scope of this paper,
but we note that, as Fig.\ref{fig:generated-angles} suggests, 
angular analysis improves the signal-background separation.
As an example, consider $30$ events due to the SM Higgs-like resonance  at 250 GeV.
If only $m_{\sss ZZ}$ is employed in the fit, the measured signal significance is 5.7 $\sigma$ 
and it increases by nearly twenty percent if also angular variables are included in the fit.
Here significance is calculated from ${2\ln({\cal L}_1/{\cal L}_2)}$ using the likelihood
ratio for two scenarios with signal and with pure background, analogous to the technique 
discussed below, and corresponds to the Gaussian probability for pure background to 
fluctuate to the observed signal yield. The only constraint on background size comes 
from data sidebands included in the fit.

\begin{figure}[t!]
\centerline{
\setlength{\epsfxsize}{0.35\linewidth}\leavevmode\epsfbox{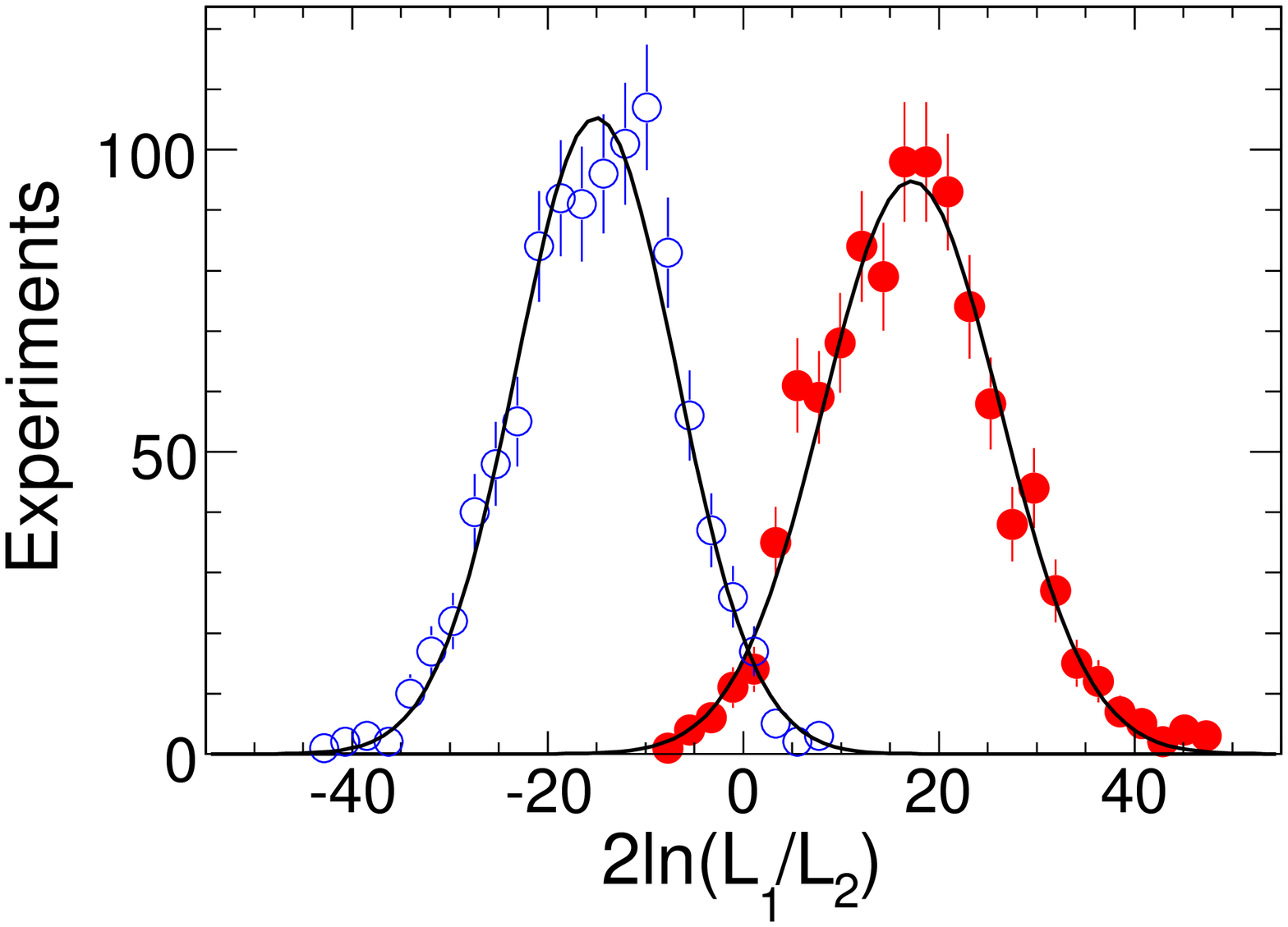}
~~~~
\setlength{\epsfxsize}{0.35\linewidth}\leavevmode\epsfbox{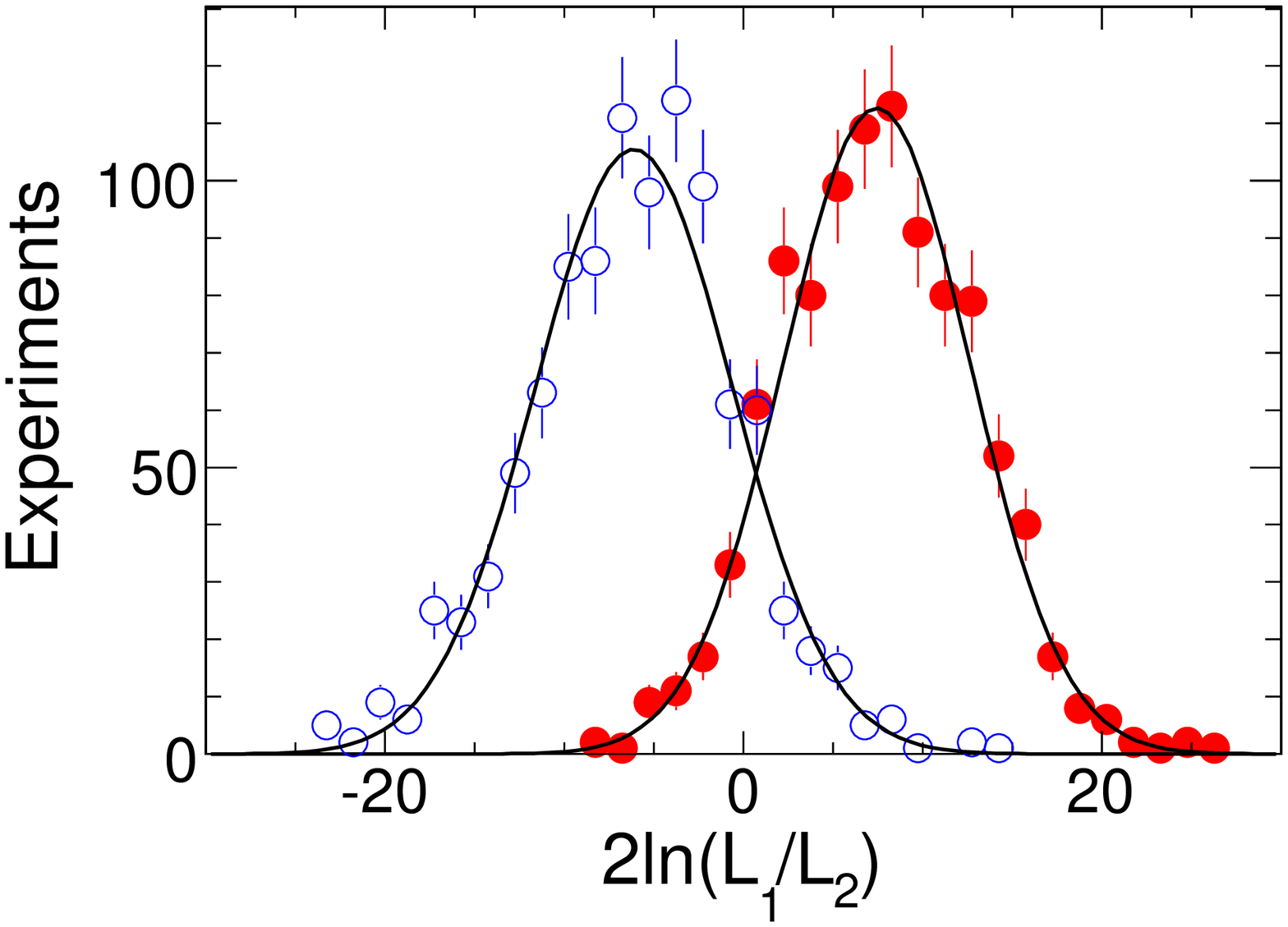}
}
\caption{
Distribution of ${2\ln({\cal L}_1/{\cal L}_2)}$ with the likelihood 
${\cal L}$ evaluated for two models $k=1,2$ and shown for 1000 generated
experiments with the MC events generated according to model one 
($k=1$, open dots) and model two ($k=2$, solid dots).
Left plot: $0^+$ vs. $0^-$; right plot: $0^+$ vs. $2_m^+$.
Effective signal hypothesis separation power ${\cal S}$ 
is 4.1 (left plot) and 2.8 (right plot).
}
\label{fig:separation_1vs2}
\end{figure}

\begin{table}[t!]
\caption{
Results of the hypothesis separation of the seven scenarios listed 
in Table~\ref{table-scenarios} for $m_{\sss X}=250$ GeV.
The five numbers quoted in each case correspond to the 1D/2D/3D/4D/5D angular information
used in the fit, as discussed in the text. The results are reported as an 
equivalent separation of two single-width Gaussian distributions in 
terms of the number of Gaussian standard deviations 
 between the two peaks.
}
\begin{tabular}{lcccccc}
\hline\hline
    & $0^-$ &  $1^+$ &  $1^-$ &  $2_m^+$ &  $2_L^+$ &  $2^-$ \\
    \hline
 $0^+$  & 0.0/0.0/3.9/4.1/4.1 & 0.8/1.0/1.8/1.9/2.3 & 0.9/1.0/2.5/2.6/2.6 & 0.8/0.9/2.4/2.5/2.8 & 2.6/2.6/0.0/2.6/2.6 & 1.6/1.7/2.4/3.0/3.3 \\
 $0^-$  &   & 0.8/1.2/2.8/3.0/3.1 & 0.9/1.0/2.5/2.8/3.0 & 0.8/0.8/1.7/2.0/2.4 & 2.9/2.9/4.1/4.8/4.8 & 1.6/1.7/2.0/2.7/2.9 \\
 $1^+$  &  &  & 0.0/1.1/1.1/1.2/2.2 & 0.1/1.2/1.3/1.4/2.6 & 2.8/2.8/1.9/3.5/3.6 & 2.5/2.4/1.2/2.7/2.9 \\
 $1^-$  &   &  &  & 0.1/0.1/1.3/1.5/1.8 & 2.8/2.9/2.5/3.8/3.8 & 2.5/2.6/0.6/2.8/3.4 \\
 $2_m^+$ &  &  &  &  & 2.9/2.9/2.6/3.6/3.8 & 2.3/2.5/0.5/2.5/3.2 \\
 $2_L^+$ &   &  &  &  &  & 3.6/3.6/2.5/4.2/4.3 \\
\hline\hline
\end{tabular}
\label{table-hypothesis-test}
\caption{
Results of the hypothesis separation of the seven scenarios for $m_{\sss X}=1$ TeV with the same
notation as in Table~\ref{table-hypothesis-test}.
}
\begin{tabular}{lcccccc}
\hline\hline
    & $0^-$ &  $1^+$ &  $1^-$ &  $2_m^+$ &  $2_L^+$ &  $2^-$ \\
\hline 
 $0^+$  & 0.0/0.0/3.9/4.0/4.1 & 0.4/0.6/2.2/2.2/2.4 & 0.3/0.5/2.2/2.1/2.4 & 1.2/1.1/2.8/2.7/2.8 & 1.0/1.0/0.0/1.2/1.2 & 1.2/1.0/2.0/2.7/2.8 \\
 $0^-$   &   & 0.5/0.9/2.8/2.9/3.1 & 0.5/0.8/2.7/2.9/3.1 & 1.4/1.5/1.4/2.1/2.1 & 1.1/1.1/4.0/4.3/4.5 & 0.8/0.9/2.9/2.8/2.9 \\
 $1^+$  &   &  & 0.0/1.1/1.2/1.2/2.3 & 0.9/0.9/2.0/2.0/2.3 & 1.3/1.4/2.1/2.4/2.6 & 1.3/1.3/1.1/1.8/2.0 \\
 $1^-$   &   &  &  & 0.9/0.9/2.0/2.0/2.1 & 1.3/1.3/2.2/2.5/2.7 & 1.4/1.6/0.0/1.4/2.2 \\
 $2_m^+$  &  &  &  &  & 1.7/1.7/3.0/3.2/3.2 & 2.1/2.0/2.1/2.6/2.6 \\ 
 $2_L^+$   &  &  &  &  &  & 1.5/1.5/2.1/3.3/3.3 \\
\hline\hline
\end{tabular}
\label{table-hypothesis-test-1tev}
\end{table}
%

We are now in position to discuss the results of our analysis. 
As an illustration of the first approach, in Fig.~\ref{fig:separation_1vs2},
we plot the quantity ${2\ln({\cal L}_1/{\cal L}_2)}$ for two sets of
MC experiments generated with the models $0^+$ and $0^-$ (left plot) or 
$0^+$ and $2_m^+$ (right plot), as listed in Table~\ref{table-scenarios}. 
Each experiment has on average 24 background and 30 signal events, 
Poisson distributed.
These events are  generated at $m_{\sss X}=250$ GeV in 
1000 statistically independent experiments. 
The likelihood ${\cal L}_k$ in Eq.~(\ref{eq:likelihood}) is evaluated 
independently for each hypothesis $k$. The  discussion of the
statistical interpretation of ${2\ln({\cal L}_1/{\cal L}_2)}$ 
was given recently in  Ref.~\cite{Cousins2005}.
The distribution of ${2\ln({\cal L}_1/{\cal L}_2)}$ is expected to peak 
at positive values for events  generated according to model one 
(on average ${\cal L}_1>{\cal L}_2$) and 
at negative values for events  generated according to model two 
(on average ${\cal L}_2>{\cal L}_1$).
From Fig.~\ref{fig:separation_1vs2}, we extract the quantity ${\cal S}$ 
in the following way. 
We find the point beyond which the right-side tail of the left 
histogram and the left-side tail of the right histogram have equal areas. 
These areas correspond to the one-sided Gaussian probability 
outside of the ${\cal S}$/2 $\sigma$ range. If the two histograms in 
Fig.~\ref{fig:separation_1vs2} were perfectly Gaussian distributed
with unit width, then ${\cal S}$ corresponds to the separation
between the peaks of the two distributions.

The above procedure is repeated for all combinations of seven hypotheses
listed in Table~\ref{table-scenarios}, for $m_{\sss X} = 250~{\rm GeV}$ 
and $m_{\sss X} = 1~{\rm TeV}$,  and results are presented
in Tables~\ref{table-hypothesis-test} and~\ref{table-hypothesis-test-1tev}.
In those tables, we show the increase in the separation power between 
two hypotheses if more information about angular distributions is 
included in the likelihood fit. It is natural to refer to a likelihood 
fit that includes $n$ angular variables as $n$-dimensional or $n$D; 
the results of the 1D, 2D, 3D, 4D, and 5D fits are shown in 
Tables~\ref{table-hypothesis-test} and ~\ref{table-hypothesis-test-1tev}.
In the 1D fit, only the $\cos\theta^*$ angular distribution  is used;
in the 2D fit, the   ($\cos\theta^*,\Phi_1$) distribution is  included;
in the 3D fit,  we have $(\cos\theta_1,\cos\theta_2,\Phi)$;
in the 4D fit,  the  $(\cos\theta^*,\cos\theta_1,\cos\theta_2,\Phi)$  distribution  is used;
and finally in 5D fit, the $(\cos\theta^*,\Phi_1,\cos\theta_1,\cos\theta_2,\Phi)$  
distribution is employed.
Statistical significance achieved in the 4D fit is already close to the statistical 
significance of the optimal analysis which employs all five angles.
We note that because we generate a finite number of experiments (1000),  
the statistical uncertainties on the values of ${\cal S}$ reported in
Tables~\ref{table-hypothesis-test} and~\ref{table-hypothesis-test-1tev}
are typically in the range $0.1-0.2$. This explains some of the small
discrepancies in the values quoted. For example, one could expect the
separation of $0^+$ and $0^-$ hypotheses to be the same in 3D, 4D, and 5D cases.
However, statistical fluctuations and better background suppression make the
4D and 5D cases look better.

It is clear  from Tables~\ref{table-hypothesis-test} and~\ref{table-hypothesis-test-1tev} 
that  the multidimensional angular analysis 
allows efficient  separation of  different spin hypotheses
and, therefore,  is  a very  powerful approach to spin determination.
While separations ${\cal S} = 2$ are typically achieved, there are important 
cases when larger separations are possible. In particular, we observe ${\cal S}=4$ 
for the separation of $0^+$ and $0^-$ hypotheses, which might be useful
even for the SM  Higgs boson if its mass exceeds $2m_{\sss Z}$.
Also,  the separation of the spin-two 
 with the minimal couplings hypothesis  from 
the spin-two  with longitudinal amplitude dominance hypothesis 
is close to ${\cal S}=3$, for any production mechanism scenario.
In all cases, the results of  the 5D fit provide the best separation power
and supersede the other fits.
Inclusion of the $\Phi_1$ angle in the 2D or 5D fit,  compared to the 1D or 4D fit, 
usually adds little information, except for the separation of the two spin-one 
hypotheses which have identical $\cos\theta^*$ distributions.
Therefore, the  separation power obtained from the angular 
analysis with all five angles is very close to the separation power 
of the 4D fit. 

A glance at Tables~\ref{table-hypothesis-test} and~\ref{table-hypothesis-test-1tev} 
is sufficient to recognize that there are cases  when different hypotheses about 
the nature of the resonance cannot be separated (${\cal S}=0$).
For example, the production angles $\cos\theta^*$ and $\Phi_1$
provide no separation between the two spin-zero scenarios ($0^+$ and $0^-$)
due to the absence of spin correlations.
At the same time, the three helicity angle $(\cos\theta_1,\cos\theta_2,\Phi)$  
distributions are identical for the $0^+$ and $2_L^+$ hypotheses, therefore
the 3D case results in ${\cal S}=0$ as well. The only source of separation 
in this case is the production angle. 
We have not considered any example with no separation
from either production or helicity angles, but such situation is possible
when spin-zero and spin-two scenarios with $f_{z1}=f_{z2}=0.4$ are considered,
as we discussed earlier. A hint at  this situation is observed in the 
{\it decreased}
${\cal S}$-value that characterizes the separation of $0^+$ and
$2^+_L$ hypotheses when the mass of the resonance increases from 
250 GeV to 1 TeV.   This is the consequence of the fact that 
at 1 TeV  we allow $q\bar{q}$ production. This leads to 
$f_{z1}=0.25$ and $f_{z2}=0.30$, which is closer to the  
unpolarized case. As the result,  the ${\cal S}$-value 
decreases dramatically.

Obviously, in the above studies we could not cover all possible scenarios.
Just as an example, one could consider a scenario of a non-SM Higgs-like
scalar with quantum numbers $0^+$ with $a_1 = 0,\, a_2 \ne 0$ in Eq.~(\ref{eq:ampl-spin0}).
The separation of this hypothesis from the SM Higgs-like scalar will
depend strongly on the mass of the resonance. As it is evident from
Eq.~(\ref{eq:helicity-ampl0}), the single $A_{00}$ amplitude dominates
for ${m_{\sss X}^2}/{m_{\sss V}^2}\gg 1$ in both cases. Therefore, separation of the
two hypotheses is impossible at high mass. However, at lower mass, there
is a sizable contribution of both $A_{++}$ and $A_{--}$ in the SM Higgs-like
case, but not in the other case. Therefore, the most optimal analysis
strategy is to fit for all the polarization parameters, and then 
interpret the result within a particular model of BSM physics. 
We describe the feasibility of such fit in the remainder of this section.

\begin{table}[t!]
\caption{
Results of the fit for the free parameters of the spin-zero hypothesis
with generated samples of SM Higgs-like $X$ corresponding to $0^+$ 
in Table~\ref{table-hypothesis-test}.
Experiments have been generated with two $X$ masses according to two
models in each case, with and without detector effects.
}
\begin{tabular}{lcccccc}
\hline\hline
\vspace{-0.1cm}
& \multicolumn{3}{c}{$m_{\sss X}=250$ GeV} & \multicolumn{3}{c}{$m_{\sss X}=1$ TeV} \\
\vspace{-0.15cm}
 & generated & \multicolumn{2}{c}{fitted}  & generated & \multicolumn{2}{c}{fitted}   \\
\vspace{0.1cm}
 &  & without detector & with detector &  & without detector & with detector    \\
\hline 
$n_{\rm sig}$   & 150 & $150 \pm 13$ & $153 \pm 15$  & 150 & $150 \pm 12$ & $152 \pm 12$ \\
$(f_{++}+f_{--})$ & 0.208 & $0.21 \pm 0.07$ & $0.23 \pm 0.08$ & 0.000 & $0.00 \pm 0.03$ & $0.00 \pm 0.03$ \\
$(f_{++}-f_{--})$ & 0.000 & $0.01 \pm 0.13$  & $0.01 \pm 0.14$ & 0.000 & $0.00 \pm 0.02$  & $0.00 \pm 0.02$  \\
$(\phi_{++}+\phi_{--})$ & $2\pi$ & $6.30 \pm 1.46$  & $6.39 \pm 1.54$  & $2\pi$ & free  & free \\
\vspace{0.1cm}
$(\phi_{++}-\phi_{--})$ & $0$    & $0.00 \pm 1.06$  & $0.01 \pm 1.09$  & $0$    & free & free  \\
\hline\hline
\end{tabular}
\label{table-fit-test1}
\caption{
Results of the fit for the free parameters of the spin-zero hypothesis
with generated samples of a pseudo-scalar $X$ corresponding to $0^-$ 
as discussed in Table~\ref{table-fit-test1}.
}
\begin{tabular}{lcccccc}
\hline\hline
\vspace{-0.1cm}
& \multicolumn{3}{c}{$m_{\sss X}=250$ GeV} & \multicolumn{3}{c}{$m_{\sss X}=1$ TeV} \\
\vspace{-0.15cm}
 & generated & \multicolumn{2}{c}{fitted}  & generated & \multicolumn{2}{c}{fitted}   \\
\vspace{0.1cm}
 &  & without detector & with detector &  & without detector & with detector    \\
\hline 
$n_{\rm sig}$   & 150 & $150 \pm 13$  & $151 \pm 15$  & 150 & $151 \pm 12$ & $150 \pm 13$  \\
$(f_{++}+f_{--})$ & 1.000 & $1.00 \pm 0.05$ &  $1.00 \pm 0.06$ & 1.000 & $1.00 \pm 0.05$ & $1.00 \pm 0.06$  \\
$(f_{++}-f_{--})$ & 0.000 & $0.00 \pm 0.35$ &  $0.00 \pm 0.40$ &  0.000 & $0.00 \pm 0.31$  & $-0.01 \pm 0.32$  \\
$(\phi_{++}+\phi_{--})$ & N/A  & free  & free   & N/A & free  & free \\
\vspace{0.1cm}
$(\phi_{++}-\phi_{--})$ & $\pi$    & $3.15 \pm 0.31$   & $3.14 \pm 0.41$  & $\pi$    & $3.15 \pm 0.31$   & $3.14 \pm 0.33$ \\
\hline\hline
\end{tabular}
\label{table-fit-test2}
\caption{
Results of the fit for the free parameters of the spin-one hypothesis
with generated samples of an exotic pseudo-vector $X$ corresponding to $1^+$ 
as discussed in Table~\ref{table-fit-test1}.
}
\begin{tabular}{lcccccc}
\hline\hline
\vspace{-0.1cm}
& \multicolumn{3}{c}{$m_{\sss X}=250$ GeV} & \multicolumn{3}{c}{$m_{\sss X}=1$ TeV} \\
\vspace{-0.15cm}
 & generated & \multicolumn{2}{c}{fitted}  & generated & \multicolumn{2}{c}{fitted}   \\
\vspace{0.1cm}
 &  & without detector & with detector &  & without detector & with detector    \\
\hline 
$n_{\rm sig}$ & 150 & $150 \pm 13$  & $152 \pm 15$ & 150 & $151 \pm 12$  & $152 \pm 13$ \\
$f_{+0}$ & 0.250 & $0.26 \pm 0.15$ & $0.26 \pm 0.17$ & 0.250 & $0.25 \pm 0.16$  & $0.25 \pm 0.15$\\
\vspace{0.1cm}
$(\phi_{+0}-\phi_{0-})$ & $0$ &  $0.02 \pm 0.73$ &  $0.01 \pm 0.86$  & $0$ &  $-0.03 \pm 0.67$ &  $-0.03 \pm 0.77$  \\
\hline\hline
\end{tabular}
\label{table-fit-test3}
\caption{
Results of the fit for the free parameters of the spin-one hypothesis
with generated samples of an exotic vector $X$ corresponding to $1^-$ 
as discussed in Table~\ref{table-fit-test1}.
}
\begin{tabular}{lcccccc}
\hline\hline
\vspace{-0.1cm}
& \multicolumn{3}{c}{$m_{\sss X}=250$ GeV} & \multicolumn{3}{c}{$m_{\sss X}=1$ TeV} \\
\vspace{-0.15cm}
 & generated & \multicolumn{2}{c}{fitted}  & generated & \multicolumn{2}{c}{fitted}   \\
\vspace{0.1cm}
 &  & without detector & with detector &  & without detector & with detector    \\
\hline 
$n_{\rm sig}$ & 150 & $150 \pm 13$ & $152 \pm 16$   & 150 & $151 \pm 12$  & $152 \pm 13$ \\
$f_{+0}$ & 0.250 & $0.24 \pm 0.15$ & $0.24 \pm 0.16$   & 0.250 & $0.24 \pm 0.17$ & $0.25 \pm 0.15$  \\
\vspace{0.1cm}
$(\phi_{+0}-\phi_{0-})$ & $\pi$ &  $3.14 \pm 0.72$ &  $3.18 \pm 0.76$  & $\pi$ &  $3.14 \pm 0.71$  &  $3.14 \pm 0.69$  \\
\hline\hline
\end{tabular}
\label{table-fit-test4}
\end{table}

\begin{table}[t!]
\caption{
Results of the fit for the free parameters of the spin-two hypothesis
with positive parity
with generated samples of an exotic $X$ corresponding to $2^+_m$ 
as discussed in Table~\ref{table-fit-test1}.
}
\begin{tabular}{lcccccc}
\hline\hline
\vspace{-0.1cm}
& \multicolumn{3}{c}{$m_{\sss X}=250$ GeV} & \multicolumn{3}{c}{$m_{\sss X}=1$ TeV} \\
\vspace{-0.15cm}
 & generated & \multicolumn{2}{c}{fitted}  & generated & \multicolumn{2}{c}{fitted}   \\
\vspace{0.1cm}
 &  & without detector & with detector &  & without detector & with detector    \\
\hline 
$n_{\rm sig}$ & 150 & $150 \pm 13$  & $151 \pm 16$  & 150 & $151 \pm 12$  & $153 \pm 13$ \\
$f_{z2}$ & 1.000 & $1.00 \pm 0.17$ & $0.84 \pm 0.17$    & 0.750 & $0.75 \pm 0.12$ & $0.80 \pm 0.10$ \\
$f_{z1}$ & 0.000 & $0.00 \pm 0.19$ & $0.00 \pm 0.25$    & 0.250 & $0.25 \pm 0.14$ & $0.16 \pm 0.15$  \\
$f_{++}$ & 0.013 & $0.01 \pm 0.04$ & $0.00 \pm 0.05$    & 0.000 & $0.00 \pm 0.05$ & $0.00 \pm 0.05$\\
$f_{+-}$ & 0.282 & $0.28 \pm 0.04$ & $0.31 \pm 0.05$    & 0.445 & $0.44 \pm 0.06$ & $0.44 \pm 0.04$ \\
$f_{+0}$ & 0.075 & $0.07 \pm 0.04$ & $0.06 \pm 0.05$    & 0.000 & $0.01 \pm 0.06$ & $0.01 \pm 0.06$ \\
\vspace{0.1cm}
$\phi_{++}$ & $0$ & $0.00 \pm 1.75$ & $0.04 \pm 1.76$ & $0$  & free & free \\
\hline\hline
\end{tabular}
\label{table-fit-test5}
\caption{
Results of the fit for the free parameters of the spin-two hypothesis
with positive parity
with generated samples of an exotic $X$ corresponding to $2^+_L$ 
as discussed in Table~\ref{table-fit-test1}.
}
\begin{tabular}{lcccccc}
\hline\hline
\vspace{-0.1cm}
& \multicolumn{3}{c}{$m_{\sss X}=250$ GeV} & \multicolumn{3}{c}{$m_{\sss X}=1$ TeV} \\
\vspace{-0.15cm}
 & generated & \multicolumn{2}{c}{fitted}  & generated & \multicolumn{2}{c}{fitted}   \\
\vspace{0.1cm}
 &  & without detector & with detector &  & without detector & with detector    \\
\hline 
$n_{\rm sig}$ & 150 & $150 \pm 13$ & $154 \pm 15$  & 150 & $151 \pm 12$ & $152 \pm 13$ \\
$f_{z2}$ & 0.400 & $0.40 \pm 0.07$  & $0.33 \pm 0.10$  & 0.300 & $0.30 \pm 0.08$ & $0.34 \pm 0.09$\\
$f_{z1}$ & 0.000 & $0.00 \pm 0.03$  & $0.00 \pm 0.04$  & 0.250 & $0.25 \pm 0.07$ & $0.23 \pm 0.07$ \\
$f_{++}$ & 0.104 & $0.10 \pm 0.04$  & $0.06 \pm 0.05$  & 0.000 & $0.00 \pm 0.01$ & $0.00 \pm 0.02$ \\
$f_{+-}$ & 0.000 & $0.00 \pm 0.04$  & $0.04 \pm 0.05$  & 0.000 & $0.00 \pm 0.01$ & $0.00 \pm 0.03$ \\
$f_{+0}$ & 0.000 & $0.00 \pm 0.02$  & $0.00 \pm 0.02$  & 0.000 & $0.00 \pm 0.02$ & $0.00 \pm 0.02$ \\
\vspace{0.1cm}
$\phi_{++}$ & $\pi$ & $3.20 \pm 0.75$ & $3.17 \pm 0.71$ & $\pi$  & free & free \\
\hline\hline
\end{tabular}
\label{table-fit-test6}
\caption{
Results of the fit for the free parameters of the spin-two hypothesis
with negative parity
with generated samples of an exotic $X$ corresponding to $2^-$ 
as discussed in Table~\ref{table-fit-test1}.
}
\begin{tabular}{lcccccc}
\hline\hline
\vspace{-0.1cm}
& \multicolumn{3}{c}{$m_{\sss X}=250$ GeV} & \multicolumn{3}{c}{$m_{\sss X}=1$ TeV} \\
\vspace{-0.15cm}
 & generated & \multicolumn{2}{c}{fitted}  & generated & \multicolumn{2}{c}{fitted}   \\
\vspace{0.1cm}
 &  & without detector & with detector &  & without detector & with detector    \\
\hline 
$n_{\rm sig}$ & 150 & $150 \pm 13$  & $152 \pm 15$ & 150 & $151 \pm 12$  & $152 \pm 13$ \\
$f_{z2}$ & 1.000 & $1.00 \pm 0.06$  & $0.97 \pm 0.12$ & 0.750 & $0.75 \pm 0.11$  & $0.81 \pm 0.15$ \\
$f_{z1}$ & 0.000 & $0.00 \pm 0.11$  & $0.05 \pm 0.20$ & 0.250 & $0.25 \pm 0.13$  & $0.19 \pm 0.24$ \\
\vspace{0.1cm}
$f_{++}$ & 0.125 & $0.12 \pm 0.06$   & $0.13 \pm 0.07$ & 0.000 & $0.00 \pm 0.07$ & $0.00 \pm 0.07$ \\
\hline\hline
\end{tabular}
\label{table-fit-test7}
\end{table}

\begin{figure}[t!]
\centerline{
\setlength{\epsfxsize}{0.30\linewidth}\leavevmode\epsfbox{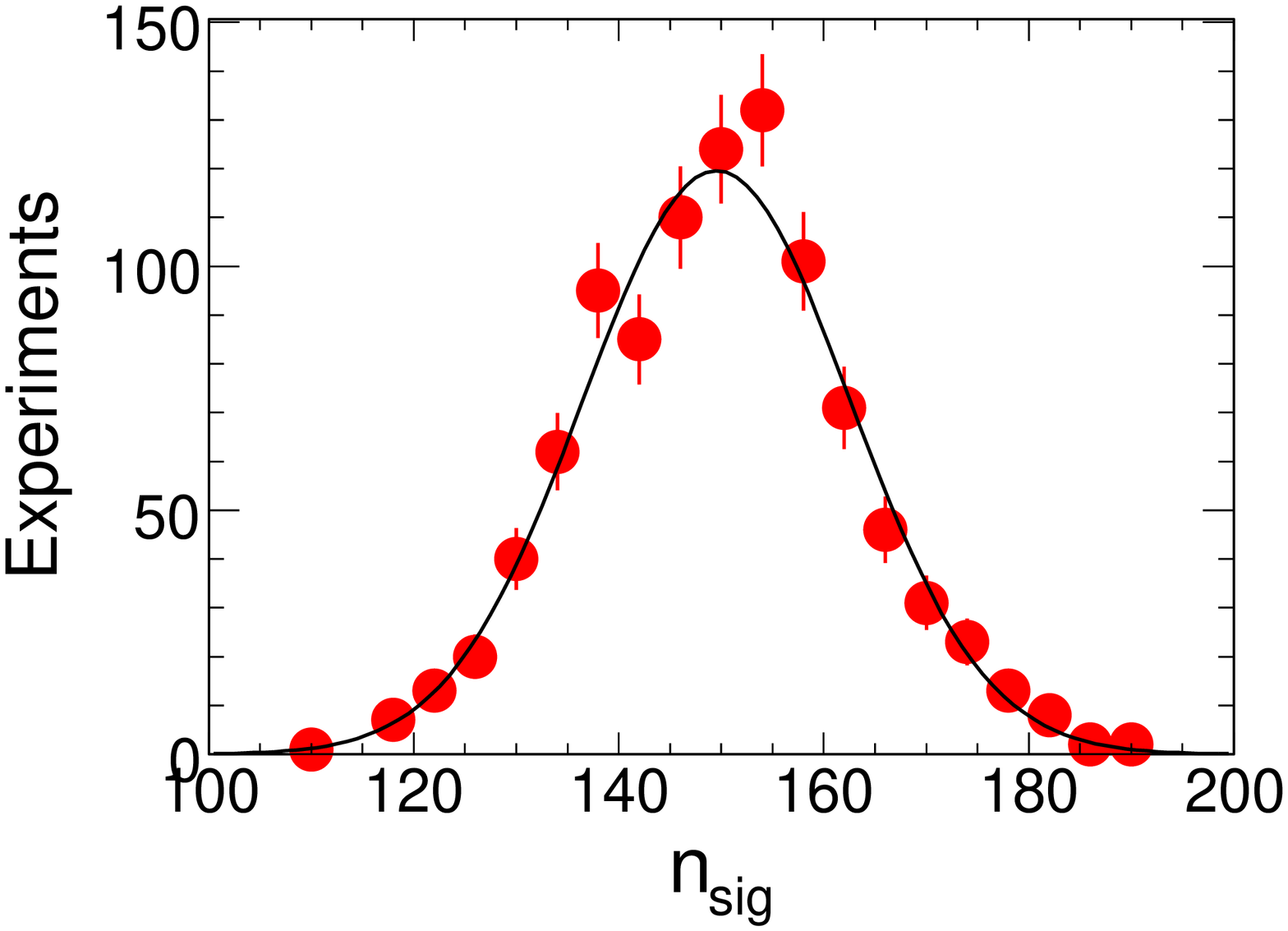}
~~~~~~~
\setlength{\epsfxsize}{0.30\linewidth}\leavevmode\epsfbox{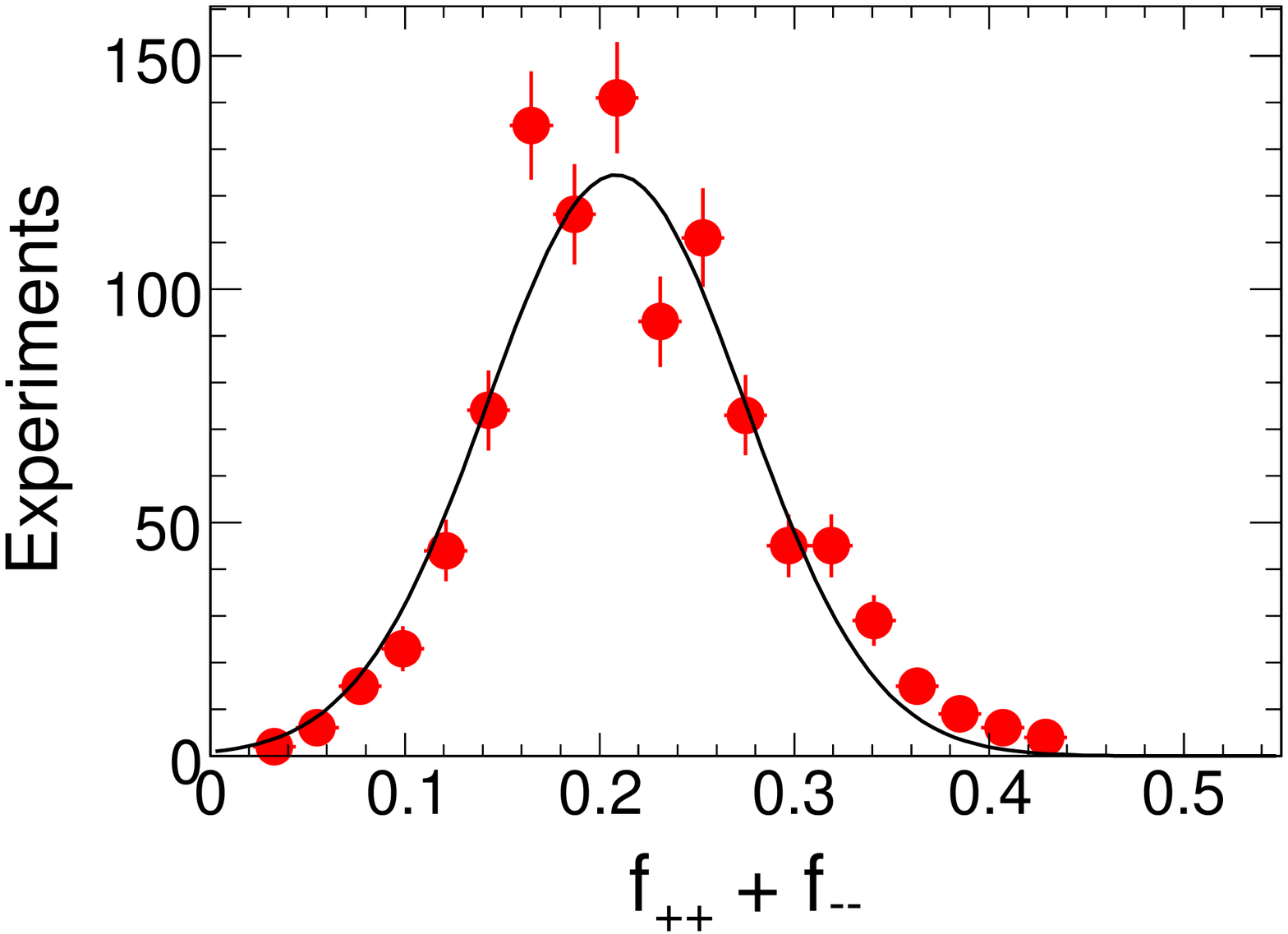}
}
\centerline{
\setlength{\epsfxsize}{0.30\linewidth}\leavevmode\epsfbox{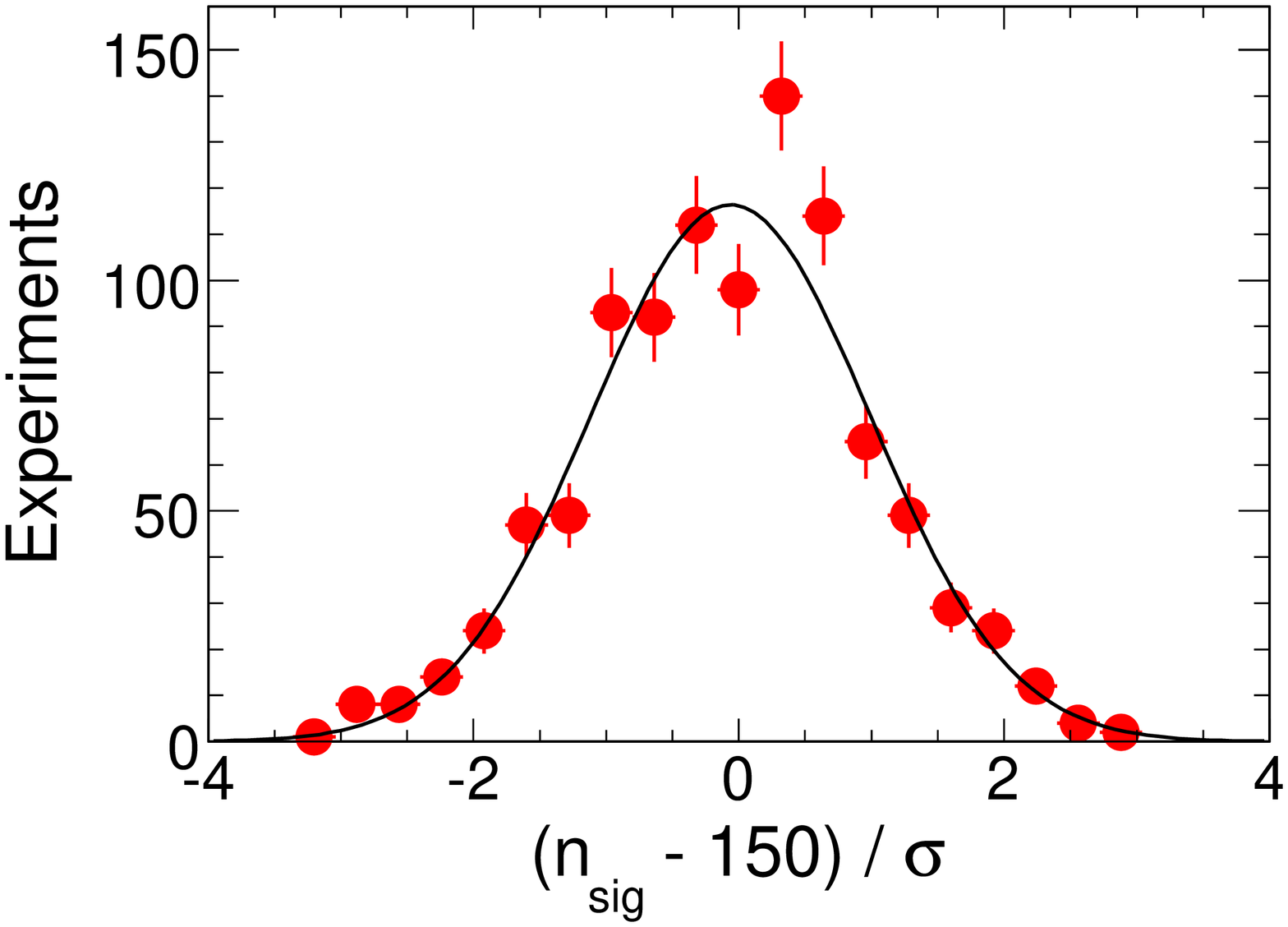}
~~~~~~~
\setlength{\epsfxsize}{0.30\linewidth}\leavevmode\epsfbox{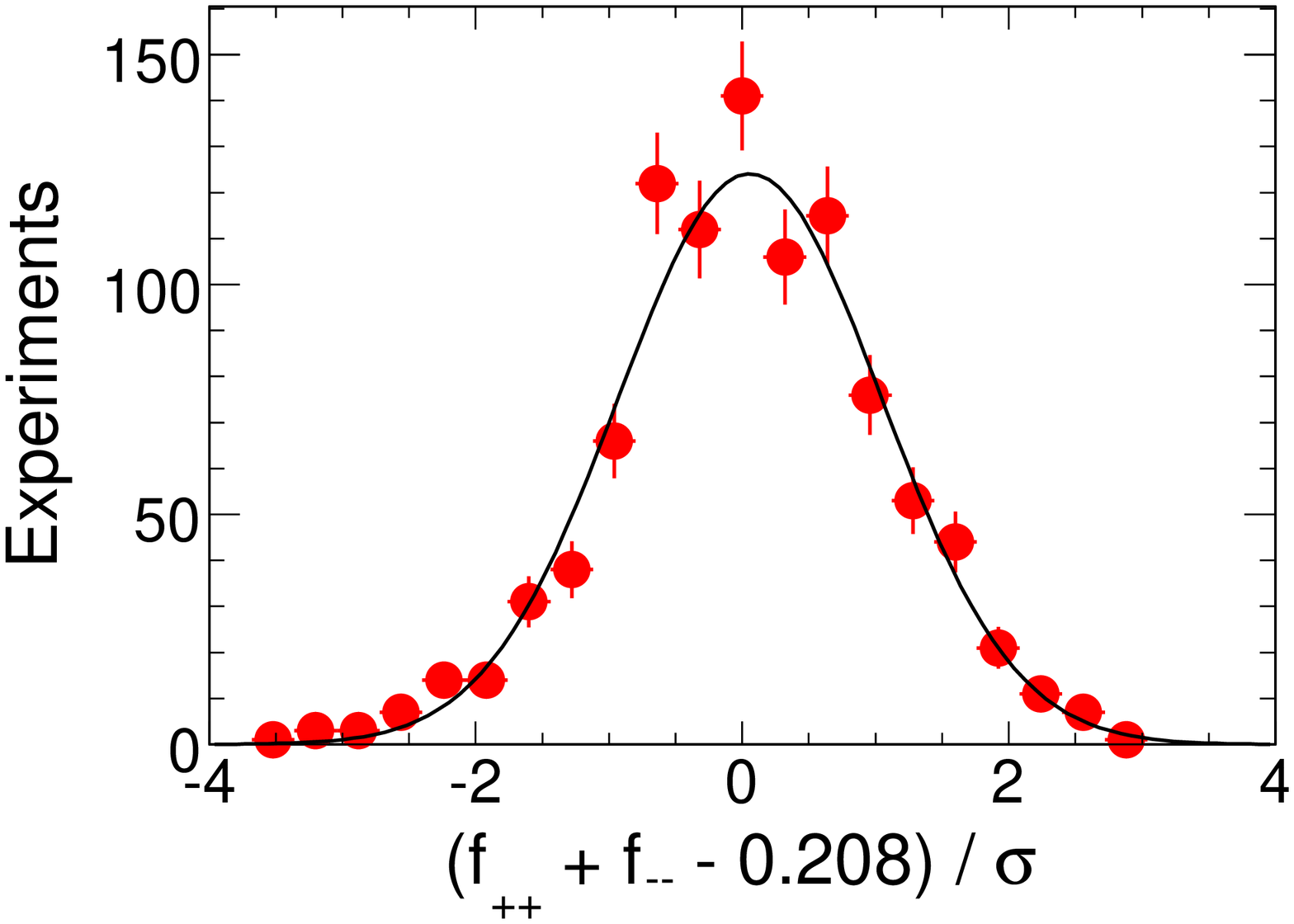}
}
\caption{
Top: distribution of the number of fitted signal events $n_{\rm sig}$ (left)
and the fraction of transverse component in the decay amplitude 
$(f_{++}+f_{--})$ (right) in 1000 generated experiments with $0^+$
hypothesis corresponding to Table~\ref{table-fit-test1}.
Bottom: distribution of the above parameters normalized by the fit errors.
}
\label{fig:nsig}
\end{figure}

The most general analysis of an observed resonance requires that  
the likelihood in Eq.~(\ref{eq:likelihood}) is  maximized with respect to 
the signal yield and all signal angular parameters, $\vec{\zeta}_{\sss J}$,
as well as with respect to unconstrained background parameters $\vec{\xi}$.
This requires a large number of free parameters even for signal alone.
The fit delivers the most probable values of these 
parameters and the covariance matrix which describes their  uncertainties and 
correlations\footnote{Knowledge of the full dependence of $-2\ln{\cal L}$ 
on all the parameters is even more beneficial.}.  
In principle, it is possible to pursue this strategy 
within our approach, but for the fit
to converge with high success rate, one may need large enough event sample.
Therefore, for the illustration of the fit  technique, we consider a limited 
procedure where  only one spin hypothesis 
is considered at a time, but all other parameters relevant 
for that particular spin hypothesis are allowed to float in the fit.

In Tables~\ref{table-fit-test1} and~\ref{table-fit-test2}, 
this technique is illustrated with the spin-zero 
hypothesis with samples generated according to 
the $0^+$ and $0^-$ scenarios, respectively. We show the results 
of the fit with both ``perfect'' and realistic detectors.
In Fig.~\ref{fig:nsig} we show distributions for two parameters in this analysis
for a set of 1000 generated experiments.
The means $\mu$ and widths $w$ of these distributions are shown in 
Table~\ref{table-fit-test1} as $\mu\pm w$. These values 
are the most probable central values and errors of a measurement, 
but they are subject to statistical fluctuation in each given experiment.
Normalized distributions in Fig.~\ref{fig:nsig} illustrate proper error estimates in the fit.
In Tables~\ref{table-fit-test3} and~\ref{table-fit-test4}, 
the result of a similar study with the spin-one hypotheses $1^+$ and $1^-$ are presented.
Finally, in Tables~\ref{table-fit-test4}, \ref{table-fit-test5}, 
and~\ref{table-fit-test6} we illustrate the fit with the spin-two hypotheses 
for $2_m^+$, $2_L^+$, and $2^-$, respectively, where we also restrict the hypothesis 
to the definite parity to reduce the number of free parameters. 
We used $1000$ generated samples of $150$ signal events each, 
Poisson distributed, and the corresponding number of background events, per sample.
In all cases the set of observables includes
$m_{\sss ZZ},\cos\theta^*,\cos\theta_1,\cos\theta_2,$ and $\Phi$.
There are several cases where contribution of a certain helicity 
amplitude  is either exactly zero (e.g. $A_{00}$ in the $0^-$ case) 
or negligible (e.g. $A_{++}$ in the $2^+_m$ case).
Since measuring the phase  of that amplitude becomes 
impossible, we leave it as ``free'' in the tables.

Coming back to the question of separating SM-like and non-SM-like $0^+$
hypotheses, it is clear from Fig.~\ref{fig:nsig} and Table~\ref{table-fit-test1}
that the hypothesis $a_1=0$ and $a_2\ne0$ in Eq.~(\ref{eq:ampl-spin0}),
corresponding to $f_{++}+f_{--}=0$, can be separated with high
confidence at 250 GeV, but no separation is possible at 1 TeV due
to the absence of a sizable transverse amplitude in both scenarios.
As another example, the separation of the $1^+$ and $1^-$ scenarios in 
Tables~\ref{table-hypothesis-test} and~\ref{table-hypothesis-test-1tev} 
is related to the measurement of $(\phi_{+0}-\phi_{0-})$ 
in Tables~\ref{table-fit-test3} and~\ref{table-fit-test4},
although there is an additional free parameter in the latter case.
The signature of the minimal coupling model of a spin-two resonance 
decay is the dominance of the $A_{+-}$ and $A_{-+}$ amplitudes, while the 
model discussed in Refs.~\cite{fitzpatrick, agashe, atwood} predicts longitudinal $A_{00}$ 
amplitude dominance. This was our motivation for comparing 
the $2^+_m$ and $2^+_L$ cases. We observe that the
angular distributions of the two models could be well separated based
on the measurement of $f_{+-}$, 
as it is also evident from Fig.~\ref{fig:generated-angles} and 
Tables~\ref{table-hypothesis-test},~\ref{table-hypothesis-test-1tev}.
There are several cases where probability   distributions 
of some parameter are  not Gaussian; typically, this occurs  when the central 
value of a  parameter is  at, or very close to, its physical boundary.  
In those cases, we still approximate the expectation value for this parameter 
and the error by  $\mu\pm w$, 
where we extrapolate the value of $w$ from the rms, but  we note 
that statistical meaning of the error estimated in such a 
way can be very different.
For this reason, the proper interpretation of the values quoted in the tables
in terms of the separation ${\cal S}$ value requires a more careful treatment 
which accounts for the non-Gaussian shape of all distributions.

The general conclusion from Tables~\ref{table-fit-test1}--\ref{table-fit-test7}
is that our analysis yields unbiased results for all the parameters that  
describe the angular distributions in the production and decay of a resonance $X$.
Indeed, it follows from those tables that, when detector effects 
are neglected,  the central values of all the parameters correspond to their 
input values. This verifies both the fit 
implementation and angular parameterizations.

Once detector effects are introduced, certain biases appear 
but they are minor. These biases come from the fact that 
we ignore angular correlations in both the detector acceptance function and 
in our parametrization of angular background distributions.
 Both of these parameterizations
are detector-specific and would need to be modeled in  each experiment.
For illustration purposes, we introduced a simplified model 
of the acceptance function in the present paper.
We pointed out that the {\it sophistication} of the 
required model for the acceptance 
function and the background parametrization depends
on the actual mass of the observed resonance. Indeed, 
it is evident from Tables~\ref{table-fit-test1}--\ref{table-fit-test7} 
that the bias is smaller at higher masses due to reduced 
angular correlations, as we explained  earlier. 
The expected statistical errors of the measurements  are  not
strongly affected by detector effects; this 
allows us to predict statistical precision in future
measurements. 

The above studies illustrate the  precision on individual parameters
that can be achieved with a sample of 150 signal events. These results 
can be easily extrapolated to an event sample of a different size. 
On the other hand, if an accurate precision estimate is required 
in a {\it particular} scenario, dedicated analysis is necessary. 
In the present paper, we  do not include the evaluation of systematic 
uncertainties beyond simplified studies of 
resolution and acceptance detector effects, but
these and other systematic effects 
are typically much smaller than statistical uncertainties 
if the signal significance is close to the discovery threshold.


\section{Conclusions}
\label{conclusions}

The  importance of the LHC physics program for the future of 
 high-energy physics requires designing  
techniques and analysis tools that are not biased towards 
a particular model of beyond the Standard Model physics. 
While it may be impractical or impossible to do so 
in many complicated cases, the case of a single resonance production 
at the LHC is sufficiently robust to aim at the most general description, 
yet simple enough to make such a description possible. 
Motivated by this, we studied production 
of a single resonance at the LHC and its decay into four-lepton final states 
in the process $pp \to X \to ZZ \to l_1 \bar l_1 l_2 \bar l_2$, 
allowing for the most general couplings of the resonance 
to Standard Model matter and gauge fields. Our goal was to understand 
if  full reconstruction of the final state  and the analysis of the 
most general 
angular  distributions can be used to  determine the spin and parity 
of the resonance and to constrain  its couplings 
to vector gauge bosons, quarks, and gluons. 

It is interesting to point out that, in some cases, generality of the 
couplings leads to new effects in the angular distributions. 
For example, imagine that the spin-two resonance is identified with the
massive graviton whose coupling to gauge and matter fields is fixed through the 
energy-momentum tensor. Production of such a resonance in gluon fusion 
occurs only with spin projections on the collision axis $J_z = \pm 2$.  
On the other hand, if a {\it general} coupling of the spin-two resonance to 
gluons is considered, production with $J_z = 0$ becomes possible as well. 
There are interesting consequences of this observation related to a possibility 
to discriminate between minimal and non-minimal couplings of the spin-two 
resonance through the analysis of angular distributions. 

Motivated by this example,  we derived the most general
angular distributions of four fermions in the process 
$pp \to X \to VV \to f_1 f^\prime_1 f_2 f^\prime_2$, considering spin-zero, spin-one,
and spin-two resonances\footnote{Angular distributions for   
two-particle decays of the $X$ bosons are  given in the appendix.}. 
We used those angular distributions to construct 
the likelihood functions for the angular analysis.  
We wrote a Monte Carlo simulation program that describes the production of the 
resonance $X$ in the process   $pp \to X \to ZZ \to l_1 \bar l_1 l_2 \bar l_2$, 
includes all spin correlations throughout the decay chains, and 
employs  the most general couplings of the resonance $X$ to matter and gauge fields 
of the Standard Model. We supplemented this analysis with a simplified (but fairly
realistic) model of the detector effects.  When all the pieces of 
our study are put together, we obtain  a powerful  analysis 
tool that allows for a realistic estimate of how much information 
the two LHC experiments will be able to extract from the study of angular 
distributions  once the resonance in the 
process $pp \to X \to ZZ \to l_1 \bar l_1 l_2 \bar l_2$ is  observed.

We find that angular distributions provide good separation of various hypotheses 
about the spin of the resonance. As we illustrate 
in Tables~\ref{table-hypothesis-test} and ~\ref{table-hypothesis-test-1tev}, 
with $30$ fully reconstructed events a typical separation of various spin hypotheses is   
${\cal S}=2-3\; \sigma$, as defined in text, but in some cases  separation as good as $4\,\sigma$ 
is achieved.  With a somewhat larger event sample, it becomes possible 
to determine helicity amplitude fractions that characterize resonance production 
and decay as well as the resonance spin from a multidimensional likelihood 
fit with decent precision. Model-independent determination of the helicity 
amplitude fractions and phases is the ultimate goal of such an analysis which 
can then be interpreted within any model of beyond the Standard Model physics. 
We provide relationships between the amplitude measurements and the fundamental
coupling constants of the resonance to matter and gauge fields.
Our studies show that such model-independent analysis is viable at the LHC. 
We look forward to its application to real LHC data.

Finally, we note that the analysis presented in this paper can be extended in a 
number of ways. A natural possibility is to allow hadronic final states (e.g. $Z \to q \bar q$) 
and/or missing energy. An obvious candidate for the latter is the decay $X \to W^+W^-$. 
We note that most of the discussion of the $X\to ZZ$ decays given in this paper also 
applies to decays $X \to W^+W^-$. However, there are important differences 
related to the fact that $W^+$ and $W^-$ are not identical particles. 
Consequently, Eq.~(\ref{eq:amplitude-identical}) does not apply in the $W^+W^-$ case 
and nine, rather than six, independent helicity amplitudes are required 
to describe $X \to W^+W^-$ decay. While the increase in the number of the 
helicity amplitudes implies that most of the formulas for angular distributions 
derived in this paper are not complete for $X \to W^+W^-$, it is interesting 
to remark that Eq.~(\ref{eq:ampl-spin0}) remains
the most general description of a spin-zero $X\to W^+W^-$ decay and 
Eq.~(\ref{eq:higgs-ang1}) is valid, with $R_1=R_2=1$.

Another aspect of the decay of the resonance $X$ to two non-identical particles 
is that forward-backward asymmetry can be generated. Of course,  
this requires that the initial state is asymmetric (e.g. $q \bar q$) as well. 
In general, the forward-backward  asymmetry manifests itself in 
the odd terms in $\cos \theta^*$ distribution. To isolate and 
measure these terms, an unambiguous definition of the {\it direction} of the 
$z$-axis is required; a suggestion on how to do that was given 
in Ref.~\cite{Dittmar1997}.  
On the experimental side, the angular  analysis of $X \to W^+W^- \to 4l$ decay  
is challenging due to the presence of two neutrinos in the final state. 
It might be beneficial to consider semileptonic final states where 
one $W$ decays hadronically and the other leptonically. However,  
this case requires detailed studies because of potentially large 
irreducible backgrounds from events with jets and missing energy. 
Similar issues should be investigated if semileptonic final states are allowed 
in $X \to ZZ$ decays. We hope to return to the discussion  
of these channels in the near future.

\bigskip
\noindent
{\bf Acknowledgments}:
We wish to thank Kaustubh Agashe for discussion of the KK graviton models.
Several of us would like to thank CMS collaboration colleagues for feedback
during the regular working group presentations of this analysis, and in particular 
Bob Cousins for discussion of the ${\cal S}$ significance estimator.
This research is partially supported by US NSF under grants PHY-0644849,  PHY-0758083, and PHY-0855365,
by the A.~P.~Sloan Foundation, and by the University Research Association. 
FNAL is operated by Fermi Research Alliance, LLC under Contract No. DE-AC02-07CH11359 with the US DOE.
We also acknowledge support by the start-up funds provided by the Johns Hopkins University (JHU). 
Calculations reported in this paper were performed on the Homewood High Performance Cluster of the JHU. 
%



\appendix

\section{Angular distributions for other decay channels}

In this appendix, the angular distributions for a resonance decaying into two particles,
including $\gamma \gamma$, $gg$,  $l^+l^-$, and $t \bar t$, are given.  
The general formula for the two-particle production and decay 
$gg$ or $q\bar{q}\to X \to P_1 P_2$ can be 
obtained from Eq.~(\ref{eq:a-15})  by  changing the spin-quantization axis 
of the $X$ to $z$ and setting $\Omega = (\Phi_1, \theta^\ast, -\Phi_1)$
and $\Omega^\ast = (0, 0, 0)$, where $\Phi_1$ is now an arbitrary azimuthal 
angle to be integrated out. The resulting formula reads
%
\begin{eqnarray}
\label{eq:production-ll}
\frac{d{\Gamma(X_J\to P_1P_2)}}{\Gamma d\cos\theta^*}
= \left(J+\frac{1}{2}\right)
\sum_{\lambda_1,\lambda_2}  f_{\lambda_1\lambda_2}
\sum_{m} f_{m}
\Bigl(d^{J}_{m,\lambda_1-\lambda_2}(\theta^*)\Bigr)^2 \,,
\label{eq:production-general}
\end{eqnarray}
%
where $\lambda_1$ and  $\lambda_2$ run over all possible helicities of
$P_1$ and $P_2$, $m$ runs over all possible $X$ spin projections,
and $f_{\lambda_1\lambda_2}=|A_{\lambda_1\lambda_2}|^2/ \sum |A_{kl}|^2$,
as defined earlier for the $X\to ZZ$ decay. 
However, this definition is more general
and includes $\lambda_i=\pm1/2$ for decays to fermions $X\to f\bar{f}$.
In the latter case, we choose the same notation as in 
Eq.~(\ref{eq:fractions}) for their parameterization with $f_{++}, f_{--}, f_{+-}, f_{-+}$, 
where we omit the 1/2 for simplicity.
Equation~(\ref{eq:production-general}) includes $f_{m}$ parameters,
where $f_{0}=f_{z0}$, $f_{\pm2}=f_{z2}/2$, and $f_{\pm1}=(f_{z1}\pm\Delta f_{z1})/2$,
allowing for a quark direction measurement in $q\bar{q}$ production
and, generally, for two non-identical particles $P_1$ and~$P_2$.

We now specialize to particular channels. 
For the $X\to\gamma\gamma$ decay channel, the angular distribution
can be obtained either from 
Eq.~(\ref{eq:production-general}) or by integrating out the
$\Phi_1$ dependence in Eq.~(\ref{eq:grav-ang2mix2d}). Note that only
$f_{++}, f_{--}$, and $f_{-+} = f_{+-}$ parameters are non-zero 
in the decay to two massless photons. Therefore
\begin{eqnarray}
\frac{16\, d\Gamma(X_{J=2}\to\gamma\gamma)}{5\, \Gamma d\cos\theta^\ast} 
&&   \!\!\!\!  = (2-2f_{z1}+f_{z2}) -6(2 -4f_{z1} -f_{z2})\cos^2\theta^\ast +3(6 -10f_{z1}-5f_{z2})\cos^4\theta^\ast \nonumber \\
&& +f_{+-}\,\left\{(2+2f_{z1}-7f_{z2})+6(2-6f_{z1}+f_{z2})\cos^2\theta^\ast-5(6-10f_{z1}-5f_{z2})\cos^4\theta^\ast \right\}\,.
\label{eq:grav-twophoton}
\end{eqnarray}
This equation describes the most general case. The special case of the minimal
coupling in both production and decay corresponds to $f_{z1}+f_{z2}=1$ and $f_{+-}=f_{-+}=1/2$. 
In this case, one obtains $(1+6\cos^2\theta^*+\cos^4\theta^*)$ for the $gg$ production mechanism
with $f_{z2}=1$ and  $(1-\cos^4\theta^*)$ for the $q\bar{q}$ production mechanism with $f_{z1}=1$.
Equation~(\ref{eq:grav-twophoton}) is also applicable to the decay to two gluon jets
$X_{J=2}\to gg$, but additional constraints could be used when the production
and decay mechanisms are the same: $(1-f_{z1}-f_{z2})/f_{z2} = (1-f_{+-}-f_{-+})/(f_{+-}+f_{-+})$.

For the decay to  a fermion-antifermion pair, we obtain
%
\begin{eqnarray}
\label{eq:production-ll-1}
\frac{8\, d{\Gamma(X_{J=1}\to f\bar{f})}}{3\, \Gamma d\cos\theta^*} = 
(f_{+-}+f_{-+}) (1+\cos^2\theta^*) +2\,(1-f_{+-}-f_{-+}) (1-\cos^2\theta^*) +2\, (f_{+-}-f_{-+})\Delta f_{z1}\cos\theta^* \,,
\end{eqnarray}
\begin{eqnarray}
\label{eq:production-ll-2}
\frac{16\, d{\Gamma(X_{J=2}\to f\bar{f})}}{5\, \Gamma d\cos\theta^*} 
&& \!\!\!\!  = 
 2\,(f_{+-}+f_{-+}) \,\left\{ (f_{z1}+f_{z2})+3(2-3f_{z1}-2f_{z2})\cos^2\theta^\ast-(6-10f_{z1}-5f_{z2})\cos^4\theta^\ast \right\}\nonumber \\
&& + (1-f_{+-}-f_{-+})\,\left\{(2-2f_{z1}+f_{z2}) -6(2 -4f_{z1} -f_{z2})\cos^2\theta^\ast +3(6 -10f_{z1}-5f_{z2})\cos^4\theta^\ast \right\} \nonumber \\
&& -4\,(f_{+-}-f_{-+}) \Delta f_{z1} (\cos\theta^\ast-2\cos^3\theta^\ast ) \,.
\end{eqnarray}
where for a massless fermion in the final state
$(f_{++}+f_{--}) = (1-f_{+-}-f_{-+})=0$, which would describe the decay $X\to l^+l^-$.
It follows from this formula that there is a forward-backward asymmetry in this 
decay, as was pointed out in Ref.~\cite{rosner1996} in the context of 
spin-one decays to a fermion pair. 
A dilution factor needs to be introduced in front of the $\Delta f_{z1}$ terms, 
which depends on the ability to measure the sign of $\cos\theta^\ast$ in an experiment.
The special case of the minimal coupling in gluon
fusion corresponds to $f_{z1}+f_{z2}=1$.


\section{Supporting material}

Supporting material for this analysis may be found in Ref.~\cite{support}, where we provide
the Monte Carlo simulation program and the most general angular distributions used in
this analysis. For completeness, we present the general angular distribution in the 
production and decay of a spin-$J$ particle $X$ in parton collisions
$ab\to X\to ZZ\to (f_1\bar{f}_1)(f_2\bar{f}_2)$. In order to simplify expressions, 
we redefine the fifth angle from $\Phi_1$ to $\Psi=\Phi_1+\Phi/2$, 
which can be interpreted as the angle between the production plane and
the average between the two decay planes shown in Fig.~\ref{fig:decay}. 
%
\begin{eqnarray}
&&\frac{{\cal{N}}_J \; d \Gamma_{J}}{ \;\Gamma d\cos\theta^\ast d\Psi d\cos\theta_1d\cos\theta_2 d\Phi } = 
 \nonumber \\
&& F^J_{00} (\theta^\ast) \times  \Bigl\{  4\, f_{00}\,\sin^2\theta_1 \sin^2\theta_2 
     +(f_{++}+f_{--})\left((1+\cos^2\theta_1)(1+\cos^2\theta_2)+4R_{1}R_{2}\cos\theta_1\cos\theta_2\right) \nonumber \\
&& ~~ ~~ ~~  ~~ ~~ ~~ ~~ -2\,(f_{++}-f_{--}) \left(R_{1}\cos\theta_1(1+\cos^2\theta_2) + R_{2}(1+\cos^2\theta_1)\cos\theta_2\right)   \nonumber \\
&& ~~ ~~ ~~  ~~ ~~ ~~ ~~ +4 \sqrt{f_{++}f_{00}} \, (R_1-\cos\theta_1)\sin\theta_1(R_2-\cos\theta_2)\sin\theta_2\cos ( \Phi + \phi_{++} ) \nonumber \\
&& ~~ ~~ ~~  ~~ ~~ ~~ ~~ +4 \sqrt{f_{--}f_{00}} \,  (R_1+\cos\theta_1)\sin\theta_1(R_2+\cos\theta_2)\sin\theta_2\cos ( \Phi - \phi_{--} ) \nonumber \\
&& ~~ ~~ ~~  ~~ ~~ ~~ ~~ +2\sqrt{f_{++}f_{--}}\sin^2\theta_1\sin^2\theta_2 \cos(2\Phi+\phi_{++}-\phi_{--})  \Bigl\} \nonumber \\
&& + 4 F^J_{11} (\theta^\ast)\times \Bigl\{ (f_{+0}+f_{0-})(1-\cos^2\theta_1\cos^2\theta_2) 
   -(f_{+0} - f_{0-})( R_1 \cos\theta_1\sin^2\theta_2 + R_2\sin^2\theta_1 \cos\theta_2)  \nonumber \\
&&  ~~ ~~  ~~  ~~ ~~ ~~ ~~ + 2  \sqrt{f_{+0}f_{0-}}\sin\theta_1\sin\theta_2(R_1R_2 - \cos\theta_1\cos\theta_2) \cos (\Phi+\phi_{+0}-\phi_{0-}) \Bigr\} \nonumber \\ 
&& +(-1)^J \times 4 F^J_{-11} (\theta^\ast)\times \Bigl\{  (f_{+0}+f_{0-})(R_1R_2+\cos\theta_1\cos\theta_2)- (f_{+0}-f_{0-})(R_1\cos\theta_2+R_2\cos\theta_1) \nonumber \\ 
&&  ~~ ~~  ~~  ~~ ~~ ~~ ~~ + 2\sqrt{f_{+0}f_{0-}}\sin\theta_1\sin\theta_2\cos(\Phi+\phi_{+0}-\phi_{0-}) \Bigr\} \sin\theta_1\sin\theta_2\cos(2\Psi)\nonumber \\
&& + 2 F^J_{22}(\theta^\ast)\times f_{+-}  \Bigl\{ (1+\cos^2\theta_1)(1+\cos^2\theta_2) - 4R_1R_2\cos\theta_1\cos\theta_2  \Bigr\} \nonumber \\
&& + (-1)^J \times 2 F^J_{-22} (\theta^\ast)\times f_{+-} \sin^2\theta_1 \sin^2\theta_2\cos (4\Psi) \nonumber  \\
&& +2 F^J_{02}(\theta^\ast) \times \Bigl\{ 2 \sqrt{f_{00}f_{+-}} ~\sin\theta_1 \sin\theta_2  \times \Bigl[ (R_1 -\cos\theta_1)(R_2+\cos\theta_2)\cos (2\Psi-\phi_{+-})  \nonumber \\
&&  ~~ ~~  ~~  ~~ ~~ ~~ ~~  ~~ ~~  ~~ + (R_1 +\cos\theta_1)(R_2-\cos\theta_2)\cos (2\Psi+\phi_{+-})\Bigr] \nonumber \\
&& ~~ ~~  ~~  ~~ ~~ ~~ ~~ + \sqrt{f_{++}f_{+-}} \Bigl[\sin^2\theta_1(1-2R_2\cos\theta_2+\cos^2\theta_2)\cos(2\Psi-\Phi+\phi_{+-}-\phi_{++}) \nonumber \\
&& ~~ ~~  ~~  ~~ ~~ ~~ ~~ ~~ ~~  ~~ +(1-2R_1\cos\theta_1+\cos^2\theta_1) \sin^2\theta_2\cos(2\Psi+\Phi-\phi_{+-}+\phi_{++}) \Bigr] \nonumber \\
&& ~~ ~~  ~~  ~~ ~~ ~~ ~~ + \sqrt{f_{--}f_{+-}} \Bigl[\sin^2\theta_1(1+2R_2\cos\theta_2+\cos^2\theta_2)\cos(2\Psi-\Phi-\phi_{+-}+\phi_{--}) \nonumber \\
&& ~~ ~~  ~~  ~~ ~~ ~~ ~~ ~~ ~~  ~~ + (1+2R_1\cos\theta_1+\cos^2\theta_1)\sin^2\theta_2\cos(2\Psi+\Phi+\phi_{+-}-\phi_{--}) \Bigr]  \Bigr\}  \nonumber \\
&& -2\sqrt{2}~F^J_{01}(\theta^\ast) \times \Bigl\{ 2\sqrt{f_{00}f_{+0}} \Bigl[ \sin\theta_1(R_1-\cos\theta_1) \sin^2\theta_2\cos(\Psi-\Phi/2-\phi_{+0}) \nonumber \\
&& ~~ ~~  ~~  ~~ ~~ ~~ ~~ ~~Ê~~ -\sin^2\theta_1\sin\theta_2(R_2-\cos\theta_2) \cos(\Psi+\Phi/2+\phi_{+0})\Bigr] \nonumber \\
&& ~~ ~~  ~~  ~~ ~~ ~~ ~~ + 2\sqrt{f_{00}f_{0-}}  \Bigl[ \sin^2\theta_1\sin\theta_2(R_2+\cos\theta_2) \cos(\Psi+\Phi/2-\phi_{0-}) \nonumber \\
&& ~~ ~~  ~~  ~~ ~~ ~~ ~~ ~~ ~~ -  \sin\theta_1(R_1 +\cos\theta_1)\sin^2\theta_2\cos(\Psi-\Phi/2+\phi_{0-}) \Bigr]  \nonumber \\
&&~~ ~~  ~~  ~~ ~~ ~~ ~~ + \sqrt{f_{++}f_{+0}} \Bigl[ (1-2R_1\cos\theta_1+\cos^2\theta_1)\sin\theta_2(R_2 -\cos\theta_2) \cos(\Psi+\Phi/2+\phi_{++}-\phi_{+0}) \nonumber \\
&& ~~ ~~  ~~  ~~ ~~ ~~ ~~ ~~ ~~ -  \sin\theta_1(R_1-\cos\theta_1)(1-2R_2\cos\theta_2+\cos^2\theta_2)\cos(\Psi-\Phi/2-\phi_{++}+\phi_{+0})  \Bigr] \nonumber \\
&& ~~ ~~  ~~  ~~ ~~ ~~ ~~ + \sqrt{f_{++}f_{0-}} \Bigl[  \sin\theta_1(R_1-\cos\theta_1)\sin^2\theta_2\cos(\Psi+3\Phi/2+\phi_{++}-\phi_{0-}) \nonumber \\
&& ~~ ~~  ~~  ~~ ~~ ~~ ~~ ~~ ~~  ~~ - \sin^2\theta_1 \sin\theta_2(R_2-\cos\theta_2) \cos(\Psi-3\Phi/2-\phi_{++}+\phi_{0-})  \Bigr] \nonumber \\
&& ~~ ~~  ~~  ~~ ~~ ~~ ~~  + \sqrt{f_{--}f_{+0}} \Bigl[  \sin^2\theta_1 \sin\theta_2(R_2+\cos\theta_2)\cos(\Psi-3\Phi/2+\phi_{--}-\phi_{+0}) \nonumber \\
&& ~~ ~~  ~~  ~~ ~~ ~~ ~~ ~~ ~~  ~~ -  \sin\theta_1(R_1+\cos\theta_1)\sin^2\theta_2\cos(\Psi+3\Phi/2-\phi_{--}+\phi_{+0})  \Bigr] \nonumber \\
&& ~~ ~~  ~~  ~~ ~~ ~~ ~~  + \sqrt{f_{--}f_{0-}} \Bigl[  \sin\theta_1 (R_1+\cos\theta_1)(1+2R_2\cos\theta_2+\cos^2\theta_2) \cos( \Psi - \Phi/2 + \phi_{--} - \phi_{0-}) \nonumber \\
&& ~~ ~~  ~~  ~~ ~~ ~~ ~~ ~~ ~~  ~~ -  (1+2R_1\cos\theta_1+\cos^2\theta_1)\sin\theta_2 (R_2+\cos\theta_2) \cos(\Psi+\Phi/2-\phi_{--}+\phi_{0-}) \Bigr] \Bigr\}  \nonumber \\
&& - 2\sqrt{2}~F^J_{12}(\theta^\ast) \times \Bigl\{ \sqrt{f_{+-}f_{+0}} \Bigl[ (1-2R_1\cos\theta_1+\cos^2\theta_1)\sin\theta_2 (R_2+\cos\theta_2)\cos(\Psi+\Phi/2-\phi_{+-}+\phi_{+0}) \nonumber \\
&& ~~ ~~  ~~  ~~ ~~ ~~ ~~  ~~ ~~  ~~ -\sin\theta_1(R_1+\cos\theta_1)(1-2R_2\cos\theta_2+\cos^2\theta_2)\cos(\Psi-\Phi/2+\phi_{+-}-\phi_{+0})\Bigr]  \nonumber \\
&& ~~ ~~  ~~  ~~ ~~ ~~ ~~  + \sqrt{f_{+-}f_{0-}} \Bigl[ \sin\theta_1 (R_1-\cos\theta_1) (1+2R_2\cos\theta_2+\cos^2\theta_2) \cos(\Psi-\Phi/2-\phi_{+-}+\phi_{0-})\nonumber \\
&& ~~ ~~  ~~  ~~ ~~ ~~ ~~ ~~ ~~  ~~  -(1+2R_1\cos\theta_1+\cos^2\theta_1)\sin\theta_2(R_2-\cos\theta_2)\cos(\Psi+\Phi/2+\phi_{+-}-\phi_{0-})\Bigr] \Bigr\} \nonumber \\
&& -(-1)^J \times 2\sqrt{2}~F^J_{-12} (\theta^\ast) \times \Bigl\{ \sqrt{f_{+-}f_{+0}} \Bigl[  \sin\theta_1  (R_1-\cos\theta_1)\sin^2\theta_2\cos(3\Psi+\Phi/2-\phi_{+-}+\phi_{+0})\nonumber \\
&& ~~ ~~  ~~  ~~ ~~ ~~ ~~ ~~ ~~  ~~ -  \sin^2\theta_1 \sin\theta_2(R_2-\cos\theta_2)\cos(3\Psi-\Phi/2+\phi_{+-}-\phi_{+0}) \Bigr]  \nonumber \\
&& ~~ ~~  ~~  ~~ ~~ ~~ ~~ + \sqrt{f_{+-}f_{0-}}\Bigl[  \sin^2\theta_1 \sin\theta_2(R_2+\cos\theta_2) \cos(3\Psi-\Phi/2-\phi_{+-}+\phi_{0-}) \nonumber \\
&& ~~ ~~  ~~  ~~ ~~ ~~ ~~ ~~ ~~  ~~ -  \sin\theta_1(R_1+\cos\theta_1)\sin^2\theta_2\cos(3\Psi+\Phi/2+\phi_{+-}-\phi_{0-})\Bigr] \Bigr\}\,,
\label{eq:mixedJtotal}
\end{eqnarray}
%
where ${\cal N}_J$ is the normalization factor which does not affect the angular 
distributions and the functions $F^{J}_{ij}(\theta^\ast)$ are defined as follows
\begin{eqnarray}
F^J_{ij} (\theta^\ast) = \sum_{m=0,\pm1,\pm2} f_m \, d^J_{m i} (\theta^\ast) d^J_{m j} (\theta^\ast) \,,
\label{eq:formfactors}
\end{eqnarray}
where $f_{\pm1}=f_{z1}/2$, $f_{\pm2}=f_{z2}/2$, and $f_{0} =f_{z0} =(1-f_{z1} -f_{z2})$. 
Note that for odd $J$ one has $f_{00}=f_{++}=f_{--}=0$, and therefore $F^{J=\rm odd}_{0j}(\theta^\ast)$ 
terms do not contribute.

Below we show $F^J_{ij} (\theta^\ast)$ explicitly for $J=0, 1,$ and $2$.
For spin-zero, we have
\begin{eqnarray}
 && F^0_{00}= 1 \,, \nonumber\\ 
 && F^0_{11}=F^0_{-11}=F^0_{22}=F^0_{-22}=F^0_{02}=F^0_{01}=F^0_{12}=F^0_{-12}=0 \,,
\label{eq:mixed-spin0}
\end{eqnarray}
where only four parameters$f_{++}$, $f_{--}$, $\phi_{++}$, and $\phi_{--}$  remain relevant,
and $f_{00}$ can be expressed as $f_{00}=(1-f_{++}-f_{--})$.
For spin-one, we have
\begin{eqnarray}
 && F^1_{11}= \frac{1}{4} (1+ \cos^2\theta^\ast)  \,, \nonumber\\ 
 && F^{1}_{-11}= \frac{1}{4} (1-\cos^2\theta^\ast)  \,, \nonumber\\ 
 && F^1_{00}=F^1_{22}=F^1_{-22}=F^1_{02}=F^1_{01}=F^1_{12}=F^1_{-12}=0 \,,
\label{eq:mixed-spin1}
\end{eqnarray}
where only two parameters $f_{+0}$ and $[\phi_{+0}-\phi_{0-}]$  remain relevant,
and $f_{0-}$ can be expressed as $f_{0-}=(1-2f_{+0})/2$.
Finally, for spin-two, we have
\begin{eqnarray}
&& F^2_{00}= \frac{1}{8}  \Bigl\{ (2-2f_{z1}+f_{z2})-6(2-4f_{z1}-f_{z2})\cos^2\theta^\ast+3(6-10f_{z1}-5f_{z2})\cos^4\theta^\ast\Bigr\} 
  \,, \nonumber\\ 
 && F^2_{11}= \frac{1}{4} \Bigl\{ (f_{z1} + f_{z2}) + 3(2-3f_{z1}-2f_{z2}) \cos^2\theta^\ast - (6-10f_{z1}-5f_{z2})  \cos^4\theta^\ast\Bigr\}
  \,, \nonumber\\ 
 && F^{2}_{-11}= -\frac{1}{4}  \Bigl\{ (f_{z1}-f_{z2}) + (6-10f_{z1}-5f_{z2}) \cos^2\theta^\ast \Bigr\} \sin^2\theta^\ast
  \,, \nonumber\\ 
  && F^2_{22}= \frac{1}{16} \Bigl\{ (6-2f_{z1}-5f_{z2})-6(2-2f_{z1}-3f_{z2}) \cos^2\theta^\ast+(6-10f_{z1}-5f_{z2}) \cos^4\theta^\ast\Bigr\}  
  \,, \nonumber\\ 
 &&  F^{2}_{-22}= \frac{1}{16} \Bigl\{ 6-10f_{z1} -5f_{z2} \Bigr\}  \sin^4\theta^\ast 
  \,, \nonumber\\ 
 &&  F^2_{02}=-\frac{1}{8} \sqrt{\frac{3}{2}}  \Bigl\{ (2-2f_{z1}-3f_{z2}) - (6-10f_{z1}-5f_{z2}) \cos^2\theta^\ast \Bigr\} \sin^2\theta^\ast
  \,, \nonumber\\ 
 &&  F^2_{01}= -\frac{\sqrt{6}}{8} \Bigl\{ (2-4f_{z1}-f_{z2})-(6-10f_{z1}-5f_{z2}) \cos^2\theta^\ast  \Bigr\} \cos\theta^\ast \sin\theta^\ast
  \,, \nonumber\\ 
 &&  F^2_{12}= \frac{1}{8} \Bigl\{ (6-6f_{z1}-9f_{z2})-(6-10f_{z1}-5f_{z2})\cos^2\theta^\ast \Bigr\} \cos\theta^\ast \sin\theta^\ast
  \,, \nonumber\\ 
 &&  F^{2}_{-12}= -\frac{1}{8}  (6-10f_{z1}-5f_{z2})\cos\theta^\ast \sin^3\theta^\ast
 \,,
\label{eq:mixed-spin2}
\end{eqnarray}
where all parameters contribute and again $f_{00}$ can be expressed as 
$f_{00}=(1-f_{++}-f_{--}-2f_{+0}-2f_{0-}-2f_{+-})$.




\end{document}